\documentclass[prl,twocolumn,superscriptaddress,nobibnotes]{revtex4-2}
\usepackage[utf8]{inputenc}
\usepackage[version=4]{mhchem}
\usepackage{multirow}
\usepackage{graphicx}
\usepackage{physics}
\usepackage{amssymb}
\usepackage{textgreek}
\usepackage{gensymb}
\usepackage{float}
\usepackage{titletoc}%!needs to be loaded before hyperref!
\usepackage{hyperref}
\usepackage[utf8]{inputenc}

\hypersetup{
    colorlinks,
    citecolor=black,
    filecolor=black,
    linkcolor=black,
    urlcolor=black
}

\makeatletter
\def\maketitle{
\@author@finish
\title@column\titleblock@produce
\suppressfloats[t]}
\makeatother

\begin{document}

\title{Measured potential profile in a quantum anomalous Hall system suggests bulk-dominated current flow}
%OR: is consistent with bulk-dominated current flow

\author{Ilan T. Rosen}
\altaffiliation[Current address: ]{Research Laboratory of Electronics, Massachusetts Institute of Technology, Cambridge, MA 02139, USA}
\email{itrosen@mit.edu}
\affiliation{Department of Applied Physics, Stanford University, Stanford, California 94305, USA}
\affiliation{Stanford Institute for Materials and Energy Sciences, SLAC National Accelerator Laboratory, Menlo Park, California 94025, USA}
\author{Molly P. Andersen}
\affiliation{Department of Materials Science and Engineering, Stanford University, Stanford, California 94305, USA}
\affiliation{Stanford Institute for Materials and Energy Sciences, SLAC National Accelerator Laboratory, Menlo Park, California 94025, USA}
\author{Linsey K. Rodenbach}
\affiliation{Department of Physics, Stanford University, Stanford, California 94305, USA}
\affiliation{Stanford Institute for Materials and Energy Sciences, SLAC National Accelerator Laboratory, Menlo Park, California 94025, USA}
\author{Lixuan Tai}
\affiliation{Department of Electrical Engineering, University of California, Los Angeles, California 90095, USA}
\author{Peng Zhang}
\affiliation{Department of Electrical Engineering, University of California, Los Angeles, California 90095, USA}
\author{Kang L. Wang}
\affiliation{Department of Electrical Engineering, University of California, Los Angeles, California 90095, USA}
\author{M. A. Kastner}
\affiliation{Department of Physics, Stanford University, Stanford, California 94305, USA}
\affiliation{Stanford Institute for Materials and Energy Sciences, SLAC National Accelerator Laboratory, Menlo Park, California 94025, USA}
\affiliation{Department of Physics, Massachusetts Institute of Technology, Cambridge, Massachusetts 02139, USA}
\author{David Goldhaber-Gordon}
\email{goldhaber-gordon@stanford.edu}
\affiliation{Department of Physics, Stanford University, Stanford, California 94305, USA}
\affiliation{Stanford Institute for Materials and Energy Sciences, SLAC National Accelerator Laboratory, Menlo Park, California 94025, USA}

\begin{abstract}
Ideally, quantum anomalous Hall systems should display zero longitudinal resistance. Yet in experimental quantum anomalous Hall systems elevated temperature can make the longitudinal resistance finite, indicating dissipative flow of electrons. Here, we show that the measured potentials at multiple locations within a device at elevated temperature are well-described by solution of Laplace's equation, assuming spatially-uniform conductivity, suggesting non-equilibrium current flows through the two-dimensional bulk. Extrapolation suggests that at even lower temperatures current may still flow primarily through the bulk rather than, as had been assumed, through edge modes. An argument for bulk current flow previously applied to quantum Hall systems supports this picture.

\end{abstract}

\maketitle

The quantum anomalous Hall (QAH) effect ideally features longitudinal resistivity $\rho_{xx}$ that vanishes as the Hall resistivity $\rho_{xy}$ approaches $\pm h/e^2$. Such quantized Hall resistivity has been confirmed with high precision in magnetically-doped films of the topological insulator \ce{(Bi,Sb)2Te3} at zero or low magnetic field~\cite{bestwick2015,chang2015,fox2018, gotz2018, okazaki2020,okazaki2021}. The onset of dissipation, where $\rho_{xx}$ becomes finite and $\rho_{xy}$ departs from $h/e^2$, may be induced through increasing the temperature, increasing the source-drain bias, or electrostatic gating. Understanding the dissipative state is crucial for quantifying and improving material quality, with the goal of engineering materials in which the anomalous Hall effect is quantized at higher temperatures.

The QAH system has been theoretically shown to host a chiral edge mode (CEM)~\cite{liu2008,yu2010} associated with the system's nonzero Chern number. Scanning measurements of local microwave impedance appear to show conducting edge modes~\cite{allen2019}. The general understanding in the field has been that non-equilibrium current---current flowing in response to applied source-drain bias, as distinct from the persistent circulating current---flows through the CEM~\cite{wang2013, kou2014, chang2015, changPRL2015,bestwick2015,fox2018,gotz2018, allen2019, okazaki2020,okazaki2021}. This understanding originates in early pictures of the quantum Hall (QH) system~\cite{halperin1982, buttiker1988}. Yet a recent study of transport in QAH Hall bar and Corbino geometries at high bias identified that dissipation occurs through the two-dimensional bulk, requiring at least some bulk current flow in dissipative regimes~\cite{rodenbach2021}. That study suggests the need for further inquiry into the conditions under which current flows in the bulk versus through the edges, and whether dissipation in the bulk is spatially uniform or is higher in particular areas of devices. 

In this work, we measure the longitudinal voltage at various points along the edge of a current-biased Hall bar in the QAH state at the onset of dissipation. We find that the longitudinal electric field is not uniform, but rather varies monotonically along each edge of the Hall bar. We find that, when dissipation is induced by increased temperature, the spatial profile of the measured potential in the QAH system nearly perfectly matches numerical simulations of Laplace's equation that assume spatially uniform longitudinal and Hall conductivity within the device geometry. Mirroring arguments borrowed from the QH literature~\cite{macdonald1983, hirai1994, komiyama1996representations, weis2011}, we deduce that the source-drain bias manifests primarily as a transverse electric field that drives current through the two-dimensional bulk. That this holds down to $\rho_{xx}\approx 0.01 h/e^2$ (the lowest we can probe with small enough uncertainties) raises the possibility that even in the dissipationless regime, non-equilibrium current in the QAH system may flow through the bulk of the material and not, as has been generally assumed, through chiral edge modes.

\begin{figure}
\centering
	\includegraphics[width=0.48\textwidth]{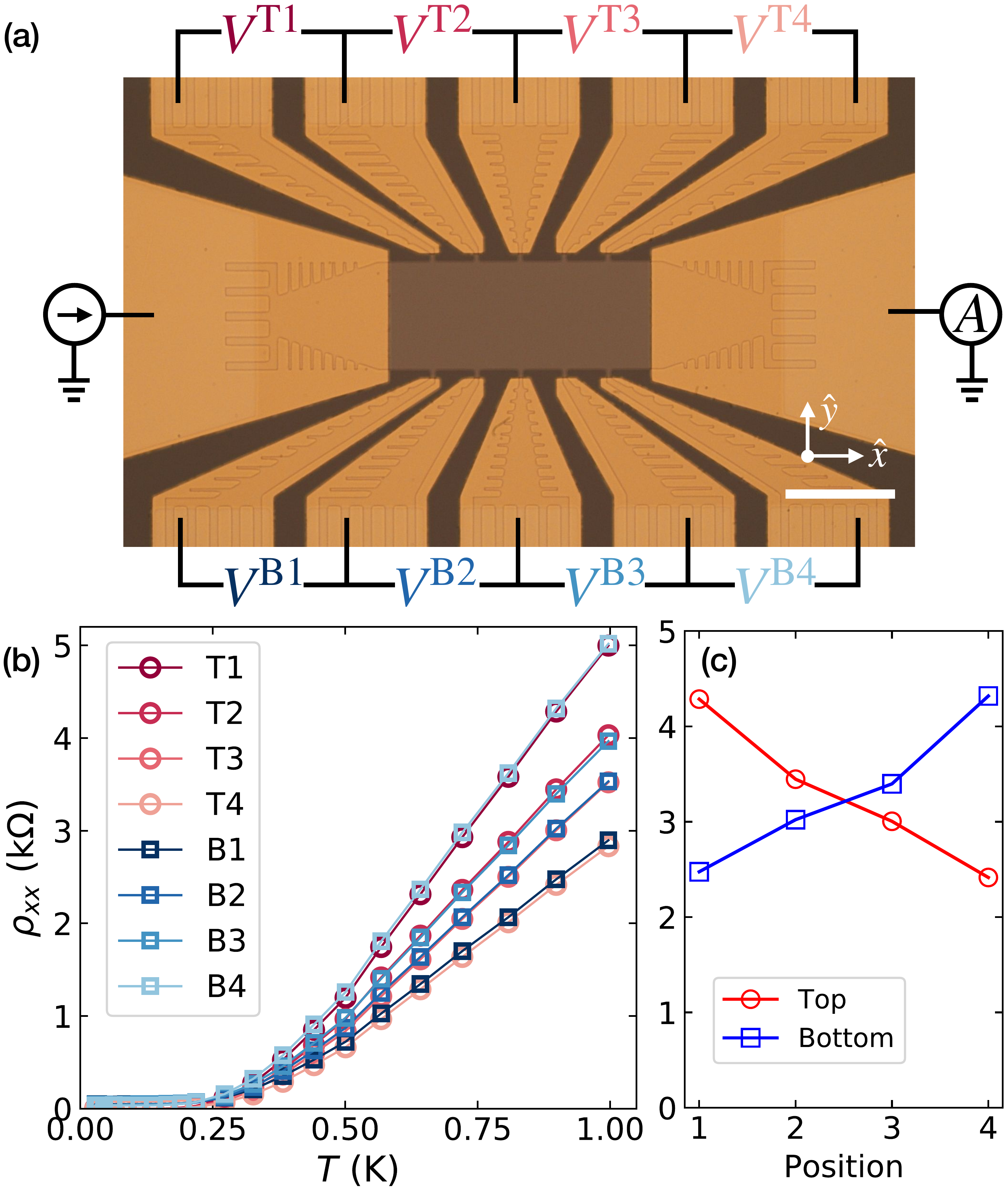}
	\caption{(a) Optical micrograph of the Hall bar prior to top gate metal deposition. Current injection at the source terminal (left) and current measurement at the drain terminal (right) are indicated, along with the eight longitudinal voltage measurements $V_{xx}^{\alpha i}$ enabled by the ten voltage taps. Coordinate axes are indicated. Scale bar, 100~$\mu$m. (b) The apparent resistivities as a function of temperature at the optimum gate voltage. (c) Apparent resistivity shown as a function of position on the top and bottom edges of the Hall bar at 898~mK.}
	\label{fig1}
\end{figure}

\section{Methods}

Here we study low-frequency electronic transport in a Hall bar fabricated from a 6~nm film of \ce{(Bi_x Sb_{1-x})2Te3}, with layers near the top and bottom heavily doped with \ce{Cr}. The fabrication process, described in Ref.~\cite{rodenbach2021}, includes an electrostatic top gate with an alumina dielectric to control the Fermi level. The Hall bar features five voltage taps spaced evenly along the top edge and five more along the bottom edge, so that the longitudinal voltage drop can be measured across four consecutive segments along each edge (Fig.~\ref{fig1}(a)). We define the apparent differential resistivities
\begin{equation}
\rho_{xx}^{\alpha i}=\frac{W}{L} \frac{dV^{\alpha i}}{dI},
\end{equation}
where $\alpha=\mathrm{T, B}$ indexes the top and bottom edges of the Hall bar, $i=1, 2, 3, 4$ indexes the four measurement segments across the Hall bar from left to right, $W=100$~$\mu$m is the width of the Hall bar, and $L=40$~$\mu$m is the length between voltage taps for each measurement segment. A 5~nA ac current bias is applied to the source terminal at the left edge of the device, and a dc current bias is added in measurements where indicated. The differential longitudinal voltages $dV^{\alpha i}$ and the ac bias current $dI$ are amplified and measured simultaneously with separate lock-in amplifiers~\cite{sup}.

Measurements are made in a dilution refrigerator with a base temperature of 28~mK at zero external field after magnetizing the film with an out-of-plane field of 0.4~T. Dissipation is induced in three ways: (1) by increasing the temperature of the system, (2) by applying nonzero dc bias, and (3) by tuning the gate voltage away from its optimal value. The temperature is increased using a heater on the mixing chamber stage of the refrigerator, and is measured by a thermometer also on the mixing chamber stage. Electronic wiring is thermalized to the mixing chamber temperature through filters attached to the mixing chamber stage so that when the mixing chamber stage is heated, the sample's electron temperature and lattice temperature should both track the mixing chamber temperature. In the Supplemental Material, we discuss additional data taken while heating in a manner designed to preferentially heat the electronic system or the lattice.

\section{Results}

At base temperature and with magnetization directed vertically upward (henceforth referred to as positive magnetization), the Hall resistance is quantized (zero longitudinal resistance and Hall resistance $-h/e^2$), within the precision of the measurement over a wide range of gate voltages $V_g$. The optimum gate voltage, that which minimizes the longitudinal resistivity at an elevated temperature, is $-1.18$~V. Here, the conductivity of the device versus temperature is well-fit by Arrhenius activation with a temperature scale of 1.40~K~\cite{sup}. All measurements, aside from those where the gate voltage is explicitly varied, are taken at $V_g=-1$~V.

\begin{figure}
\centering
	\includegraphics[width=0.48\textwidth]{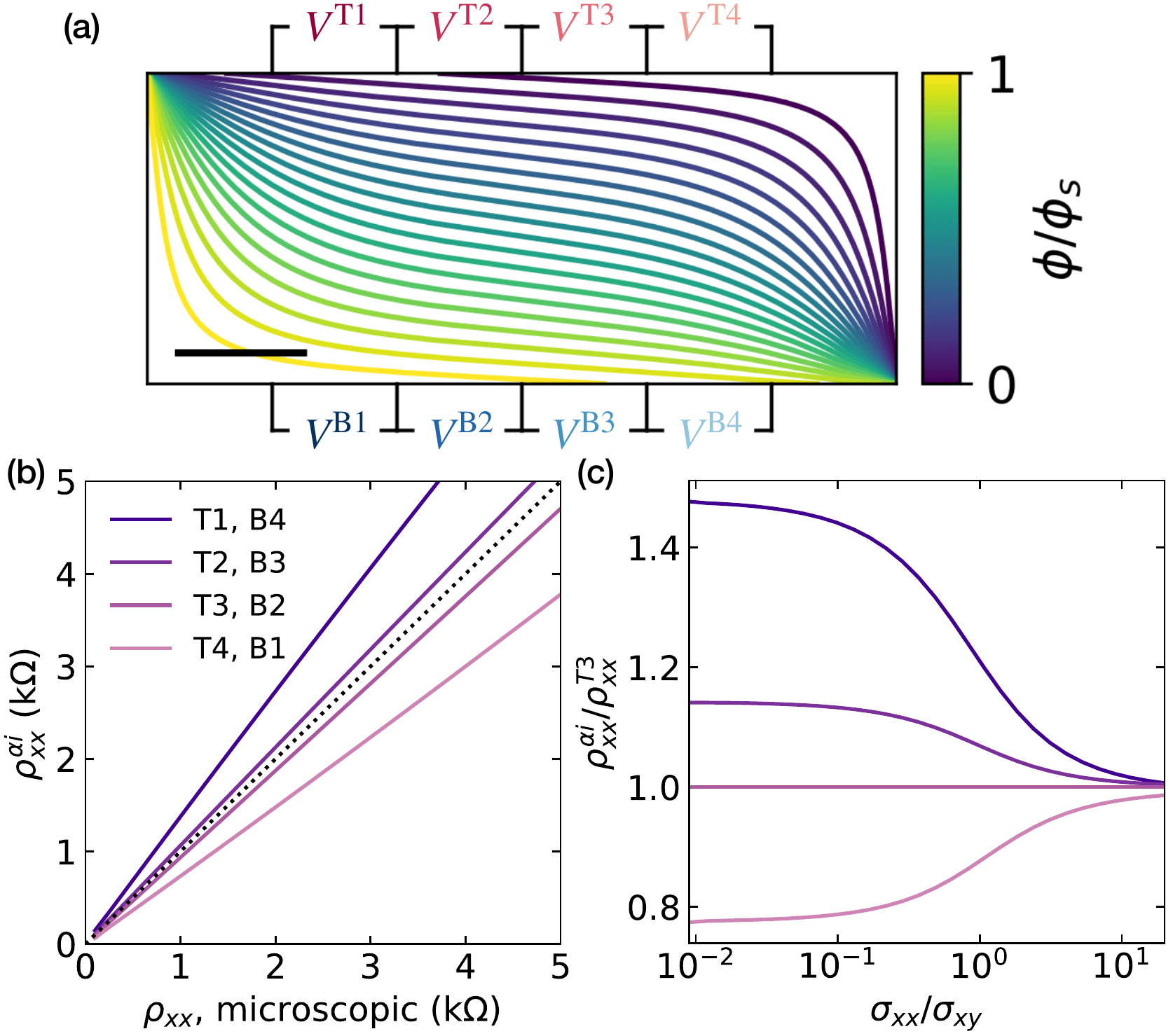}
	\caption{(a) Contour plot of equipotentials, normalized to the source potential $\phi_s$, in the Hall bar simulated at $\sigma_{xx}/\sigma_{xy}=0.05$, with positive magnetization. The electric field is concentrated near the top left and bottom right corners. Scale bar, 40~$\mu$m. (b) Simulated apparent resistivities $\rho_{xx}^{\alpha i}$ as a function of the spatially-homogeneous microscopic resistivity of the material. The dotted line with a slope of 1 is included for reference. (c) The ratio of simulated resistivities $\rho_{xx}^{\alpha i}/\rho_{xx}^{\mathrm{T3}}$.}
	\label{fig2}
\end{figure}

Fig.~1(b) shows $\rho_{xx}^{\alpha i}$ as a function of temperature. At a given temperature, the apparent resistivity is not constant throughout the device, but follows a pattern that is clarified in Fig.~1(c) by plotting the resistivities as a function of position. The apparent resistivity decreases (increases) monotonically moving rightwards across the top (bottom) edge of the device. The apparent resistivities on opposite ends of opposite edges are approximately equal, so that the largest are $\rho_{xx}^{\mathrm{T1, B4}}$.

This pattern is reminiscent of the classical Hall effect, where the electric potential $\phi$ satisfies Laplace's equation $\nabla^2 \phi=0$ subject to Dirichlet boundary conditions at the source and drain contacts and the condition
\begin{equation}
\frac{d\phi}{dy}=\frac{\sigma_{xy}}{\sigma_{xx}} \frac{d\phi}{dx}
\end{equation}
at edges of the device without contacts (the top and bottom edges of the Hall bar)~\cite{moelter1998}. The latter conditions are a statement of current continuity at the device edges. To compare our measurements with solutions of Laplace's equation, we numerically solve Laplace's equation in the specific geometry of our Hall bar for a variety of values of the parameter $\sigma_{xx}/\sigma_{xy}$. An example result is shown in Fig.~2(a), demonstrating concentration of the electric field $\mathbf{E}=\nabla \phi$ in the top left and bottom right corners.

The simulated apparent resistivities, shown in Fig.~2(b) as a function of the spatially-homogeneous microscopic resistivity~\cite{sup}, reproduce several key features of our measurements. For nonzero $\sigma_{xx}$ and $\sigma_{xy}$, resistivities measured at different locations are unequal. The apparent resistivity along the top (bottom) edge increases monotonically nearer to the left (right) corner. Centrosymmetry is intact: the apparent resistivities on opposite sides of opposite edges are equal.

\begin{figure}
\centering
	\includegraphics[width=0.48\textwidth]{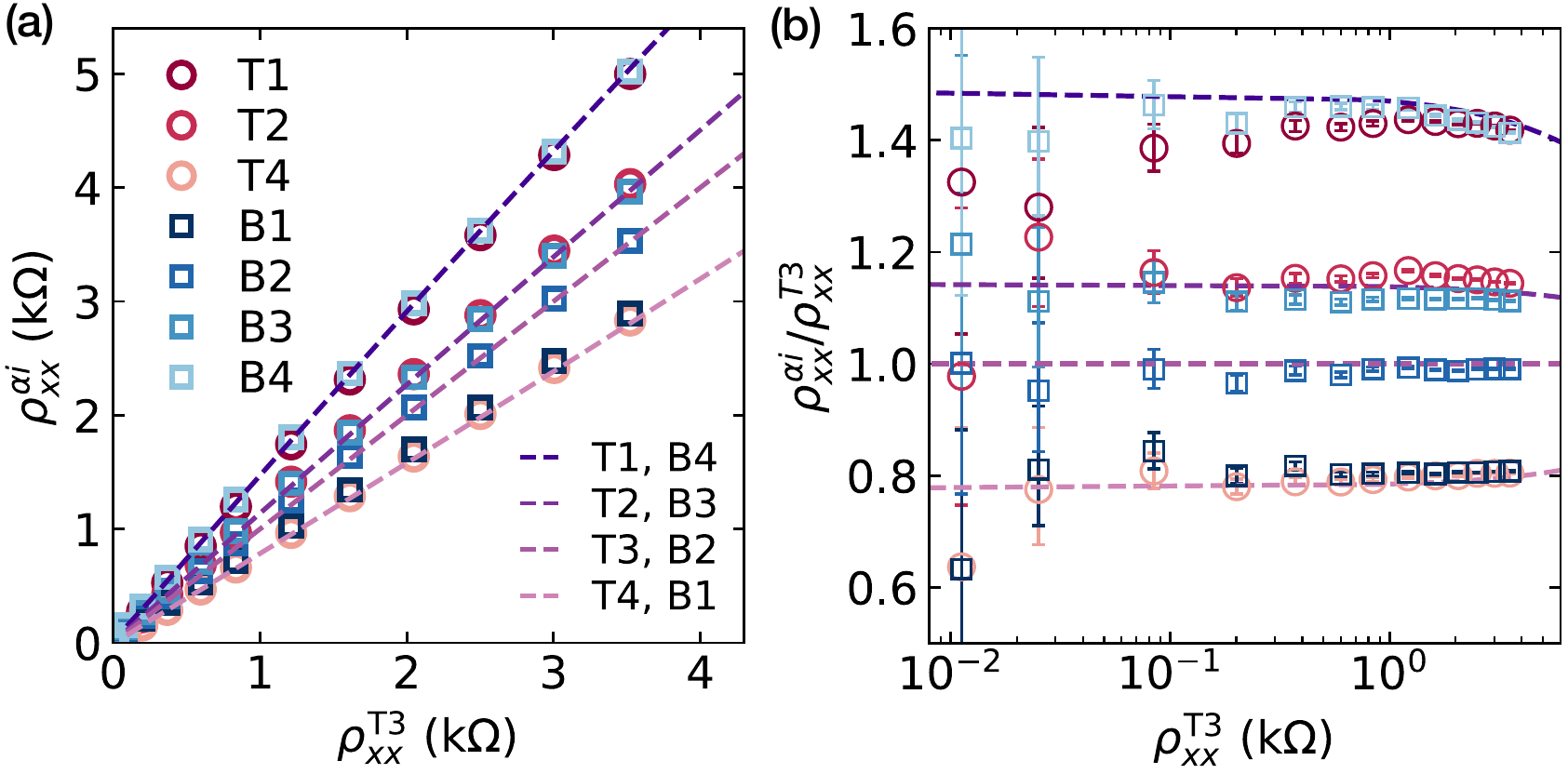}
	\caption{(a) The apparent resistivity at multiple locations and multiple temperatures (data from Fig.~1(b)), plotted parametrically versus measurement T3 at the same temperature. The simulated behavior (Fig.~2(b)) is shown by the dashed violet lines. (b) The ratio of measured apparent resistivities $\rho_{xx}^{\alpha i}/\rho_{xx}^{\mathrm{T3}}$, shown as a function of measured $\rho_{xx}^{\mathrm{T3}}$. Simulated behavior (Fig.~2(c)) is shown by the dashed violet lines. Data in (b) are corrected for finite input impedance of our voltage preamplifiers. The corrections and error budget are described in the Supplemental Material.}
	\label{fig3}
\end{figure}

We next quantitatively compare the simulations to the measurements. The microscopic resistivity cannot be directly determined from the measurements. We therefore parametrically plot $\rho_{xx}^{\alpha i}$ versus its value at one arbitrarily-chosen location $\rho_{xx}^{\mathrm{T3}}$, providing a parameter-free measure of the spatial variation of the electric field. The data from Fig.~1(b) are shown in this manner in Fig.~3(a). The concurrence between simulations and measurements is striking, and, as emphasized in Fig.~3(b), holds in the low-dissipation limit to $\rho_{xx}\sim 200$~$\Omega \approx 0.01h/e^2$, below which measurement errors become so large as to prevent meaningful comparison to our model.

When the magnetization of the film is reversed, switching the sign of the Hall conductance, the aforementioned pattern reverses, so that $\rho_{xx}^{\text{T4, B1}}$ are highest. Measurements for negative magnetization are shown in the Supplemental Material, and are also in good agreement with simulations.

%Our results indicate that the spatial profile of the electric potential of the QAH system is nearly identical to that of a homogeneous classical Hall system, which features current propagation through its bulk and lacks a CEM. This concurrence suggests the possibility that non-equilibrium current flows through the bulk in the QAH system as well.

\section{Discussion}

Voltage contacts measure the local electrochemical potential, which in this experiment is the sum of the chemical potential of the CEM and the electrostatic potential. Our experiment shows that the electrochemical potential along the edge of a dissipative-regime quantum anomalous Hall system satisfies Laplace's equation in two dimensions. From this, we will next deduce that there is a transverse electric field $E_y$ across the device and, in turn, that current flows through the bulk.

Just as voltage contacts measure local electrochemical potential, applied voltage imposes an electrochemical potential difference between source and drain. The way the electrochemical potential falls across the sample may in general be split between electrostatic and chemical potential gradients, the latter of which drives current through the CEM~\cite{halperin1982}. Let us consider whether one of those components is dominant in our measurements, or if both play roles. The electrostatic portion should satisfy Laplace's equation (the electrostatic potential should always satisfy Poisson's equation; that it should satisfy Laplace's equation follows from the assertion that Ohm's law $\mathbf{j}=\boldsymbol{\sigma}\mathbf{E}$ and conservation of charge $\nabla\mathbf{j}=0$ hold~\cite{sup}). In contrast, absent elaborate fine-tuning, models for dissipative-regime edge mode transport yield chemical potential profiles that {\em do not} satisfy Laplace's equation in two dimensions, as we demonstrate in the Supplemental Material. As Laplace's equation is linear, the sum of electrostatic and electrochemical potentials thus should not be expected to satisfy Laplace's equation unless the electrochemical component is negligible. Our experimental finding that electrochemical potential {\em does} satisfy Laplace's equation along the sample edge then strongly suggests that the source-drain bias primarily drives an electrostatic potential gradient $E_y$ throughout the bulk, not a chemical potential gradient across the CEM. That the potential profile is electrostatic could not have been concluded simply based on measuring current from source to drain: net current can flow from source to drain through the CEM in response to a chemical potential difference, even with no electric field anywhere in the sample~\cite{halperin1982}.

Next, we consider the relationship between the electrostatic potential and the chemical potential throughout the bulk, noting that whereas electrostatic potential is uniquely defined (up to a constant) at every location in space, the chemical potential of the bulk does not necessarily equilibrate with that of the CEM. When the electrostatic potential in a small area of the bulk changes, charge enters the region according to its geometric capacitance, in turn proportionally modifying the chemical potential. This relationship may be represented as
\begin{equation}
\frac{\Delta\mu(x,y)}{e}=\frac{C_g}{C_q}\Delta \phi(x,y),
\end{equation}
where $\Delta\mu$ and $\Delta \phi$ are the changes to the chemical and electrostatic potentials from the source-drain bias, respectively, $C_g$ is the 2D bulk's geometric capacitance per unit area, and $C_q$ is the 2D bulk's quantum capacitance per unit area, which here is determined by the density of localized states within the gap between the conducting surface state bands. In our device, $C_g$ is dominated by the gate capacitance. We place an upper bound on $C_g/C_q$ as follows: at base temperature, we observe a well-quantized QAH effect in this device across a gate voltage range of 6~V. Since the bulk conduction is small in this range, this gate voltage swing cannot change the chemical potential by more than the intrinsic gap, which has been measured in similar materials to be roughly 30~meV~\cite{lee2015, chong2020} (this is a conservative upper bound; note that the aforementioned transport gap $k_B\times 1.40\ \mathrm{K}=121\ \mu\mathrm{eV}$ is orders of magnitude smaller). We thus bound $C_g/C_q\leq 1/200$. This calculation establishes that the electrochemical potential difference across the bulk of the device is almost entirely manifested as an electrostatic potential gradient $\nabla \phi$, not a chemical potential gradient $\nabla \mu$. We conclude that a source-drain bias is mostly manifested as an electrostatic potential difference, creating a transverse electric field within the device's 2D bulk.

Since there is a transverse electric field $E_y$, and we have $\sigma_{xx} \ll \abs{\sigma_{xy}}\approx e^2/h$, it follows from Ohm's law $\mathbf{j}=\boldsymbol{\sigma}\mathbf{E}$ that the non-equilibrium current flows primarily through the two-dimensional bulk of the device. The ideal QAH system at zero temperature should have quantized conductivity $\sigma_{xx}=0$, $\sigma_{xy}=\pm e^2/h$ (a consequence of the system's Chern number $C=\pm 1$, and derived via the Kubo formula~\cite{yu2010}). The analysis technique presented in our present work requires measurable resistance. We have observed that the electrochemical potential continues to satisfy Laplace's equation as dissipation is reduced down to $\sigma_{xx}\sim 0.01 h/e^2$ (Fig.~3(b)), the lowest we can access with acceptable resolution. Though we cannot make direct claims about the regime of even lower dissipation, this observation suggests that source-drain bias continues to manifest as a transverse electric field, and therefore that non-equilibrium current flows through the bulk, even in the limit $\sigma_{xx}\rightarrow 0$, where this current flows without dissipation and Hall resistance is precisely quantized.

We emphasize that we are here discussing the non-equilibrium portion of the the current, that is, the difference in total current with versus without a source-drain bias; even without bias a persistent current (from occupied states of the CEM) should also circulate around the edge of the device without contributing to a net source-drain current. Our experiment cannot discern through which two-dimensional states bulk currents flow---surface states or quantum well states derived from three-dimensional states---although we would predict the former. We also note that when $\sigma_{xx}= 0$, $\sigma_{xy}\neq 0$, in contrast to the dissipative regime we directly probe, resistances predicted by the Landauer-B\"uttiker formalism, based on the edge-current picture, are identical to those derived from Laplace's equation, even in nonlocal geometries~\cite{sup}.

Though our findings depart from the extant QAH literature, which generally presents dissipationless current flow as a circulating edge current, a strand of the QH literature has long recognized that the amount of non-equilibrium current flowing through the bulk versus the edge depends on the extent to which the source-drain bias manifests as an electrostatic or a chemical potential difference, respectively, and that the drop is often primarily electrostatic, implying bulk current flow~\cite{macdonald1983, fontein1991, hirai1994, ando1994, komiyama1996representations, wiegers1999, ahlswede2001, weis2011, suddards2012, panos2014}. We present further comparison of the QAH and QH systems in the Supplemental Material.

\begin{figure}
\centering
	\includegraphics[width=0.48\textwidth]{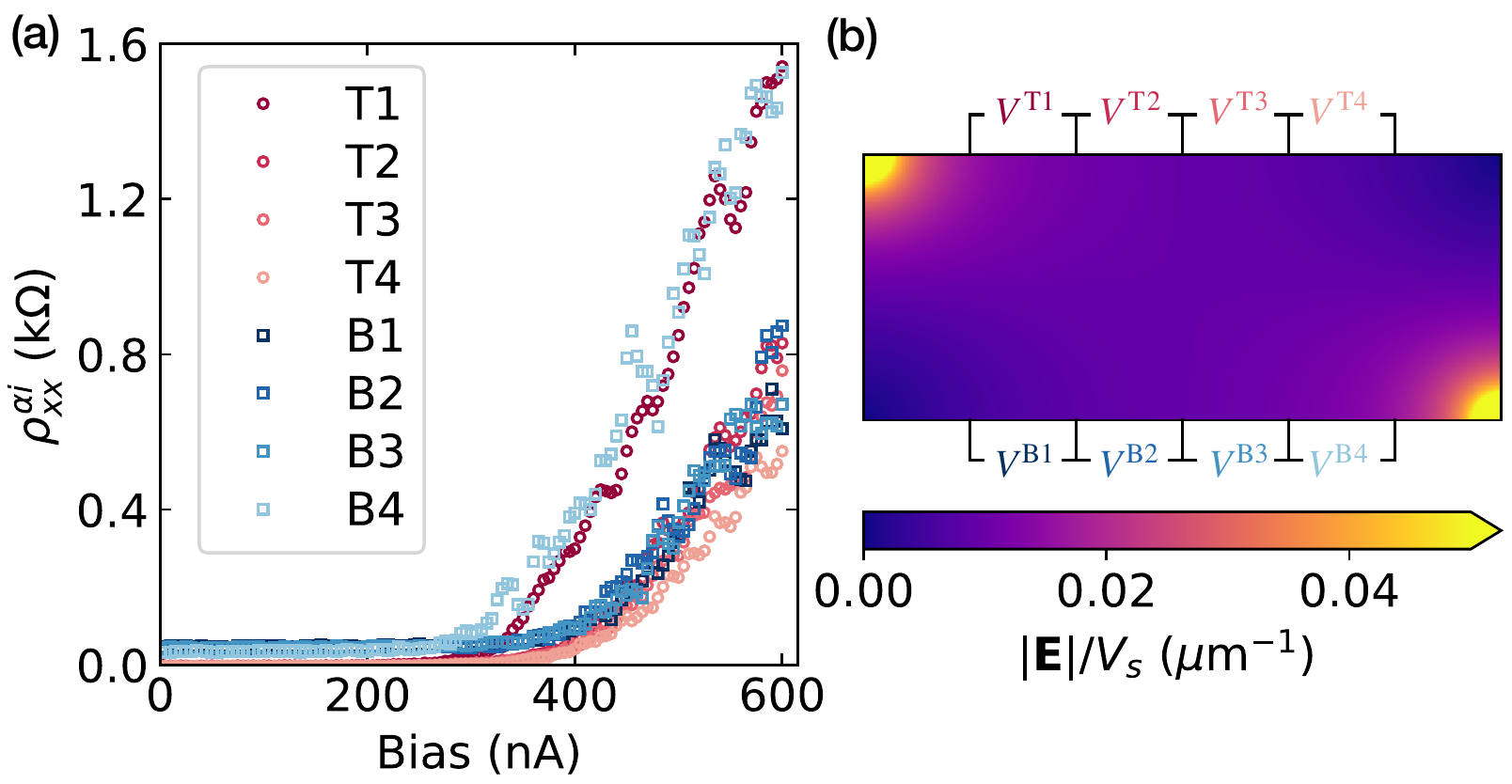}
	\caption{(a) Resistivity as a function of applied dc bias at optimum gate voltage and $T\approx 28$~mK. (b) The electric field normalized to the source-drain bias, simulated using Laplace's equation with appropriate boundary conditions at edges of the device. For this simulation, conductivity is assumed to be uniform over the entire device. In fact, we know the conductivity is a strong function of electric field at high electric field, and since the simulation shows that electric field is non-uniform within the device, our model should be inaccurate at high current bias.}
	\label{fig4}
\end{figure}

Having so far focused on linear conductance, we now consider the QAH device subjected to a large current bias. The resulting large transverse electric field across the device induces dissipation and eventually breakdown of the QAH state~\cite{fox2018,gotz2018,rodenbach2021}. Because the electric field is highest in two corners of the device (called the ``hot spots" in this context)~\cite{ahlswede2001}, we expect that the longitudinal conductivity becomes nonuniform, taking higher values near the hot spots. Solutions to Laplace's equation assuming uniform conductivity should thus no longer describe the potential throughout the device. In the Supplemental Material the apparent resistivities shown as a function of current bias (Fig.~4(a) of the main text) are compared to simulations of Laplace's equation. Indeed, the data no longer match uniform-conductivity solutions to Laplace's equation. Near the hot spots (measurements T1 and B4), breakdown occurs at a lower bias and the resistivity increases more rapidly after breakdown. This effect is invariant to the sign of dc bias, but reverses, so that $\rho_{xx}^{\text{T4, B1}}$ become highest, when the magnetization of the film is switched (thus switching which corners host the hot spots)~\cite{sup}.

\section{Conclusion}

We have here shown that the resistivity in Hall-geometry QAH devices in the dissipative regime varies as a function of where in the device the resistivity is measured. When the dissipation is induced by increased temperature, the spatial dependence of resistivity quantitatively matches solutions to Laplace's equation with finite and spatially-uniform longitudinal and Hall conductivities. This result is consistent with flow of non-equilibrium current primarily through the bulk, rather than the edges, of devices, which, we argue, should be expected in QAH devices. Our analysis extends that of Ref.~\cite{rodenbach2021}, which showed that {\em dissipative} currents mostly flow through the bulk, but did not establish that the non-equilibrium current mostly flows through the bulk {\em even in the limit of low dissipation}.

The analysis presented in this work is limited to regimes with finite longitudinal conductivity, and we have discussed the dissipationless regime only by extrapolation from the low-dissipation limit. Using high-precision, high-input-impedance voltmeters~\cite{fox2018} could allow similar analysis at even lower levels of dissipation, $\rho_{xx}\sim 10$~m$\Omega$. Other measurements, such as scanning probe measurements of current flow or local potential, would enable study of the system's electronic behavior in the regime of vanishing longitudinal conductivity. Scanning impedance measurements under large current bias should also be able to detect local variation in the conductivity near the hot spots.

A recent study claimed the observation of chiral current flow in a modified Corbino device~\cite{fijalkowski2021}. In the Supplemental Material, we reproduce with no free parameters the main features of the data using our simulations of Laplace's equation, which feature bulk-only current flow.

While preparing this manuscript, we became aware of work with similar conclusions being prepared by G. M. Ferguson {\em et al.}~\cite{ferguson2021}.

\section{Acknowledgements}

P. Z., L. T., and K. L. W. developed and grew the Cr-doped \ce{(Bi_x Sb_{1-x})2Te3} film. I. T. R. and M. P. A. fabricated the devices. I. T. R. conducted low-frequency measurements and simulated the classical Hall effect. I. T. R., L. K. R., M. P. A., M. A. K., and D. G.-G. analyzed the data. I. T. R. wrote the manuscript with contributions from all authors.

We thank B. I. Halperin and R. B. Laughlin for enlightening discussions. I. T. R. acknowledges M. A. Zaman, upon whose code (available at \url{https://github.com/zaman13/poisson-solver-2D}) the simulations in this work are based, and enlightening discussions at the early stages of this work with G. M. Ferguson and K. C. Nowack. I. T. R., M. P. A., and L. K. R. were supported by the U.S. Department of Energy, Office of Science, Basic Energy Sciences, Materials Sciences and Engineering Division, under Contract DE-AC02-76SF00515. I. T. R. additionally acknowledges support from the ARCS foundation. P. Z., L. T., and K. L. W. were supported by the U.S. Army Research Office MURI program under Grants No. W911NF-16-1-0472. Infrastructure and cryostat support were funded in part by the Gordon and Betty Moore Foundation through Grant No. GBMF3429. We thank NF Corporation for providing low-noise, high-input-impedance voltage preamplifiers. We acknowledge measurement assistance from colleagues at the National Institute of Standards and Technology. Part of this work was performed at the nano@Stanford labs, supported by the National Science Foundation under award ECCS-2026822. 

Datasets and data analysis code are provisioned at \url{https://doi.org/10.5281/zenodo.5826196}. The simulation package is available at \url{https://github.com/itrosen/hall-solver}.

\nocite{van1988, klab1991, komiyama1996, komiyama1996representations, dolan1999,delahaye2003, granger2009, wang2013, checkelsky2014, lachman2015visualization, lee2015, rosen2017, wang2018direct, fox2018, rodenbach2021, fijalkowski2021}

\bibliography{ref.bib}

%apsrev4-2.bst 2019-01-14 (MD) hand-edited version of apsrev4-1.bst
%Control: key (0)
%Control: author (72) initials jnrlst
%Control: editor formatted (1) identically to author
%Control: production of article title (-1) disabled
%Control: page (0) single
%Control: year (1) truncated
%Control: production of eprint (0) enabled
\begin{thebibliography}{42}%
\makeatletter
\providecommand \@ifxundefined [1]{%
 \@ifx{#1\undefined}
}%
\providecommand \@ifnum [1]{%
 \ifnum #1\expandafter \@firstoftwo
 \else \expandafter \@secondoftwo
 \fi
}%
\providecommand \@ifx [1]{%
 \ifx #1\expandafter \@firstoftwo
 \else \expandafter \@secondoftwo
 \fi
}%
\providecommand \natexlab [1]{#1}%
\providecommand \enquote  [1]{``#1''}%
\providecommand \bibnamefont  [1]{#1}%
\providecommand \bibfnamefont [1]{#1}%
\providecommand \citenamefont [1]{#1}%
\providecommand \href@noop [0]{\@secondoftwo}%
\providecommand \href [0]{\begingroup \@sanitize@url \@href}%
\providecommand \@href[1]{\@@startlink{#1}\@@href}%
\providecommand \@@href[1]{\endgroup#1\@@endlink}%
\providecommand \@sanitize@url [0]{\catcode `\\12\catcode `\$12\catcode
  `\&12\catcode `\#12\catcode `\^12\catcode `\_12\catcode `\%12\relax}%
\providecommand \@@startlink[1]{}%
\providecommand \@@endlink[0]{}%
\providecommand \url  [0]{\begingroup\@sanitize@url \@url }%
\providecommand \@url [1]{\endgroup\@href {#1}{\urlprefix }}%
\providecommand \urlprefix  [0]{URL }%
\providecommand \Eprint [0]{\href }%
\providecommand \doibase [0]{https://doi.org/}%
\providecommand \selectlanguage [0]{\@gobble}%
\providecommand \bibinfo  [0]{\@secondoftwo}%
\providecommand \bibfield  [0]{\@secondoftwo}%
\providecommand \translation [1]{[#1]}%
\providecommand \BibitemOpen [0]{}%
\providecommand \bibitemStop [0]{}%
\providecommand \bibitemNoStop [0]{.\EOS\space}%
\providecommand \EOS [0]{\spacefactor3000\relax}%
\providecommand \BibitemShut  [1]{\csname bibitem#1\endcsname}%
\let\auto@bib@innerbib\@empty
%</preamble>
\bibitem [{\citenamefont {Bestwick}\ \emph {et~al.}(2015)\citenamefont
  {Bestwick}, \citenamefont {Fox}, \citenamefont {Kou}, \citenamefont {Pan},
  \citenamefont {Wang},\ and\ \citenamefont {Goldhaber-Gordon}}]{bestwick2015}%
  \BibitemOpen
  \bibfield  {author} {\bibinfo {author} {\bibfnamefont {A.~J.}\ \bibnamefont
  {Bestwick}}, \bibinfo {author} {\bibfnamefont {E.~J.}\ \bibnamefont {Fox}},
  \bibinfo {author} {\bibfnamefont {X.}~\bibnamefont {Kou}}, \bibinfo {author}
  {\bibfnamefont {L.}~\bibnamefont {Pan}}, \bibinfo {author} {\bibfnamefont
  {K.~L.}\ \bibnamefont {Wang}},\ and\ \bibinfo {author} {\bibfnamefont
  {D.}~\bibnamefont {Goldhaber-Gordon}},\ }\href@noop {} {\bibfield  {journal}
  {\bibinfo  {journal} {Phys. Rev. Lett.}\ }\textbf {\bibinfo {volume} {114}},\
  \bibinfo {pages} {187201} (\bibinfo {year} {2015})}\BibitemShut {NoStop}%
\bibitem [{\citenamefont {Chang}\ \emph
  {et~al.}(2015{\natexlab{a}})\citenamefont {Chang}, \citenamefont {Zhao},
  \citenamefont {Kim}, \citenamefont {Zhang}, \citenamefont {Assaf},
  \citenamefont {Heiman}, \citenamefont {Zhang}, \citenamefont {Liu},
  \citenamefont {Chan},\ and\ \citenamefont {Moodera}}]{chang2015}%
  \BibitemOpen
  \bibfield  {author} {\bibinfo {author} {\bibfnamefont {C.-Z.}\ \bibnamefont
  {Chang}}, \bibinfo {author} {\bibfnamefont {W.}~\bibnamefont {Zhao}},
  \bibinfo {author} {\bibfnamefont {D.~Y.}\ \bibnamefont {Kim}}, \bibinfo
  {author} {\bibfnamefont {H.}~\bibnamefont {Zhang}}, \bibinfo {author}
  {\bibfnamefont {B.~A.}\ \bibnamefont {Assaf}}, \bibinfo {author}
  {\bibfnamefont {D.}~\bibnamefont {Heiman}}, \bibinfo {author} {\bibfnamefont
  {S.-C.}\ \bibnamefont {Zhang}}, \bibinfo {author} {\bibfnamefont
  {C.}~\bibnamefont {Liu}}, \bibinfo {author} {\bibfnamefont {M.~H.~W.}\
  \bibnamefont {Chan}},\ and\ \bibinfo {author} {\bibfnamefont {J.~S.}\
  \bibnamefont {Moodera}},\ }\href@noop {} {\bibfield  {journal} {\bibinfo
  {journal} {Nat. Mater.}\ }\textbf {\bibinfo {volume} {14}},\ \bibinfo {pages}
  {473} (\bibinfo {year} {2015}{\natexlab{a}})}\BibitemShut {NoStop}%
\bibitem [{\citenamefont {Fox}\ \emph {et~al.}(2018)\citenamefont {Fox},
  \citenamefont {Rosen}, \citenamefont {Yang}, \citenamefont {Jones},
  \citenamefont {Elmquist}, \citenamefont {Kou}, \citenamefont {Pan},
  \citenamefont {Wang},\ and\ \citenamefont {Goldhaber-Gordon}}]{fox2018}%
  \BibitemOpen
  \bibfield  {author} {\bibinfo {author} {\bibfnamefont {E.~J.}\ \bibnamefont
  {Fox}}, \bibinfo {author} {\bibfnamefont {I.~T.}\ \bibnamefont {Rosen}},
  \bibinfo {author} {\bibfnamefont {Y.}~\bibnamefont {Yang}}, \bibinfo {author}
  {\bibfnamefont {G.~R.}\ \bibnamefont {Jones}}, \bibinfo {author}
  {\bibfnamefont {R.~E.}\ \bibnamefont {Elmquist}}, \bibinfo {author}
  {\bibfnamefont {X.}~\bibnamefont {Kou}}, \bibinfo {author} {\bibfnamefont
  {L.}~\bibnamefont {Pan}}, \bibinfo {author} {\bibfnamefont {K.~L.}\
  \bibnamefont {Wang}},\ and\ \bibinfo {author} {\bibfnamefont
  {D.}~\bibnamefont {Goldhaber-Gordon}},\ }\href@noop {} {\bibfield  {journal}
  {\bibinfo  {journal} {Phys. Rev. B}\ }\textbf {\bibinfo {volume} {98}},\
  \bibinfo {pages} {075145} (\bibinfo {year} {2018})}\BibitemShut {NoStop}%
\bibitem [{\citenamefont {G{\"o}tz}\ \emph {et~al.}(2018)\citenamefont
  {G{\"o}tz}, \citenamefont {Fijalkowski}, \citenamefont {Pesel}, \citenamefont
  {Hartl}, \citenamefont {Schreyeck}, \citenamefont {Winnerlein}, \citenamefont
  {Grauer}, \citenamefont {Scherer}, \citenamefont {Brunner}, \citenamefont
  {Gould} \emph {et~al.}}]{gotz2018}%
  \BibitemOpen
  \bibfield  {author} {\bibinfo {author} {\bibfnamefont {M.}~\bibnamefont
  {G{\"o}tz}}, \bibinfo {author} {\bibfnamefont {K.~M.}\ \bibnamefont
  {Fijalkowski}}, \bibinfo {author} {\bibfnamefont {E.}~\bibnamefont {Pesel}},
  \bibinfo {author} {\bibfnamefont {M.}~\bibnamefont {Hartl}}, \bibinfo
  {author} {\bibfnamefont {S.}~\bibnamefont {Schreyeck}}, \bibinfo {author}
  {\bibfnamefont {M.}~\bibnamefont {Winnerlein}}, \bibinfo {author}
  {\bibfnamefont {S.}~\bibnamefont {Grauer}}, \bibinfo {author} {\bibfnamefont
  {H.}~\bibnamefont {Scherer}}, \bibinfo {author} {\bibfnamefont
  {K.}~\bibnamefont {Brunner}}, \bibinfo {author} {\bibfnamefont
  {C.}~\bibnamefont {Gould}}, \emph {et~al.},\ }\href@noop {} {\bibfield
  {journal} {\bibinfo  {journal} {Appl. Phys. Lett.}\ }\textbf {\bibinfo
  {volume} {112}},\ \bibinfo {pages} {072102} (\bibinfo {year}
  {2018})}\BibitemShut {NoStop}%
\bibitem [{\citenamefont {Okazaki}\ \emph {et~al.}(2020)\citenamefont
  {Okazaki}, \citenamefont {Oe}, \citenamefont {Kawamura}, \citenamefont
  {Yoshimi}, \citenamefont {Nakamura}, \citenamefont {Takada}, \citenamefont
  {Mogi}, \citenamefont {Takahashi}, \citenamefont {Tsukazaki}, \citenamefont
  {Kawasaki} \emph {et~al.}}]{okazaki2020}%
  \BibitemOpen
  \bibfield  {author} {\bibinfo {author} {\bibfnamefont {Y.}~\bibnamefont
  {Okazaki}}, \bibinfo {author} {\bibfnamefont {T.}~\bibnamefont {Oe}},
  \bibinfo {author} {\bibfnamefont {M.}~\bibnamefont {Kawamura}}, \bibinfo
  {author} {\bibfnamefont {R.}~\bibnamefont {Yoshimi}}, \bibinfo {author}
  {\bibfnamefont {S.}~\bibnamefont {Nakamura}}, \bibinfo {author}
  {\bibfnamefont {S.}~\bibnamefont {Takada}}, \bibinfo {author} {\bibfnamefont
  {M.}~\bibnamefont {Mogi}}, \bibinfo {author} {\bibfnamefont {K.~S.}\
  \bibnamefont {Takahashi}}, \bibinfo {author} {\bibfnamefont {A.}~\bibnamefont
  {Tsukazaki}}, \bibinfo {author} {\bibfnamefont {M.}~\bibnamefont {Kawasaki}},
  \emph {et~al.},\ }\href@noop {} {\bibfield  {journal} {\bibinfo  {journal}
  {Appl. Phys. Lett.}\ }\textbf {\bibinfo {volume} {116}},\ \bibinfo {pages}
  {143101} (\bibinfo {year} {2020})}\BibitemShut {NoStop}%
\bibitem [{\citenamefont {Okazaki}\ \emph {et~al.}(2021)\citenamefont
  {Okazaki}, \citenamefont {Oe}, \citenamefont {Kawamura}, \citenamefont
  {Yoshimi}, \citenamefont {Nakamura}, \citenamefont {Takada}, \citenamefont
  {Mogi}, \citenamefont {Takahashi}, \citenamefont {Tsukaxaki}, \citenamefont
  {Kawasaki}, \citenamefont {Tokura},\ and\ \citenamefont
  {Kaneko}}]{okazaki2021}%
  \BibitemOpen
  \bibfield  {author} {\bibinfo {author} {\bibfnamefont {Y.}~\bibnamefont
  {Okazaki}}, \bibinfo {author} {\bibfnamefont {T.}~\bibnamefont {Oe}},
  \bibinfo {author} {\bibfnamefont {M.}~\bibnamefont {Kawamura}}, \bibinfo
  {author} {\bibfnamefont {R.}~\bibnamefont {Yoshimi}}, \bibinfo {author}
  {\bibfnamefont {S.}~\bibnamefont {Nakamura}}, \bibinfo {author}
  {\bibfnamefont {S.}~\bibnamefont {Takada}}, \bibinfo {author} {\bibfnamefont
  {M.}~\bibnamefont {Mogi}}, \bibinfo {author} {\bibfnamefont {K.~S.}\
  \bibnamefont {Takahashi}}, \bibinfo {author} {\bibfnamefont {A.}~\bibnamefont
  {Tsukaxaki}}, \bibinfo {author} {\bibfnamefont {M.}~\bibnamefont {Kawasaki}},
  \bibinfo {author} {\bibfnamefont {Y.}~\bibnamefont {Tokura}},\ and\ \bibinfo
  {author} {\bibfnamefont {N.-H.}\ \bibnamefont {Kaneko}},\ }\bibfield
  {journal} {\bibinfo  {journal} {Nat. Phys.}\ }\href
  {https://doi.org/doi:10.1038/s41567-021-01424-8}
  {doi:10.1038/s41567-021-01424-8} (\bibinfo {year} {2021})\BibitemShut
  {NoStop}%
\bibitem [{\citenamefont {Liu}\ \emph {et~al.}(2008)\citenamefont {Liu},
  \citenamefont {Qi}, \citenamefont {Dai}, \citenamefont {Fang},\ and\
  \citenamefont {Zhang}}]{liu2008}%
  \BibitemOpen
  \bibfield  {author} {\bibinfo {author} {\bibfnamefont {C.-X.}\ \bibnamefont
  {Liu}}, \bibinfo {author} {\bibfnamefont {X.-L.}\ \bibnamefont {Qi}},
  \bibinfo {author} {\bibfnamefont {X.}~\bibnamefont {Dai}}, \bibinfo {author}
  {\bibfnamefont {Z.}~\bibnamefont {Fang}},\ and\ \bibinfo {author}
  {\bibfnamefont {S.-C.}\ \bibnamefont {Zhang}},\ }\href@noop {} {\bibfield
  {journal} {\bibinfo  {journal} {Phys. Rev. Lett.}\ }\textbf {\bibinfo
  {volume} {101}},\ \bibinfo {pages} {146802} (\bibinfo {year}
  {2008})}\BibitemShut {NoStop}%
\bibitem [{\citenamefont {Yu}\ \emph {et~al.}(2010)\citenamefont {Yu},
  \citenamefont {Zhang}, \citenamefont {Zhang}, \citenamefont {Zhang},
  \citenamefont {Dai},\ and\ \citenamefont {Fang}}]{yu2010}%
  \BibitemOpen
  \bibfield  {author} {\bibinfo {author} {\bibfnamefont {R.}~\bibnamefont
  {Yu}}, \bibinfo {author} {\bibfnamefont {W.}~\bibnamefont {Zhang}}, \bibinfo
  {author} {\bibfnamefont {H.-J.}\ \bibnamefont {Zhang}}, \bibinfo {author}
  {\bibfnamefont {S.-C.}\ \bibnamefont {Zhang}}, \bibinfo {author}
  {\bibfnamefont {X.}~\bibnamefont {Dai}},\ and\ \bibinfo {author}
  {\bibfnamefont {Z.}~\bibnamefont {Fang}},\ }\href@noop {} {\bibfield
  {journal} {\bibinfo  {journal} {science}\ }\textbf {\bibinfo {volume}
  {329}},\ \bibinfo {pages} {61} (\bibinfo {year} {2010})}\BibitemShut
  {NoStop}%
\bibitem [{\citenamefont {Allen}\ \emph {et~al.}(2019)\citenamefont {Allen},
  \citenamefont {Cui}, \citenamefont {Yue~Ma}, \citenamefont {Mogi},
  \citenamefont {Kawamura}, \citenamefont {Fulga}, \citenamefont
  {Goldhaber-Gordon}, \citenamefont {Tokura},\ and\ \citenamefont
  {Shen}}]{allen2019}%
  \BibitemOpen
  \bibfield  {author} {\bibinfo {author} {\bibfnamefont {M.}~\bibnamefont
  {Allen}}, \bibinfo {author} {\bibfnamefont {Y.}~\bibnamefont {Cui}}, \bibinfo
  {author} {\bibfnamefont {E.}~\bibnamefont {Yue~Ma}}, \bibinfo {author}
  {\bibfnamefont {M.}~\bibnamefont {Mogi}}, \bibinfo {author} {\bibfnamefont
  {M.}~\bibnamefont {Kawamura}}, \bibinfo {author} {\bibfnamefont {I.~C.}\
  \bibnamefont {Fulga}}, \bibinfo {author} {\bibfnamefont {D.}~\bibnamefont
  {Goldhaber-Gordon}}, \bibinfo {author} {\bibfnamefont {Y.}~\bibnamefont
  {Tokura}},\ and\ \bibinfo {author} {\bibfnamefont {Z.-X.}\ \bibnamefont
  {Shen}},\ }\href@noop {} {\bibfield  {journal} {\bibinfo  {journal} {Proc.
  Natl. Acad. Sci.}\ }\textbf {\bibinfo {volume} {116}},\ \bibinfo {pages}
  {14511} (\bibinfo {year} {2019})}\BibitemShut {NoStop}%
\bibitem [{\citenamefont {Wang}\ \emph {et~al.}(2013)\citenamefont {Wang},
  \citenamefont {Lian}, \citenamefont {Zhang},\ and\ \citenamefont
  {Zhang}}]{wang2013}%
  \BibitemOpen
  \bibfield  {author} {\bibinfo {author} {\bibfnamefont {J.}~\bibnamefont
  {Wang}}, \bibinfo {author} {\bibfnamefont {B.}~\bibnamefont {Lian}}, \bibinfo
  {author} {\bibfnamefont {H.}~\bibnamefont {Zhang}},\ and\ \bibinfo {author}
  {\bibfnamefont {S.-C.}\ \bibnamefont {Zhang}},\ }\href@noop {} {\bibfield
  {journal} {\bibinfo  {journal} {Phys. Rev. Lett.}\ }\textbf {\bibinfo
  {volume} {111}},\ \bibinfo {pages} {086803} (\bibinfo {year}
  {2013})}\BibitemShut {NoStop}%
\bibitem [{\citenamefont {Kou}\ \emph {et~al.}(2014)\citenamefont {Kou},
  \citenamefont {Guo}, \citenamefont {Fan}, \citenamefont {Pan}, \citenamefont
  {Lang}, \citenamefont {Jiang}, \citenamefont {Shao}, \citenamefont {Nie},
  \citenamefont {Murata}, \citenamefont {Tang}, \citenamefont {Wang},
  \citenamefont {He}, \citenamefont {Lee}, \citenamefont {Lee},\ and\
  \citenamefont {Wang}}]{kou2014}%
  \BibitemOpen
  \bibfield  {author} {\bibinfo {author} {\bibfnamefont {X.}~\bibnamefont
  {Kou}}, \bibinfo {author} {\bibfnamefont {S.-T.}\ \bibnamefont {Guo}},
  \bibinfo {author} {\bibfnamefont {Y.}~\bibnamefont {Fan}}, \bibinfo {author}
  {\bibfnamefont {L.}~\bibnamefont {Pan}}, \bibinfo {author} {\bibfnamefont
  {M.}~\bibnamefont {Lang}}, \bibinfo {author} {\bibfnamefont {Y.}~\bibnamefont
  {Jiang}}, \bibinfo {author} {\bibfnamefont {Q.}~\bibnamefont {Shao}},
  \bibinfo {author} {\bibfnamefont {T.}~\bibnamefont {Nie}}, \bibinfo {author}
  {\bibfnamefont {K.}~\bibnamefont {Murata}}, \bibinfo {author} {\bibfnamefont
  {J.}~\bibnamefont {Tang}}, \bibinfo {author} {\bibfnamefont {Y.}~\bibnamefont
  {Wang}}, \bibinfo {author} {\bibfnamefont {L.}~\bibnamefont {He}}, \bibinfo
  {author} {\bibfnamefont {T.-K.}\ \bibnamefont {Lee}}, \bibinfo {author}
  {\bibfnamefont {W.-L.}\ \bibnamefont {Lee}},\ and\ \bibinfo {author}
  {\bibfnamefont {K.~L.}\ \bibnamefont {Wang}},\ }\href@noop {} {\bibfield
  {journal} {\bibinfo  {journal} {Phys. Rev. Lett.}\ }\textbf {\bibinfo
  {volume} {113}},\ \bibinfo {pages} {137201} (\bibinfo {year}
  {2014})}\BibitemShut {NoStop}%
\bibitem [{\citenamefont {Chang}\ \emph
  {et~al.}(2015{\natexlab{b}})\citenamefont {Chang}, \citenamefont {Zhao},
  \citenamefont {Kim}, \citenamefont {Wei}, \citenamefont {Jain}, \citenamefont
  {Liu}, \citenamefont {Chan},\ and\ \citenamefont {Moodera}}]{changPRL2015}%
  \BibitemOpen
  \bibfield  {author} {\bibinfo {author} {\bibfnamefont {C.-Z.}\ \bibnamefont
  {Chang}}, \bibinfo {author} {\bibfnamefont {W.}~\bibnamefont {Zhao}},
  \bibinfo {author} {\bibfnamefont {D.~Y.}\ \bibnamefont {Kim}}, \bibinfo
  {author} {\bibfnamefont {P.}~\bibnamefont {Wei}}, \bibinfo {author}
  {\bibfnamefont {J.~K.}\ \bibnamefont {Jain}}, \bibinfo {author}
  {\bibfnamefont {C.}~\bibnamefont {Liu}}, \bibinfo {author} {\bibfnamefont
  {M.~H.~W.}\ \bibnamefont {Chan}},\ and\ \bibinfo {author} {\bibfnamefont
  {J.~S.}\ \bibnamefont {Moodera}},\ }\href@noop {} {\bibfield  {journal}
  {\bibinfo  {journal} {Phys. Rev. Lett.}\ }\textbf {\bibinfo {volume} {115}},\
  \bibinfo {pages} {057206} (\bibinfo {year} {2015}{\natexlab{b}})}\BibitemShut
  {NoStop}%
\bibitem [{\citenamefont {Halperin}(1982)}]{halperin1982}%
  \BibitemOpen
  \bibfield  {author} {\bibinfo {author} {\bibfnamefont {B.~I.}\ \bibnamefont
  {Halperin}},\ }\href {https://doi.org/10.1103/PhysRevB.25.2185} {\bibfield
  {journal} {\bibinfo  {journal} {Phys. Rev. B}\ }\textbf {\bibinfo {volume}
  {25}},\ \bibinfo {pages} {2185} (\bibinfo {year} {1982})}\BibitemShut
  {NoStop}%
\bibitem [{\citenamefont {B\"uttiker}(1988)}]{buttiker1988}%
  \BibitemOpen
  \bibfield  {author} {\bibinfo {author} {\bibfnamefont {M.}~\bibnamefont
  {B\"uttiker}},\ }\href@noop {} {\bibfield  {journal} {\bibinfo  {journal}
  {Phys. Rev. B}\ }\textbf {\bibinfo {volume} {38}},\ \bibinfo {pages} {9375}
  (\bibinfo {year} {1988})}\BibitemShut {NoStop}%
\bibitem [{\citenamefont {Rodenbach}\ \emph {et~al.}(2021)\citenamefont
  {Rodenbach}, \citenamefont {Rosen}, \citenamefont {Fox}, \citenamefont
  {Zhang}, \citenamefont {Pan}, \citenamefont {Wang}, \citenamefont {Kastner},\
  and\ \citenamefont {Goldhaber-Gordon}}]{rodenbach2021}%
  \BibitemOpen
  \bibfield  {author} {\bibinfo {author} {\bibfnamefont {L.~K.}\ \bibnamefont
  {Rodenbach}}, \bibinfo {author} {\bibfnamefont {I.~T.}\ \bibnamefont
  {Rosen}}, \bibinfo {author} {\bibfnamefont {E.~J.}\ \bibnamefont {Fox}},
  \bibinfo {author} {\bibfnamefont {P.}~\bibnamefont {Zhang}}, \bibinfo
  {author} {\bibfnamefont {L.}~\bibnamefont {Pan}}, \bibinfo {author}
  {\bibfnamefont {K.~L.}\ \bibnamefont {Wang}}, \bibinfo {author}
  {\bibfnamefont {M.~A.}\ \bibnamefont {Kastner}},\ and\ \bibinfo {author}
  {\bibfnamefont {D.}~\bibnamefont {Goldhaber-Gordon}},\ }\href@noop {}
  {\bibfield  {journal} {\bibinfo  {journal} {APL Mater.}\ }\textbf {\bibinfo
  {volume} {9}},\ \bibinfo {pages} {081116} (\bibinfo {year}
  {2021})}\BibitemShut {NoStop}%
\bibitem [{\citenamefont {MacDonald}\ \emph {et~al.}(1983)\citenamefont
  {MacDonald}, \citenamefont {Rice},\ and\ \citenamefont
  {Brinkman}}]{macdonald1983}%
  \BibitemOpen
  \bibfield  {author} {\bibinfo {author} {\bibfnamefont {A.~H.}\ \bibnamefont
  {MacDonald}}, \bibinfo {author} {\bibfnamefont {T.~M.}\ \bibnamefont
  {Rice}},\ and\ \bibinfo {author} {\bibfnamefont {W.~F.}\ \bibnamefont
  {Brinkman}},\ }\href {https://doi.org/10.1103/PhysRevB.28.3648} {\bibfield
  {journal} {\bibinfo  {journal} {Phys. Rev. B}\ }\textbf {\bibinfo {volume}
  {28}},\ \bibinfo {pages} {3648} (\bibinfo {year} {1983})}\BibitemShut
  {NoStop}%
\bibitem [{\citenamefont {Hirai}\ and\ \citenamefont
  {Komiyama}(1994)}]{hirai1994}%
  \BibitemOpen
  \bibfield  {author} {\bibinfo {author} {\bibfnamefont {H.}~\bibnamefont
  {Hirai}}\ and\ \bibinfo {author} {\bibfnamefont {S.}~\bibnamefont
  {Komiyama}},\ }\href@noop {} {\bibfield  {journal} {\bibinfo  {journal}
  {Physical Review B}\ }\textbf {\bibinfo {volume} {49}},\ \bibinfo {pages}
  {14012} (\bibinfo {year} {1994})}\BibitemShut {NoStop}%
\bibitem [{\citenamefont {Komiyama}\ and\ \citenamefont
  {Hirai}(1996)}]{komiyama1996representations}%
  \BibitemOpen
  \bibfield  {author} {\bibinfo {author} {\bibfnamefont {S.}~\bibnamefont
  {Komiyama}}\ and\ \bibinfo {author} {\bibfnamefont {H.}~\bibnamefont
  {Hirai}},\ }\href@noop {} {\bibfield  {journal} {\bibinfo  {journal}
  {Physical Review B}\ }\textbf {\bibinfo {volume} {54}},\ \bibinfo {pages}
  {2067} (\bibinfo {year} {1996})}\BibitemShut {NoStop}%
\bibitem [{\citenamefont {Weis}\ and\ \citenamefont
  {Von~Klitzing}(2011)}]{weis2011}%
  \BibitemOpen
  \bibfield  {author} {\bibinfo {author} {\bibfnamefont {J.}~\bibnamefont
  {Weis}}\ and\ \bibinfo {author} {\bibfnamefont {K.}~\bibnamefont
  {Von~Klitzing}},\ }\href@noop {} {\bibfield  {journal} {\bibinfo  {journal}
  {Philos. Trans. R. Soc. Lond. A}\ }\textbf {\bibinfo {volume} {369}},\
  \bibinfo {pages} {3954} (\bibinfo {year} {2011})}\BibitemShut {NoStop}%
\bibitem [{sup()}]{sup}%
  \BibitemOpen
  \href@noop {} {\bibinfo  {journal} {See Supplemental Information}\
  }\BibitemShut {NoStop}%
\bibitem [{\citenamefont {Moelter}\ \emph {et~al.}(1998)\citenamefont
  {Moelter}, \citenamefont {Evans}, \citenamefont {Elliott},\ and\
  \citenamefont {Jackson}}]{moelter1998}%
  \BibitemOpen
\bibfield  {journal} {  }\bibfield  {author} {\bibinfo {author} {\bibfnamefont
  {M.~J.}\ \bibnamefont {Moelter}}, \bibinfo {author} {\bibfnamefont
  {J.}~\bibnamefont {Evans}}, \bibinfo {author} {\bibfnamefont
  {G.}~\bibnamefont {Elliott}},\ and\ \bibinfo {author} {\bibfnamefont
  {M.}~\bibnamefont {Jackson}},\ }\href@noop {} {\bibfield  {journal} {\bibinfo
   {journal} {Am. J. Phys.}\ }\textbf {\bibinfo {volume} {66}},\ \bibinfo
  {pages} {668} (\bibinfo {year} {1998})}\BibitemShut {NoStop}%
\bibitem [{\citenamefont {Lee}\ \emph {et~al.}(2015)\citenamefont {Lee},
  \citenamefont {Kim}, \citenamefont {Lee}, \citenamefont {Billinge},
  \citenamefont {Zhong}, \citenamefont {Schneeloch}, \citenamefont {Liu},
  \citenamefont {Valla}, \citenamefont {Tranquada}, \citenamefont {Gu} \emph
  {et~al.}}]{lee2015}%
  \BibitemOpen
  \bibfield  {author} {\bibinfo {author} {\bibfnamefont {I.}~\bibnamefont
  {Lee}}, \bibinfo {author} {\bibfnamefont {C.~K.}\ \bibnamefont {Kim}},
  \bibinfo {author} {\bibfnamefont {J.}~\bibnamefont {Lee}}, \bibinfo {author}
  {\bibfnamefont {S.~J.}\ \bibnamefont {Billinge}}, \bibinfo {author}
  {\bibfnamefont {R.}~\bibnamefont {Zhong}}, \bibinfo {author} {\bibfnamefont
  {J.~A.}\ \bibnamefont {Schneeloch}}, \bibinfo {author} {\bibfnamefont
  {T.}~\bibnamefont {Liu}}, \bibinfo {author} {\bibfnamefont {T.}~\bibnamefont
  {Valla}}, \bibinfo {author} {\bibfnamefont {J.~M.}\ \bibnamefont
  {Tranquada}}, \bibinfo {author} {\bibfnamefont {G.}~\bibnamefont {Gu}}, \emph
  {et~al.},\ }\href@noop {} {\bibfield  {journal} {\bibinfo  {journal}
  {Proceedings of the National Academy of Sciences}\ }\textbf {\bibinfo
  {volume} {112}},\ \bibinfo {pages} {1316} (\bibinfo {year}
  {2015})}\BibitemShut {NoStop}%
\bibitem [{\citenamefont {Chong}\ \emph {et~al.}(2020)\citenamefont {Chong},
  \citenamefont {Liu}, \citenamefont {Sharma}, \citenamefont {Kostin},
  \citenamefont {Gu}, \citenamefont {Fujita}, \citenamefont {Davis},\ and\
  \citenamefont {Sprau}}]{chong2020}%
  \BibitemOpen
  \bibfield  {author} {\bibinfo {author} {\bibfnamefont {Y.~X.}\ \bibnamefont
  {Chong}}, \bibinfo {author} {\bibfnamefont {X.}~\bibnamefont {Liu}}, \bibinfo
  {author} {\bibfnamefont {R.}~\bibnamefont {Sharma}}, \bibinfo {author}
  {\bibfnamefont {A.}~\bibnamefont {Kostin}}, \bibinfo {author} {\bibfnamefont
  {G.}~\bibnamefont {Gu}}, \bibinfo {author} {\bibfnamefont {K.}~\bibnamefont
  {Fujita}}, \bibinfo {author} {\bibfnamefont {J.~S.}\ \bibnamefont {Davis}},\
  and\ \bibinfo {author} {\bibfnamefont {P.~O.}\ \bibnamefont {Sprau}},\
  }\href@noop {} {\bibfield  {journal} {\bibinfo  {journal} {Nano Letters}\
  }\textbf {\bibinfo {volume} {20}},\ \bibinfo {pages} {8001} (\bibinfo {year}
  {2020})}\BibitemShut {NoStop}%
\bibitem [{\citenamefont {Fontein}\ \emph {et~al.}(1991)\citenamefont
  {Fontein}, \citenamefont {Kleinen}, \citenamefont {Hendriks}, \citenamefont
  {Blom}, \citenamefont {Wolter}, \citenamefont {Lochs}, \citenamefont
  {Driessen}, \citenamefont {Giling},\ and\ \citenamefont
  {Beenakker}}]{fontein1991}%
  \BibitemOpen
  \bibfield  {author} {\bibinfo {author} {\bibfnamefont {P.}~\bibnamefont
  {Fontein}}, \bibinfo {author} {\bibfnamefont {J.}~\bibnamefont {Kleinen}},
  \bibinfo {author} {\bibfnamefont {P.}~\bibnamefont {Hendriks}}, \bibinfo
  {author} {\bibfnamefont {F.}~\bibnamefont {Blom}}, \bibinfo {author}
  {\bibfnamefont {J.}~\bibnamefont {Wolter}}, \bibinfo {author} {\bibfnamefont
  {H.}~\bibnamefont {Lochs}}, \bibinfo {author} {\bibfnamefont
  {F.}~\bibnamefont {Driessen}}, \bibinfo {author} {\bibfnamefont
  {L.}~\bibnamefont {Giling}},\ and\ \bibinfo {author} {\bibfnamefont
  {C.}~\bibnamefont {Beenakker}},\ }\href@noop {} {\bibfield  {journal}
  {\bibinfo  {journal} {Physical Review B}\ }\textbf {\bibinfo {volume} {43}},\
  \bibinfo {pages} {12090} (\bibinfo {year} {1991})}\BibitemShut {NoStop}%
\bibitem [{\citenamefont {Ando}(1994)}]{ando1994}%
  \BibitemOpen
  \bibfield  {author} {\bibinfo {author} {\bibfnamefont {T.}~\bibnamefont
  {Ando}},\ }\href@noop {} {\bibfield  {journal} {\bibinfo  {journal} {Physica
  B: Condensed Matter}\ }\textbf {\bibinfo {volume} {201}},\ \bibinfo {pages}
  {331} (\bibinfo {year} {1994})}\BibitemShut {NoStop}%
\bibitem [{\citenamefont {Wiegers}\ \emph {et~al.}(1999)\citenamefont
  {Wiegers}, \citenamefont {Lok}, \citenamefont {Jeuken}, \citenamefont
  {Zeitler}, \citenamefont {Maan},\ and\ \citenamefont {Henini}}]{wiegers1999}%
  \BibitemOpen
  \bibfield  {author} {\bibinfo {author} {\bibfnamefont {S.}~\bibnamefont
  {Wiegers}}, \bibinfo {author} {\bibfnamefont {J.}~\bibnamefont {Lok}},
  \bibinfo {author} {\bibfnamefont {M.}~\bibnamefont {Jeuken}}, \bibinfo
  {author} {\bibfnamefont {U.}~\bibnamefont {Zeitler}}, \bibinfo {author}
  {\bibfnamefont {J.}~\bibnamefont {Maan}},\ and\ \bibinfo {author}
  {\bibfnamefont {M.}~\bibnamefont {Henini}},\ }\href@noop {} {\bibfield
  {journal} {\bibinfo  {journal} {Physical Review B}\ }\textbf {\bibinfo
  {volume} {59}},\ \bibinfo {pages} {7323} (\bibinfo {year}
  {1999})}\BibitemShut {NoStop}%
\bibitem [{\citenamefont {Ahlswede}\ \emph {et~al.}(2001)\citenamefont
  {Ahlswede}, \citenamefont {Weitz}, \citenamefont {Weis}, \citenamefont {{von
  Klitzing}},\ and\ \citenamefont {Eberl}}]{ahlswede2001}%
  \BibitemOpen
  \bibfield  {author} {\bibinfo {author} {\bibfnamefont {E.}~\bibnamefont
  {Ahlswede}}, \bibinfo {author} {\bibfnamefont {P.}~\bibnamefont {Weitz}},
  \bibinfo {author} {\bibfnamefont {J.}~\bibnamefont {Weis}}, \bibinfo {author}
  {\bibfnamefont {K.}~\bibnamefont {{von Klitzing}}},\ and\ \bibinfo {author}
  {\bibfnamefont {K.}~\bibnamefont {Eberl}},\ }\href@noop {} {\bibfield
  {journal} {\bibinfo  {journal} {Phys. B: Condens. Matter}\ }\textbf {\bibinfo
  {volume} {298}},\ \bibinfo {pages} {562} (\bibinfo {year}
  {2001})}\BibitemShut {NoStop}%
\bibitem [{\citenamefont {Suddards}\ \emph {et~al.}(2012)\citenamefont
  {Suddards}, \citenamefont {Baumgartner}, \citenamefont {Henini},\ and\
  \citenamefont {Mellor}}]{suddards2012}%
  \BibitemOpen
  \bibfield  {author} {\bibinfo {author} {\bibfnamefont {M.~E.}\ \bibnamefont
  {Suddards}}, \bibinfo {author} {\bibfnamefont {A.}~\bibnamefont
  {Baumgartner}}, \bibinfo {author} {\bibfnamefont {M.}~\bibnamefont
  {Henini}},\ and\ \bibinfo {author} {\bibfnamefont {C.~J.}\ \bibnamefont
  {Mellor}},\ }\href@noop {} {\bibfield  {journal} {\bibinfo  {journal} {New J.
  Phys.}\ }\textbf {\bibinfo {volume} {14}},\ \bibinfo {pages} {083015}
  (\bibinfo {year} {2012})}\BibitemShut {NoStop}%
\bibitem [{\citenamefont {Panos}\ \emph {et~al.}(2014)\citenamefont {Panos},
  \citenamefont {Gerhardts}, \citenamefont {Weis},\ and\ \citenamefont
  {Von~Klitzing}}]{panos2014}%
  \BibitemOpen
  \bibfield  {author} {\bibinfo {author} {\bibfnamefont {K.}~\bibnamefont
  {Panos}}, \bibinfo {author} {\bibfnamefont {R.}~\bibnamefont {Gerhardts}},
  \bibinfo {author} {\bibfnamefont {J.}~\bibnamefont {Weis}},\ and\ \bibinfo
  {author} {\bibfnamefont {K.}~\bibnamefont {Von~Klitzing}},\ }\href@noop {}
  {\bibfield  {journal} {\bibinfo  {journal} {New Journal of Physics}\ }\textbf
  {\bibinfo {volume} {16}},\ \bibinfo {pages} {113071} (\bibinfo {year}
  {2014})}\BibitemShut {NoStop}%
\bibitem [{\citenamefont {Fijalkowski}\ \emph {et~al.}(2021)\citenamefont
  {Fijalkowski}, \citenamefont {Liu}, \citenamefont {Mandal}, \citenamefont
  {Schreyeck}, \citenamefont {Brunner}, \citenamefont {Gould},\ and\
  \citenamefont {Molenkamp}}]{fijalkowski2021}%
  \BibitemOpen
  \bibfield  {author} {\bibinfo {author} {\bibfnamefont {K.~M.}\ \bibnamefont
  {Fijalkowski}}, \bibinfo {author} {\bibfnamefont {N.}~\bibnamefont {Liu}},
  \bibinfo {author} {\bibfnamefont {P.}~\bibnamefont {Mandal}}, \bibinfo
  {author} {\bibfnamefont {S.}~\bibnamefont {Schreyeck}}, \bibinfo {author}
  {\bibfnamefont {K.}~\bibnamefont {Brunner}}, \bibinfo {author} {\bibfnamefont
  {C.}~\bibnamefont {Gould}},\ and\ \bibinfo {author} {\bibfnamefont {L.~W.}\
  \bibnamefont {Molenkamp}},\ }\href@noop {} {\bibfield  {journal} {\bibinfo
  {journal} {Nat. Commun.}\ }\textbf {\bibinfo {volume} {12}},\ \bibinfo
  {pages} {5599} (\bibinfo {year} {2021})}\BibitemShut {NoStop}%
\bibitem [{\citenamefont {Ferguson}\ \emph {et~al.}(2021)\citenamefont
  {Ferguson}, \citenamefont {Xiao}, \citenamefont {Richardella}, \citenamefont
  {Low}, \citenamefont {Samarth},\ and\ \citenamefont {Nowack}}]{ferguson2021}%
  \BibitemOpen
  \bibfield  {author} {\bibinfo {author} {\bibfnamefont {G.}~\bibnamefont
  {Ferguson}}, \bibinfo {author} {\bibfnamefont {R.}~\bibnamefont {Xiao}},
  \bibinfo {author} {\bibfnamefont {A.~R.}\ \bibnamefont {Richardella}},
  \bibinfo {author} {\bibfnamefont {D.}~\bibnamefont {Low}}, \bibinfo {author}
  {\bibfnamefont {N.}~\bibnamefont {Samarth}},\ and\ \bibinfo {author}
  {\bibfnamefont {K.~C.}\ \bibnamefont {Nowack}},\ }\href@noop {} {\bibfield
  {journal} {\bibinfo  {journal} {arXiv preprint arXiv:2112.13122}\ } (\bibinfo
  {year} {2021})}\BibitemShut {NoStop}%
\bibitem [{\citenamefont {Van Der~Wel}\ \emph {et~al.}(1988)\citenamefont {Van
  Der~Wel}, \citenamefont {Harmans},\ and\ \citenamefont {Mooij}}]{van1988}%
  \BibitemOpen
  \bibfield  {author} {\bibinfo {author} {\bibfnamefont {W.}~\bibnamefont {Van
  Der~Wel}}, \bibinfo {author} {\bibfnamefont {C.}~\bibnamefont {Harmans}},\
  and\ \bibinfo {author} {\bibfnamefont {J.}~\bibnamefont {Mooij}},\
  }\href@noop {} {\bibfield  {journal} {\bibinfo  {journal} {J. Phys. C: Solid
  State Phys}\ }\textbf {\bibinfo {volume} {21}},\ \bibinfo {pages} {L171}
  (\bibinfo {year} {1988})}\BibitemShut {NoStop}%
\bibitem [{\citenamefont {Klaß}\ \emph {et~al.}(1991)\citenamefont {Klaß},
  \citenamefont {Dietsche}, \citenamefont {von Klitzing},\ and\ \citenamefont
  {Ploog}}]{klab1991}%
  \BibitemOpen
  \bibfield  {author} {\bibinfo {author} {\bibfnamefont {U.}~\bibnamefont
  {Klaß}}, \bibinfo {author} {\bibfnamefont {W.}~\bibnamefont {Dietsche}},
  \bibinfo {author} {\bibfnamefont {K.}~\bibnamefont {von Klitzing}},\ and\
  \bibinfo {author} {\bibfnamefont {K.}~\bibnamefont {Ploog}},\ }\href@noop {}
  {\bibfield  {journal} {\bibinfo  {journal} {Zeitschrift für Physik B
  Condensed Matter}\ }\textbf {\bibinfo {volume} {82}},\ \bibinfo {pages} {351
  } (\bibinfo {year} {1991})}\BibitemShut {NoStop}%
\bibitem [{\citenamefont {Komiyama}\ \emph {et~al.}(1996)\citenamefont
  {Komiyama}, \citenamefont {Kawaguchi}, \citenamefont {Osada},\ and\
  \citenamefont {Shiraki}}]{komiyama1996}%
  \BibitemOpen
  \bibfield  {author} {\bibinfo {author} {\bibfnamefont {S.}~\bibnamefont
  {Komiyama}}, \bibinfo {author} {\bibfnamefont {Y.}~\bibnamefont {Kawaguchi}},
  \bibinfo {author} {\bibfnamefont {T.}~\bibnamefont {Osada}},\ and\ \bibinfo
  {author} {\bibfnamefont {Y.}~\bibnamefont {Shiraki}},\ }\href@noop {}
  {\bibfield  {journal} {\bibinfo  {journal} {Phys. Rev. Lett.}\ }\textbf
  {\bibinfo {volume} {77}},\ \bibinfo {pages} {558} (\bibinfo {year}
  {1996})}\BibitemShut {NoStop}%
\bibitem [{\citenamefont {Dolan}(1999)}]{dolan1999}%
  \BibitemOpen
  \bibfield  {author} {\bibinfo {author} {\bibfnamefont {B.~P.}\ \bibnamefont
  {Dolan}},\ }\href@noop {} {\bibfield  {journal} {\bibinfo  {journal} {Nuc.
  Phys. B}\ }\textbf {\bibinfo {volume} {554}},\ \bibinfo {pages} {487}
  (\bibinfo {year} {1999})}\BibitemShut {NoStop}%
\bibitem [{\citenamefont {Delahaye}\ and\ \citenamefont
  {Jeckelmann}(2003)}]{delahaye2003}%
  \BibitemOpen
  \bibfield  {author} {\bibinfo {author} {\bibfnamefont {F.}~\bibnamefont
  {Delahaye}}\ and\ \bibinfo {author} {\bibfnamefont {B.}~\bibnamefont
  {Jeckelmann}},\ }\href@noop {} {\bibfield  {journal} {\bibinfo  {journal}
  {Metrologia}\ }\textbf {\bibinfo {volume} {40}},\ \bibinfo {pages} {217}
  (\bibinfo {year} {2003})}\BibitemShut {NoStop}%
\bibitem [{\citenamefont {Granger}\ \emph {et~al.}(2009)\citenamefont
  {Granger}, \citenamefont {Eisenstein},\ and\ \citenamefont
  {Reno}}]{granger2009}%
  \BibitemOpen
  \bibfield  {author} {\bibinfo {author} {\bibfnamefont {G.}~\bibnamefont
  {Granger}}, \bibinfo {author} {\bibfnamefont {J.}~\bibnamefont
  {Eisenstein}},\ and\ \bibinfo {author} {\bibfnamefont {J.}~\bibnamefont
  {Reno}},\ }\href@noop {} {\bibfield  {journal} {\bibinfo  {journal} {Phys.
  Rev. Lett.}\ }\textbf {\bibinfo {volume} {102}},\ \bibinfo {pages} {086803}
  (\bibinfo {year} {2009})}\BibitemShut {NoStop}%
\bibitem [{\citenamefont {Checkelsky}\ \emph {et~al.}(2014)\citenamefont
  {Checkelsky}, \citenamefont {Yoshimi}, \citenamefont {Tsukazaki},
  \citenamefont {Takahashi}, \citenamefont {Kozuka}, \citenamefont {Falson},
  \citenamefont {Kawasaki},\ and\ \citenamefont {Tokura}}]{checkelsky2014}%
  \BibitemOpen
  \bibfield  {author} {\bibinfo {author} {\bibfnamefont {J.}~\bibnamefont
  {Checkelsky}}, \bibinfo {author} {\bibfnamefont {R.}~\bibnamefont {Yoshimi}},
  \bibinfo {author} {\bibfnamefont {A.}~\bibnamefont {Tsukazaki}}, \bibinfo
  {author} {\bibfnamefont {K.}~\bibnamefont {Takahashi}}, \bibinfo {author}
  {\bibfnamefont {Y.}~\bibnamefont {Kozuka}}, \bibinfo {author} {\bibfnamefont
  {J.}~\bibnamefont {Falson}}, \bibinfo {author} {\bibfnamefont
  {M.}~\bibnamefont {Kawasaki}},\ and\ \bibinfo {author} {\bibfnamefont
  {Y.}~\bibnamefont {Tokura}},\ }\href@noop {} {\bibfield  {journal} {\bibinfo
  {journal} {Nat. Phys.}\ }\textbf {\bibinfo {volume} {10}},\ \bibinfo {pages}
  {731} (\bibinfo {year} {2014})}\BibitemShut {NoStop}%
\bibitem [{\citenamefont {Lachman}\ \emph {et~al.}(2015)\citenamefont
  {Lachman}, \citenamefont {Young}, \citenamefont {Richardella}, \citenamefont
  {Cuppens}, \citenamefont {Naren}, \citenamefont {Anahory}, \citenamefont
  {Meltzer}, \citenamefont {Kandala}, \citenamefont {Kempinger}, \citenamefont
  {Myasoedov} \emph {et~al.}}]{lachman2015visualization}%
  \BibitemOpen
  \bibfield  {author} {\bibinfo {author} {\bibfnamefont {E.~O.}\ \bibnamefont
  {Lachman}}, \bibinfo {author} {\bibfnamefont {A.~F.}\ \bibnamefont {Young}},
  \bibinfo {author} {\bibfnamefont {A.}~\bibnamefont {Richardella}}, \bibinfo
  {author} {\bibfnamefont {J.}~\bibnamefont {Cuppens}}, \bibinfo {author}
  {\bibfnamefont {H.}~\bibnamefont {Naren}}, \bibinfo {author} {\bibfnamefont
  {Y.}~\bibnamefont {Anahory}}, \bibinfo {author} {\bibfnamefont {A.~Y.}\
  \bibnamefont {Meltzer}}, \bibinfo {author} {\bibfnamefont {A.}~\bibnamefont
  {Kandala}}, \bibinfo {author} {\bibfnamefont {S.}~\bibnamefont {Kempinger}},
  \bibinfo {author} {\bibfnamefont {Y.}~\bibnamefont {Myasoedov}}, \emph
  {et~al.},\ }\href@noop {} {\bibfield  {journal} {\bibinfo  {journal} {Science
  advances}\ }\textbf {\bibinfo {volume} {1}},\ \bibinfo {pages} {e1500740}
  (\bibinfo {year} {2015})}\BibitemShut {NoStop}%
\bibitem [{\citenamefont {Rosen}\ \emph {et~al.}(2017)\citenamefont {Rosen},
  \citenamefont {Fox}, \citenamefont {Kou}, \citenamefont {Pan}, \citenamefont
  {Wang},\ and\ \citenamefont {Goldhaber-Gordon}}]{rosen2017}%
  \BibitemOpen
  \bibfield  {author} {\bibinfo {author} {\bibfnamefont {I.~T.}\ \bibnamefont
  {Rosen}}, \bibinfo {author} {\bibfnamefont {E.~J.}\ \bibnamefont {Fox}},
  \bibinfo {author} {\bibfnamefont {X.}~\bibnamefont {Kou}}, \bibinfo {author}
  {\bibfnamefont {L.}~\bibnamefont {Pan}}, \bibinfo {author} {\bibfnamefont
  {K.~L.}\ \bibnamefont {Wang}},\ and\ \bibinfo {author} {\bibfnamefont
  {D.}~\bibnamefont {Goldhaber-Gordon}},\ }\href@noop {} {\bibfield  {journal}
  {\bibinfo  {journal} {npj Quantum Materials}\ }\textbf {\bibinfo {volume}
  {2}},\ \bibinfo {pages} {1} (\bibinfo {year} {2017})}\BibitemShut {NoStop}%
\bibitem [{\citenamefont {Wang}\ \emph {et~al.}(2018)\citenamefont {Wang},
  \citenamefont {Ou}, \citenamefont {Liu}, \citenamefont {Wang}, \citenamefont
  {He}, \citenamefont {Xue},\ and\ \citenamefont {Wu}}]{wang2018direct}%
  \BibitemOpen
  \bibfield  {author} {\bibinfo {author} {\bibfnamefont {W.}~\bibnamefont
  {Wang}}, \bibinfo {author} {\bibfnamefont {Y.}~\bibnamefont {Ou}}, \bibinfo
  {author} {\bibfnamefont {C.}~\bibnamefont {Liu}}, \bibinfo {author}
  {\bibfnamefont {Y.}~\bibnamefont {Wang}}, \bibinfo {author} {\bibfnamefont
  {K.}~\bibnamefont {He}}, \bibinfo {author} {\bibfnamefont {Q.-K.}\
  \bibnamefont {Xue}},\ and\ \bibinfo {author} {\bibfnamefont {W.}~\bibnamefont
  {Wu}},\ }\href@noop {} {\bibfield  {journal} {\bibinfo  {journal} {Nature
  Physics}\ }\textbf {\bibinfo {volume} {14}},\ \bibinfo {pages} {791}
  (\bibinfo {year} {2018})}\BibitemShut {NoStop}%
\bibitem [{\citenamefont {Chen}\ \emph {et~al.}(2008)\citenamefont {Chen},
  \citenamefont {Jang}, \citenamefont {Adam}, \citenamefont {Fuhrer},
  \citenamefont {Williams},\ and\ \citenamefont {Ishigami}}]{chen2008}%
  \BibitemOpen
  \bibfield  {author} {\bibinfo {author} {\bibfnamefont {J.-H.}\ \bibnamefont
  {Chen}}, \bibinfo {author} {\bibfnamefont {C.}~\bibnamefont {Jang}}, \bibinfo
  {author} {\bibfnamefont {S.}~\bibnamefont {Adam}}, \bibinfo {author}
  {\bibfnamefont {M.}~\bibnamefont {Fuhrer}}, \bibinfo {author} {\bibfnamefont
  {E.~D.}\ \bibnamefont {Williams}},\ and\ \bibinfo {author} {\bibfnamefont
  {M.}~\bibnamefont {Ishigami}},\ }\href@noop {} {\bibfield  {journal}
  {\bibinfo  {journal} {Nat. Phys.}\ }\textbf {\bibinfo {volume} {4}},\
  \bibinfo {pages} {377} (\bibinfo {year} {2008})}\BibitemShut {NoStop}%
\end{thebibliography}%

\clearpage
% \onecolumngrid
\title{Supplemental material for ‘‘Measured potential profile in a quantum anomalous Hall system suggests bulk-dominated current flow’’}
\maketitle
\begin{center}
\onecolumngrid
% \vspace*{0.4cm}
% \begin{center}
% {\large\bf Supplementary information}\\
% \vspace*{0.4cm}
\end{center}

\renewcommand{\thetable}{S\arabic{table}}
\renewcommand{\thefigure}{S\arabic{figure}}
\renewcommand{\thesection}{S\arabic{section}}
\renewcommand{\thesubsection}{\roman{subsection}}
\renewcommand{\theequation}{S\arabic{equation}}

\setcounter{secnumdepth}{3}
\setcounter{equation}{0}
\setcounter{figure}{0}

\startcontents[SMrefs]

\titlecontents{psection}[3em]
{} {\contentslabel{2em}} {} {\titlerule*[1pc]{.}\contentspage}
\titlecontents{psubsection}[5.5em]
{} {\contentslabel{2em}} {} {\titlerule*[1pc]{.}\contentspage}
\vspace{5mm}
\printcontents[SMrefs]{p}{1}[2]{\textbf{Table of contents:\medskip
}}

\clearpage

\section{Details of experimental methodology}

\subsection{Device geometry}

\begin{figure}[h!]
\centering
	\includegraphics[width=0.5\textwidth]{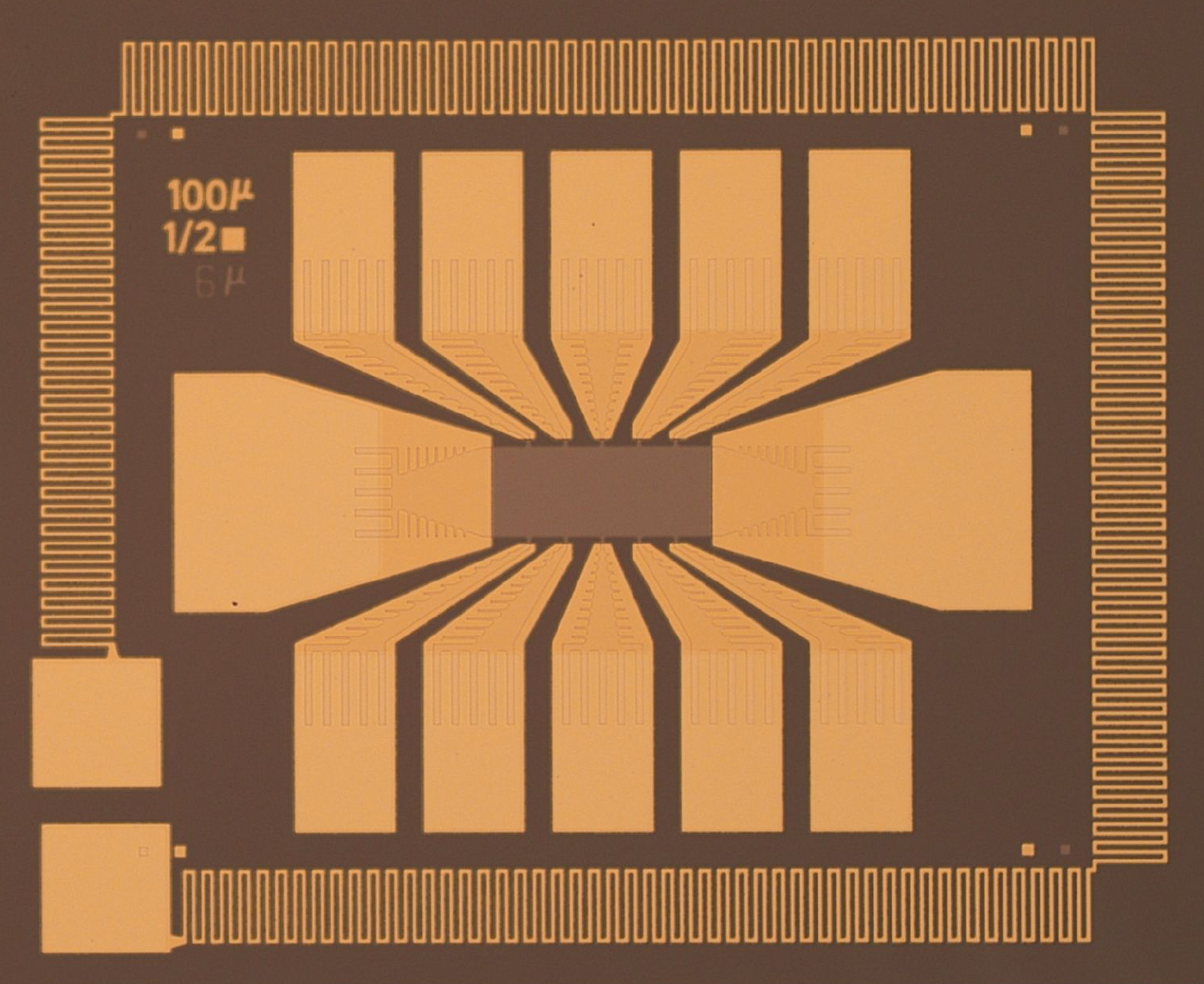}
	\caption{Optical micrograph of Device~1. The ``lattice heater" (see Section~\ref{sec:heaters}) is the Au meander surrounding the Hall bar.}
	\label{sfig_claret}
\end{figure}

The device presented in the main text, which will be referred to in the supplemental material as Device~1, is a Hall bar with a total length (distance between source and drain contacts) of 240~$\mu$m. The source and drain contacts span the full 100~$\mu$m width of the device. The five voltage terminals on each edge are spaced 40~$\mu$m center-to-center apart, and the first (last) voltage terminal is 40~$\mu$m center-to-edge from the source (drain) contact. Each voltage terminal is formed by a 6~$\mu$m wide channel of the QAH insulator, which leads to a metal contact. The metal contact of each voltage terminal has a setback of a few microns from the edge of the Hall bar to prevent unintentional doping from the work function mismatch of the contact from affecting the electronic system in the primary Hall bar region. Device~1 is shown in Fig.~\ref{sfig_claret}.

\subsection{Electrical measurements}\label{sec:electrical}

The eight voltages $V^{\alpha i}$ measured in Device~1 were measured simultaneously using separate Stanford Research Systems 830 and 860 lock-in amplifiers. Voltages are amplified using differential dc preamplifiers prior to lock-in amplification. Voltages $V^{\mathrm{T1-T4}}$ were preamplified with separate NF LI-75A preamplifiers, which have an input impedance of 100~M$\Omega$. Voltages $V^{\mathrm{B1-B4}}$ were preamplified with a single NF multi-channel preamplifier, with an input impedance of 1~M$\Omega$. Current is amplified with an Ithaco 1211 current preamplifier at $10^6$~V/A gain with 20~$\Omega$ input impedance. Lock-in measurements use a frequency of roughly 7~Hz.

Leakage current through the voltage preamplifiers leads to offsets in the measured four-terminal resistances~\cite{rodenbach2021}. In Fig.~3(b) of the main text and Fig.~\ref{sfig_nB_data}(c),we correct for the leakage current by subtracting the mean of measurements taken below 90~mK (temperatures low enough that the measurement is not sensitive enough to detect thermally-induced dissipation) from the data. This correction is not made to any of the other data we present. The ranges indicated by the error bars in Fig.~3(b) of the main text and Fig.~\ref{sfig_nB_data}(c) are determined by computing the standard deviation of measurements taken below 90~mK, and calculating the extreme values obtained when adding and subtracting the standard deviation from the numerator and denominator of the ratio of resistivities $\rho_{xx}^{\alpha i}/\rho_{xx}^{\mathrm{T3}}$.

\subsection{Material characterization}

\begin{figure}[H]
\centering
	\includegraphics[width=0.8\textwidth]{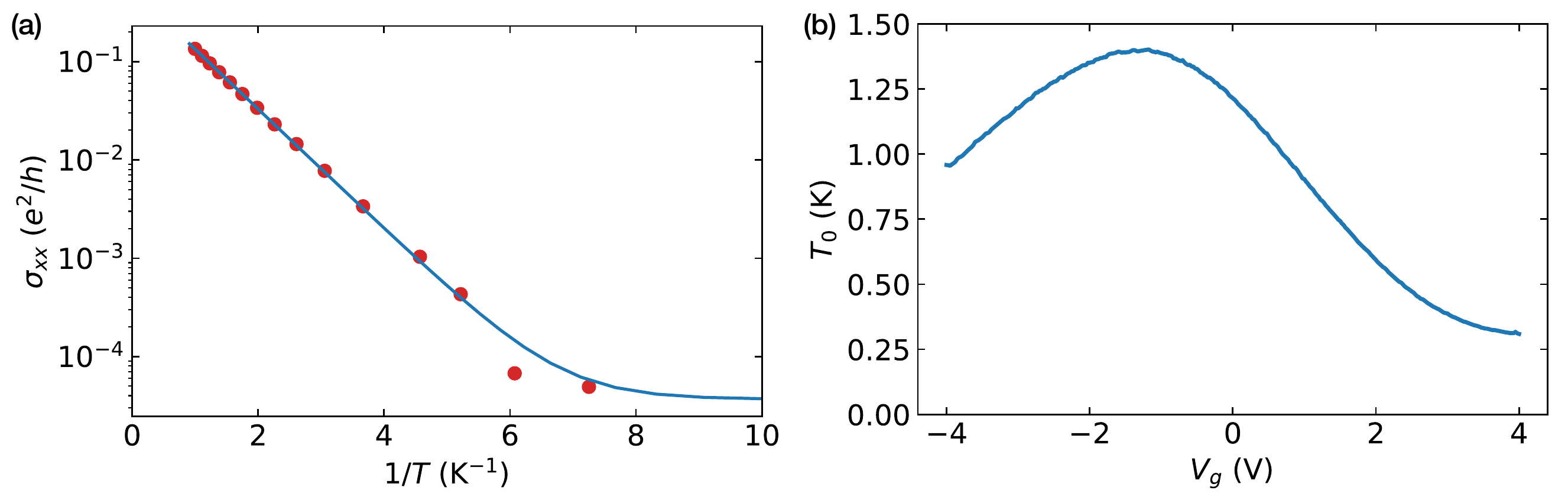}
	\caption{(a) Arrhenius plot of $\sigma_{xx}$ as a function of $1/T$ at optimal gate voltage $V_g=-1.18$~V. Measurements are shown in red circles and and fit ($\sigma_{xx} = a + be^{-T_0/T}$ for constants $a$ and $b$) as a blue line. (b) Fitted temperature scale $T_0$ for thermally-activated conduction as a function of gate voltage. Maximum temperature scale $T_0=1.40$~K occurs at $V_g=-1.18$~V.}
	\label{sfig-arrhenius}
\end{figure}

The temperature scale for thermally activated conduction, $T_0$, is extracted by fitting $\sigma_{xx} \propto e^{-T_0/T}$ for each value of gate voltage. As shown in Fig.~\ref{sfig-arrhenius}b, the maximum value of $T_0=1.40$~K at $V_g=-1.18$~V; this temperature scale provides an effective size of the magnetic exchange gap. Most of the data in this work were taken at $V_g=-1$~V, at which $T_0=1.39$~K. Saturation of $\sigma_{xx}$ at the lowest temperatures of Fig.~\ref{sfig-arrhenius}a -- where electron temperature deviates from the measured lattice temperature or Arrhenius behavior is cut off due to the strong electric field from the current bias -- is modeled by the inclusion of a temperature-independent offset to the Arrhenius form. 

\clearpage
\section{Further discussion of current flow}

In the main text, we present an argument for why the source-drain current should flow entirely through the bulk in the dissipationless QAH scenario. Here we expand on these ideas. 

First, in the main text, we argued that the electrostatic potential should satisfy Laplace's equation $\nabla^2 V=0$, not just Poisson's equation $\nabla^2 V=-\rho /\epsilon$, based on conservation of charge $\nabla \mathbf{j}=0$ and Ohm's law $\mathbf{j}=\boldsymbol{\sigma}\mathbf{E}$. We qualify that this argument requires the conductivity tensor $\boldsymbol{\sigma}$ to be nonzero, which is here the case as $\sigma_{xy}\approx e^2/h$. This argument also requires $\boldsymbol{\sigma}$ to be homogenous, which is here not strictly the case due to fluctuations in the gap from disorder. Yet we expect the length scale of fluctuations to be tens to hundreds of nm~\cite{lachman2015visualization, lee2015, wang2018direct}, significantly smaller than our hundred micron-scale device. So, on the length scales across which we measure, we expect fluctuations in the conductivity to average out so that Laplace's equation approximately holds.

\begin{figure}[h!]
\centering
	\includegraphics[width=0.7\textwidth]{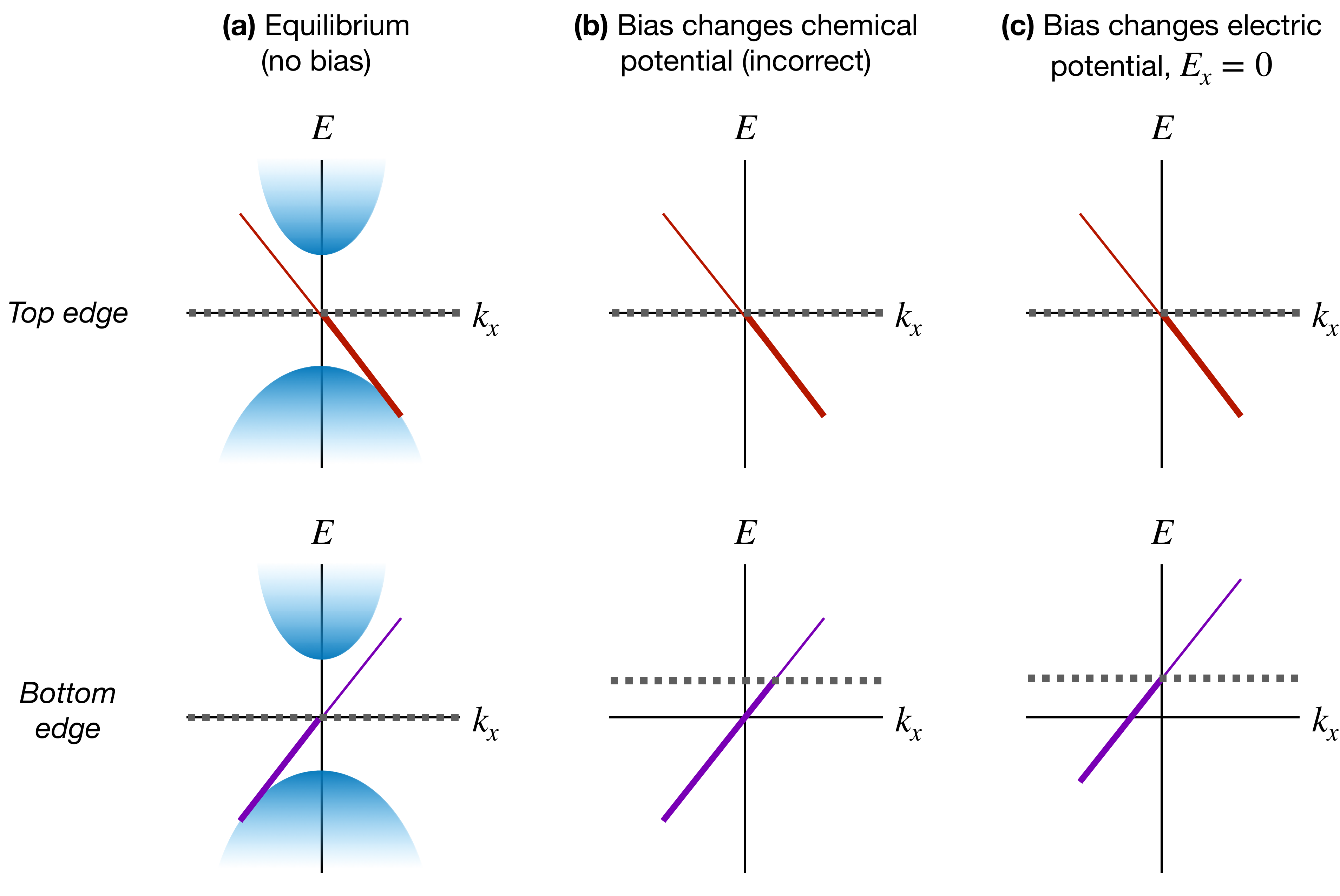}
	\caption{Sketches of how an electrochemical bias alters band structure of the QAH CEM, here treating contributions from the electrostatic and chemical potentials separately. The horizontal axis represents momentum parallel to the edge, and the vertical axis represents electrochemical energy. The CEM, shown in red (purple) for the top (bottom) edge has opposite phase velocity on the two edges of a Hall bar. The QAH state is established when the Fermi level or electrochemical potential (dotted line) is in the exchange gap between surface state bands (blue; omitted in (b,~c) for clarity). (a) In the absence of source-drain bias, the electrochemical potential is the same on both edges; here we define it as zero and, for simplicity, we take it to intersect the CEM at $k=0$. Since the CEM has occupied $k_x$ states on the top edge and ($-k_x$ states on the bottom) edge, a persistent current (but no net left-to-right current) flows around the device in equilibrium. (b) A difference in chemical potential between the CEM on the two sides of the device means there are more occupied states on one edge than the other, leading to a net current carried by the edge mode. (c) A difference in electrostatic potential also causes the electrochemical potential on one edge to rise, yet the number of occupied edge states is unchanged. Thus no non-equilibrium current flows through the CEM. Non-equilibrium current still flows, but it is carried by filled bands in the two-dimensional bulk.}
	\label{sfig_band_struct}
\end{figure}

Next, we expand on the alternative limiting view of current flow: the Landauer-B\"uttiker picture, where non-equilibrium (source-drain) current travels through the chiral edge mode (CEM). Charge carriers on the top and bottom edges have opposing velocities, the sign of which are determined by the system's chirality. For a net source-drain current to flow through the device, a different number of electrons must flow on the top edge than on the bottom edge, (unless the electrons on one edge differ in velocity from those on the other edge). This scenario occurs when the electrochemical bias between the source and drain contact manifests as a chemical potential difference in the edge mode. The fraction of current carried in this manner has been called ``Fermi-surface current" because the net current is carried by states near the Fermi level~\cite{komiyama1996representations}. We present a schematic of this contribution in Fig.~\ref{sfig_band_struct}(b).

In the main text, we showed that our experiment suggests that the source-drain bias is primarily manifested as an electrostatic potential gradient. Changing the electric potential raises the electrochemical energy of the chiral edge mode but does not change the number of occupied states, as shown in Fig.~\ref{sfig_band_struct}(c). Instead, the bulk, which is incompressible, carries the current according to $j_x = \sigma_{xy}E_y = (e^2/h) E_y$, which produces a quantized response
\begin{equation}
    I_x = \frac{e^2}{h} \int_0^W E_y dy= \frac{e^2}{h} V_S.
\end{equation}
The bulk may be viewed as gapped to single-particle excitations but carrying a dissipationless current as a displacement of its center of momentum. Note that the net current through the bulk depends only on the total electrostatic potential difference between the two edges $V_S$, and does not depend on the particular field profile $E_y(x, y)$.

While our experiment indicates that dissipationless current flows mostly through the bulk, not the edge, of devices, it does not elucidate the nature of the electronic states responsible for carrying the current. Quantization of the anomalous Hall effect requires that the bulk---consisting of both the three-dimensional bands (as confined to a film of sub-10nm-thickness) and the two-dimensional topological surface states---is fully gapped. In principle, non-equilibrium current should flow through the occupied bands that have a non-zero Chern number.

We now consider the dissipative regime, when $\sigma_{xx}$ is nonzero. A longitudinal electric field develops, so the longitudinal current through the bulk includes a dissipative component $\sigma_{xx}E_x$. There should also be a transverse current in the bulk $\sigma_{xy}E_x + \sigma_{xx}E_y$ where we assume isotropic conductivity $\sigma_{xx}=\sigma_{yy}$. Our experiment focused on the small dissipation regime $\sigma_{xx}\ll\sigma_{xy}$, $E_x\ll E_y$, where we expect the transverse current to be small. As a thought experiment, however, we speculate that transverse currents would become substantial when $\sigma_{xx}\sim\sigma_{xy}$. In this regime, the transverse current would transfer charge between the two edges of the Hall bar (as is the case in the classical Hall effect). Assuming a finite scattering rate between the bulk and the CEM, this would cause chemical potential gradients along the CEM, and therefore non-equilibrium currents in the CEM. Though our experiment did not explore a regime where dissipation is high enough for these currents to be substantial, and our experimental technique is not well-equipped to study situations with high chemical potential gradients (since the electrochemical potential of the contacts may couple to the chemical potentials of both the CEM and bulk states), we speculate that an experiment that can directly image the current density may detect non-equilibrium current through the CEM in the regime of high dissipation (but not in the nearly-dissipationless regime).

\clearpage
\section{Comparison with quantum Hall systems}

The QAH and QH systems feature similar phenomenology ($\sigma_{xy} = (\nu)e^2/h$) owing to their common topology (nonzero Chern number). Here, we wish to discuss three interesting microscopic differences between the two systems.

First, the QH state is realized at finite doping, so electron-hole symmetry is broken. In contrast, at optimum gate voltage, which corresponds (roughly~\cite{chen2008}) to the center of the magnetic exchange gap, the QAH system is approximately undoped. Excursions of the gate voltage away from $V_g^\mathrm{opt}$ induce dissipation in the QAH system. Intriguingly, we observe that gate voltage excursion breaks the two-fold rotational symmetry of the measured apparent resistivities, e.g., $\rho_{xx}^{\mathrm{T1}}$ and $\rho_{xx}^{\mathrm{B4}}$ differ slightly (data shown in Section~\ref{sec:gate}). We speculate that the approximate electron-hole symmetry in the QAH system is related to this centrosymmetry we observe in apparent resistivity. When electron-hole symmetry is broken---in the QAH system away from $V_g^\mathrm{opt}$, as we observe, or in the QH system, as observed in previous studies~\cite{klab1991, komiyama1996}---centrosymmetric apparent resistivities are no longer guaranteed.

Second, in the semiconductor quantum wells in which the QH effect is best-known CEMs are confined near edges, which typically present an edge depletion potential---a strong in-plane electrostatic potential gradient transverse to the edge. The QAH effect is here realized in a three-dimensional, albeit thin, insulator surrounded by two-dimensional topological surface states gapped by magnetic exchange. The CEM is expected to have support on and near the sidewalls, or at internal magnetic domain walls if such are present~\cite{rosen2017}, but there is not a close analogue of an edge depletion potential.

Third, whereas QH states can host multiple Landau levels, forming a series of compressible stripes along the edges of a device where each Landau level intersects the edge depletion potential, the QAH state has no Landau levels. It is now understood that non-equilibrium current in the QH system flows primarily through incompressible stripes near the edges and/or incompressible portions of the bulk, depending on microscopic details~\cite{fontein1991, wiegers1999, ahlswede2001, weis2011, suddards2012, panos2014}. The dissipationless propagation of current through the QH system is, in this picture, related to the incompressibility of the regions through which it flows. Since the QAH system lacks interleaved compressible and incompressible stripes, by analogy it is then natural to expect current flow through the incompressible bulk.

\clearpage
\section{Reversing the magnetization}

In the main text, we show data taken when the magnetization of the QAH film is set to be positive by applying a positive applied field larger than the film's coercive field, and then ramping the applied field back to zero. In this section, we explore the transport when the film has negative magnetization.

Reversing the magnetization of the film changes the sign of the Hall conductivity, which, in turn, changes the chirality of the CEM and (for high current bias) changes the corners in which the hot spots lie. For positive magnetization, the CEM has counter-clockwise chirality and the hot spots are in the top left and bottom right corners, meaning that the top edge of the device sits at the drain potential. For negative magnetization, the CEM has clockwise chirality and the hot spots are in the top left and bottom right corners, meaning that the top edge of the device sits at the source potential.

\begin{figure}[h!]
\centering
	\includegraphics[width=0.6\textwidth]{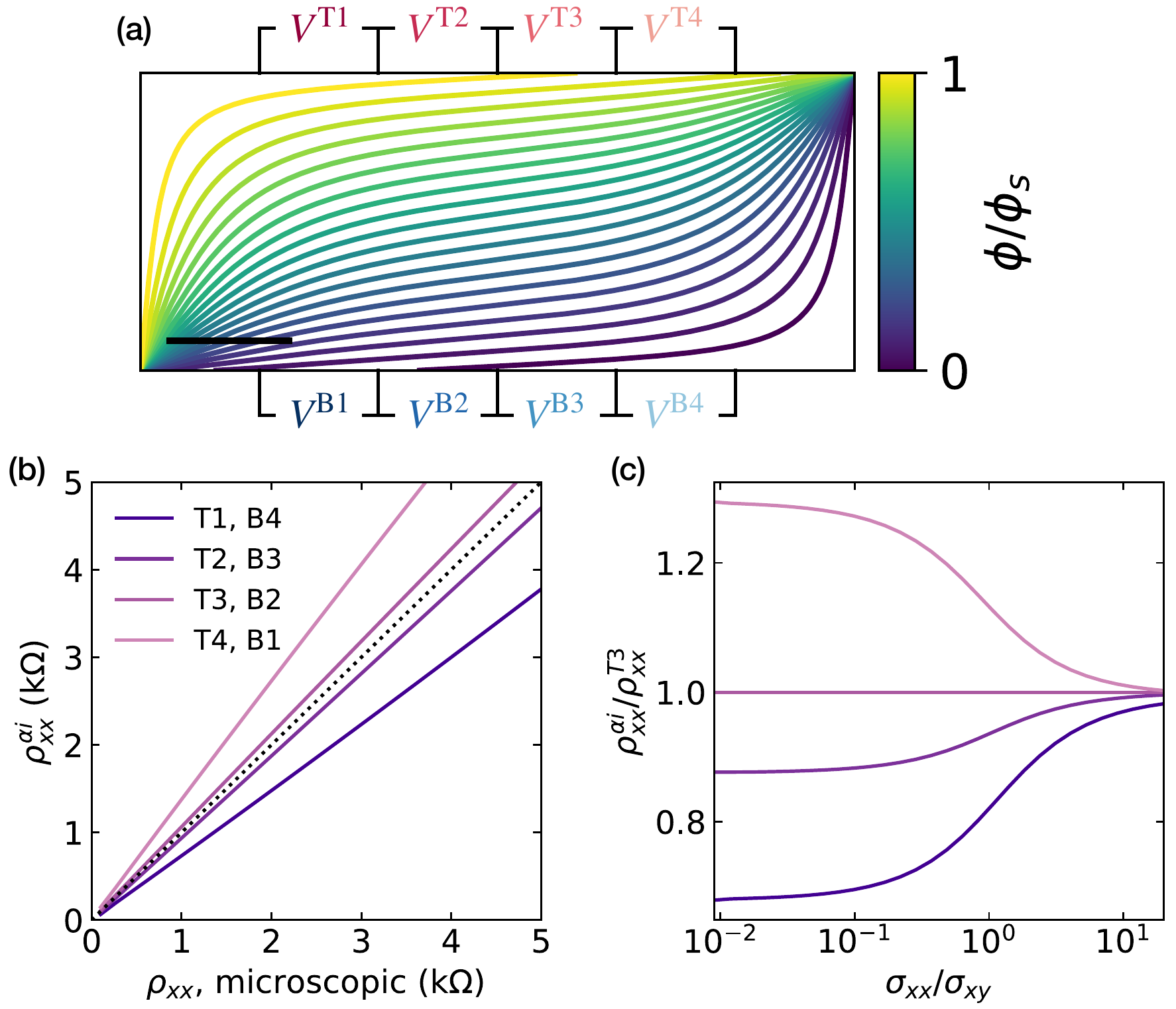}
	\caption{Simulations of the potential with negative magnetization. (a) Contour plot of equipotentials in the Hall bar simulated at $\sigma_{xx}/\sigma_{xy}=0.05$, with negative magnetization (corresponding to clockwise chirality). The electric field is concentrated near the top right and bottom left corners. Scale bar, 40~$\mu$m. (b) Simulated apparent resistivities $\rho_{xx}^{\alpha i}$ as a function of the spatially-homogeneous microscopic resistivity of the material. The dotted line with a slope of 1 is shown for reference. (c) The ratio of simulated resistivities $\rho_{xx}^{\alpha i}/\rho_{xx}^{\mathrm{T3}}$. Each such ratio becomes only weakly dependent on $\sigma_{xx}/\sigma_{xy}$ as that ratio approaches zero.}
	\label{sfig_nB_sim}
\end{figure}

\begin{figure}[h!]
\centering
	\includegraphics[width=0.9\textwidth]{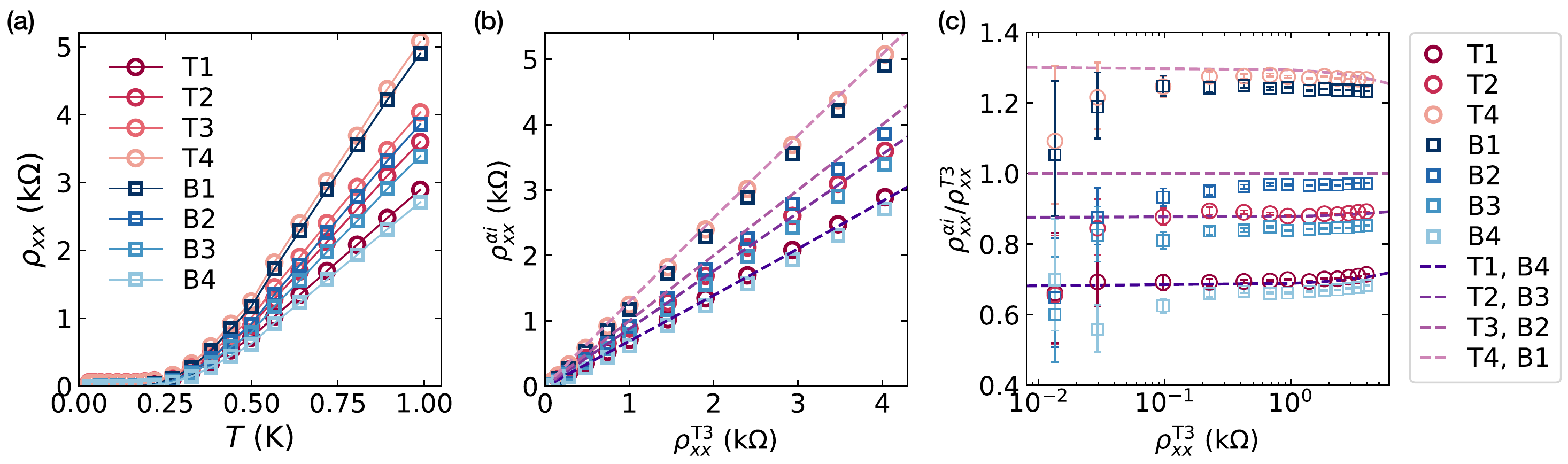}
	\caption{Measured apparent resistivities with negative magnetization. (a) The apparent resistivities as a function of temperature. (a) The apparent resistivity as a function of temperature, plotted parametrically versus measurement T3. The simulated behavior (Fig.~\ref{sfig_nB_sim}(b)) is shown by the dashed violet lines. (b) The ratio of measured apparent resistivities $\rho_{xx}^{\alpha i}/\rho_{xx}^{\mathrm{T3}}$, shown as a function of measured $\rho_{xx}^{\mathrm{T3}}$. Simulated behavior (Fig.~\ref{sfig_nB_sim}(c)) is shown by the dashed violet lines.}
	\label{sfig_nB_data}
\end{figure}

Since the sign of the Hall conductivity is reversed, we expect the monotonic change of the apparent resistivities to reverse direction, as shown in Fig.~\ref{sfig_nB_sim}, so that now $\rho_{xx}^\mathrm{T4, B1}$ are the highest rather than $\rho_{xx}^\mathrm{T1, B4}$. Otherwise, we do not expect different behavior when the magnetization is negative versus positive. Measurements with negative magnetization align with these expectations~\ref{sfig_nB_data}. Note that, with negative magnetization, leakage currents (Section~\ref{sec:electrical}) primarily affect measurements along the top edge of the device (since the voltage of these contacts tracks the source voltage, whereas with positive magnetization the voltage of the top contacts tracks the drain voltage and the voltage of the bottom contacts tracks the source voltage).

\clearpage
\section{Bias dependence}

\begin{figure}[h!]
\centering
	\includegraphics[width=0.9\textwidth]{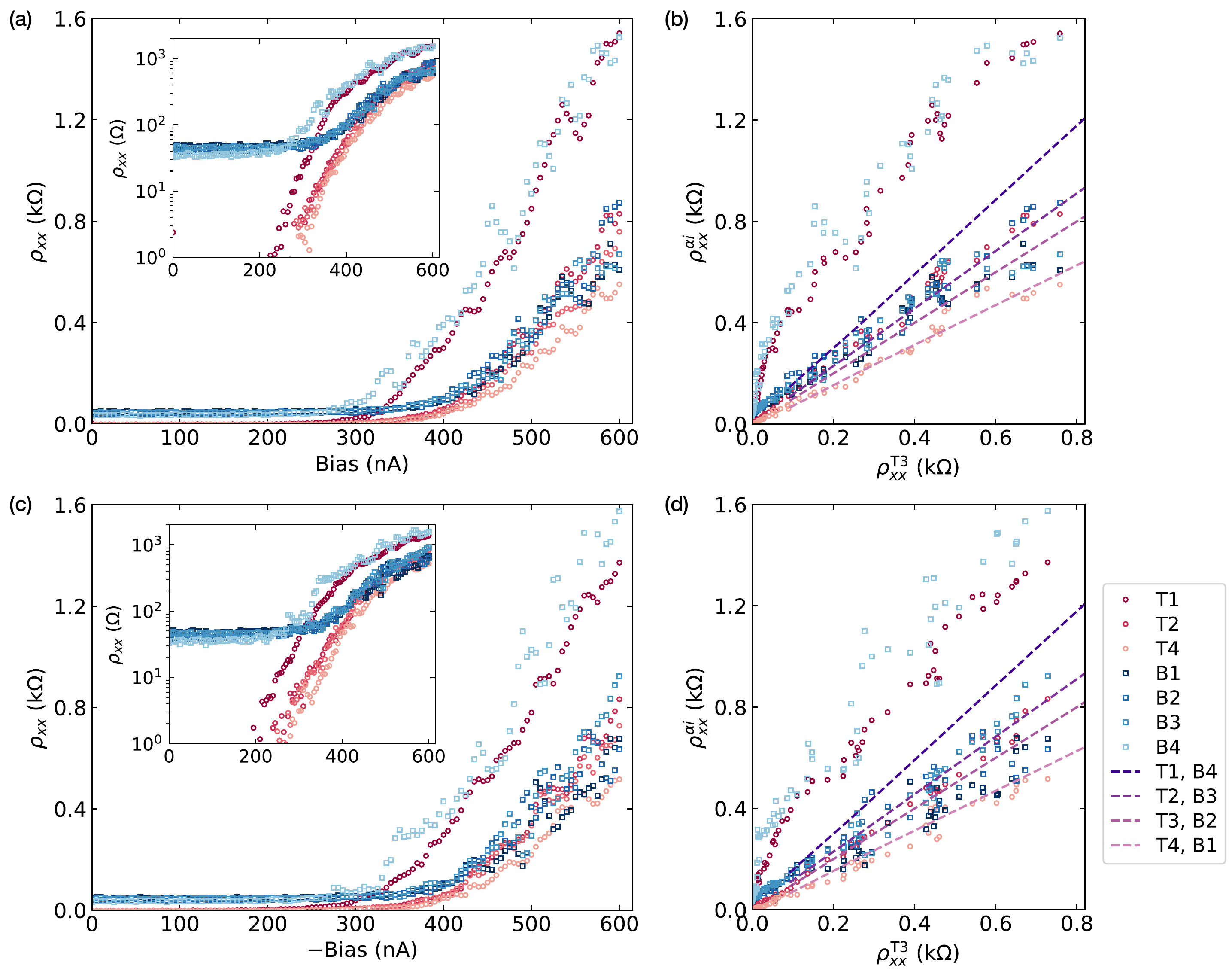}
	\caption{Independence of measured resistivities from sign of current bias. (a,~b) Positive current bias. Data is reproduced from Fig.~3 of the main text, showing resistivity as a function of positive current bias (a) on a linear (inset: logarithmic) axis and (b) shown parametrically, with simulation results indicated by dashed lines. (c,~d) Negative current bias. Resistivity is shown as a function of negative current bias (c) on a linear (inset: logarithmic) axis and (d) shown parametrically, with simulation results indicated by dashed lines.}
	\label{sfig_neg_bias}
\end{figure}

In Fig.~4(a) of the main text, we show the apparent resistivities as a function of dc current bias. In this section, we further explore transport at finite current bias. Data from the main text, where the dc current bias is positively signed, are displayed in the left column Fig.~\ref{sfig_neg_bias}, and data with negative dc current bias are displayed in the right column. There are no salient differences, demonstrating that transport is invariant to the sign of current flow. This is different from QH systems, where the magnitude of the voltage gradient in a device changes based on the sign of the applied bias---particularly near the hot spots, and particularly in the breakdown regime~\cite{komiyama1996}.
Note that these data are differential measurements: a small ac signal is added to the dc bias, and only the ac component of voltage is picked out by lock-in amplification and plotted here.

Data are shown on a logarithmic scale in the insets of Fig.~\ref{sfig_neg_bias}, revealing that the apparent resistivities on the bottom edge of the device are offset by around 40~$\Omega$. This offset is a consequence of leakage currents through the voltage preamplifiers, and is discussed at length in Ref.~\cite{rodenbach2021}.

Data are plotted parametrically in Fig.~\ref{sfig_neg_bias}(c,~d), along with the simulated dependence from the Laplace equation. The data are not well described by the Laplace equation: while both feature higher apparent resistivity near the hot spots than the rest of the device, the difference is much sharper in the data. We interpret this as a consequence of the nonuniform electric field throughout the device leading to a spatially-varying value of the conductivity.
%Breakdown of the QAH state, is driven by the high electric field induced by the source-drain bias~\cite{fox2018, rodenbach2021}, and is characterized by a rapid increase in the longitudinal conductivity (as well as a reduction in the Hall conductivity).

According to the Laplace equation, the electric field is not uniform, but is highest near the two hot spots. Since the conductivity depends on the electric field~\cite{fox2018, rodenbach2021}, we expect that the conductivity should in turn become nonuniform, with higher longitudinal conductivity nearer to the hot spots. Since our simulations assume a uniform conductivity, we do not expect our measurements of the apparent resistivities versus bias to quantitatively agree with our model. Future work could attempt to model the electric field profile at high bias by creating a map $\sigma(\textbf{E})$ between electric field and conductivity, and then finding a self-consistent solution of the Laplace equation
\begin{equation}
    0 = \nabla\left(\sigma(\textbf{E})\nabla V \right),
\end{equation}
where $V$ is the electric potential, with the non-uniform conductivity tensor $\sigma(\textbf{E})=\sigma(\nabla V(x,y))$ satisfying this mapping.

Up to this point, we have presented apparent resistivity measurements at various applied dc current biases at base temperature $T\approx 28$~mK. However, temperature tunes global longitudinal resistivity (Fig. ~1(b)). At elevated temperatures we therefore expect the film resistivity to recover toward uniformity; although resistivities in both the hot spots and in the bulk should be tuned by temperature, the difference should be greater in the bulk, allowing the bulk resistivity to approach the electric field-enhanced resisitivity at the hot spots.  At higher temperatures, longitudinal resistivity data acquired as a function of bias should thus agree with our simulations that assume spatially homogeneous conductivity.

\begin{figure}[h!]
\centering
	\includegraphics[width=0.7\textwidth]{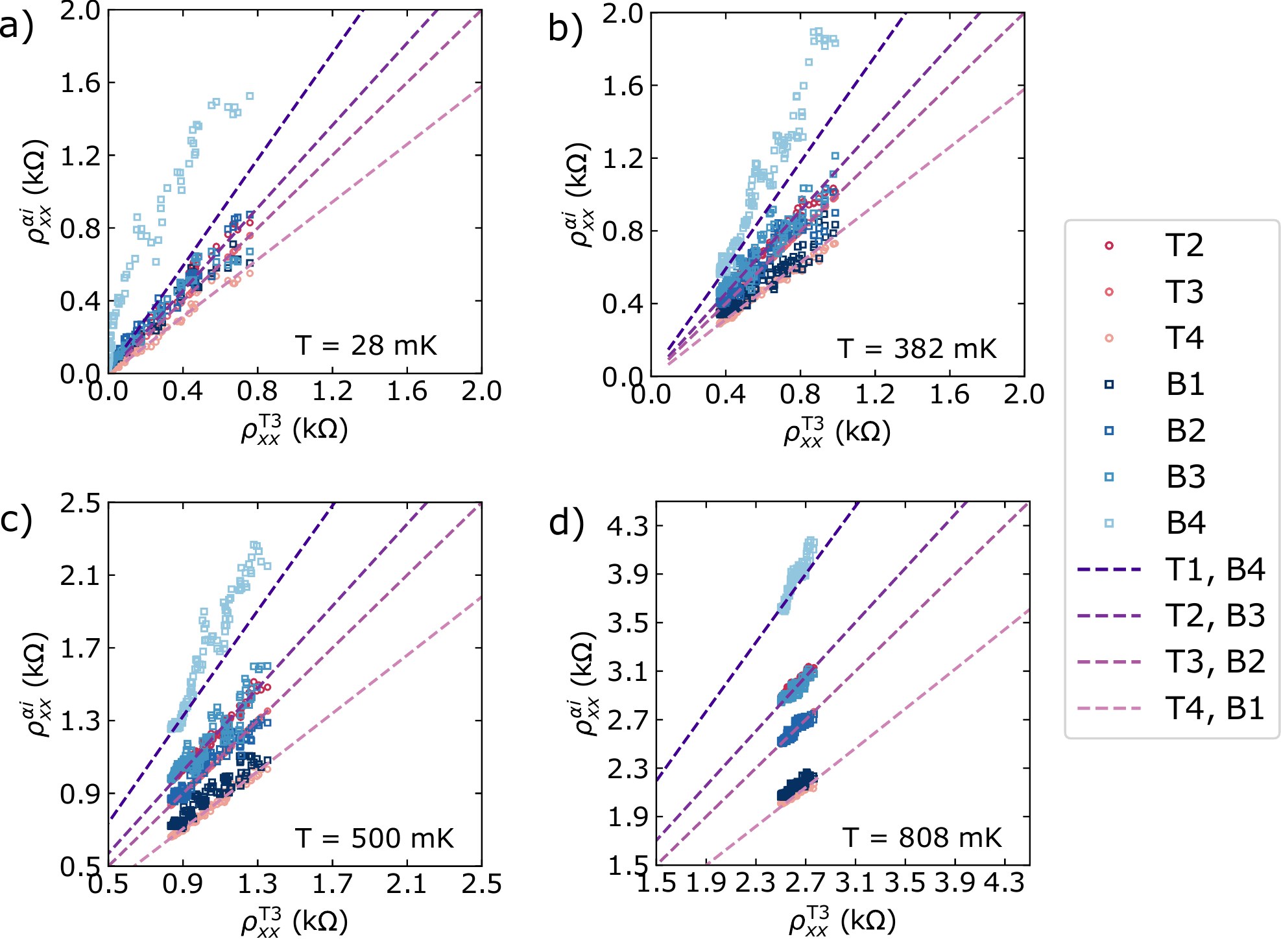}
	\caption{Parametric plots of apparent resistivities as current bias is varied at elevated temperatures. Temperature is modulated via the mixing chamber heater described in the main text. (a) $T = 28$ mK, reproduced from Fig.~\ref{sfig_neg_bias}(c). (b) $T = 382$ mK. (c) $T = 500$ mK. (d) $T = 808$ mK.}
	\label{sfig_bias_highT}
\end{figure}

Fig. ~\ref{sfig_bias_highT} shows additional resistivity data collected in the same manner as Fig.~4(a) but at elevated temperatures. At each successive temperature, measurements match simulations more closely. At the highest temperatures shown, measurements quantitatively agree with simulations. This supports our assertion that, at high current bias and low temperature, deviations of data from simulations (Fig. ~\ref{sfig_neg_bias}) are driven by local conductivity variations, which can be blurred out at higher temperatures.

\clearpage

\section{Gate voltage dependence}\label{sec:gate}

In the main text, we discussed that QH systems at $\nu\neq 0$ lack electron-hole symmetry, while QAH systems can in principle be undoped and have approximate electron-hole symmetry. In practice in experiments, this symmetry appears to be intact when $V_g=V_g^\mathrm{opt}$, where the chemical potential is understood to be approximately in the center of the magnetic exchange gap opened around the Dirac point of the surface states. Note that the symmetry should be approximate because the hole surface-state band is believed to be heavier than the electron surface-state band. We believe that the centrosymmetry (two-fold rotational symmetry, i.e. $\rho_{xx}^\mathrm{T1}\approx \rho_{xx}^\mathrm{B4}$ etc.) of the apparent resistivity in our device is related to this approximate electron-hole symmetry.

\begin{figure}[h!]
\centering
	\includegraphics[width=0.7\textwidth]{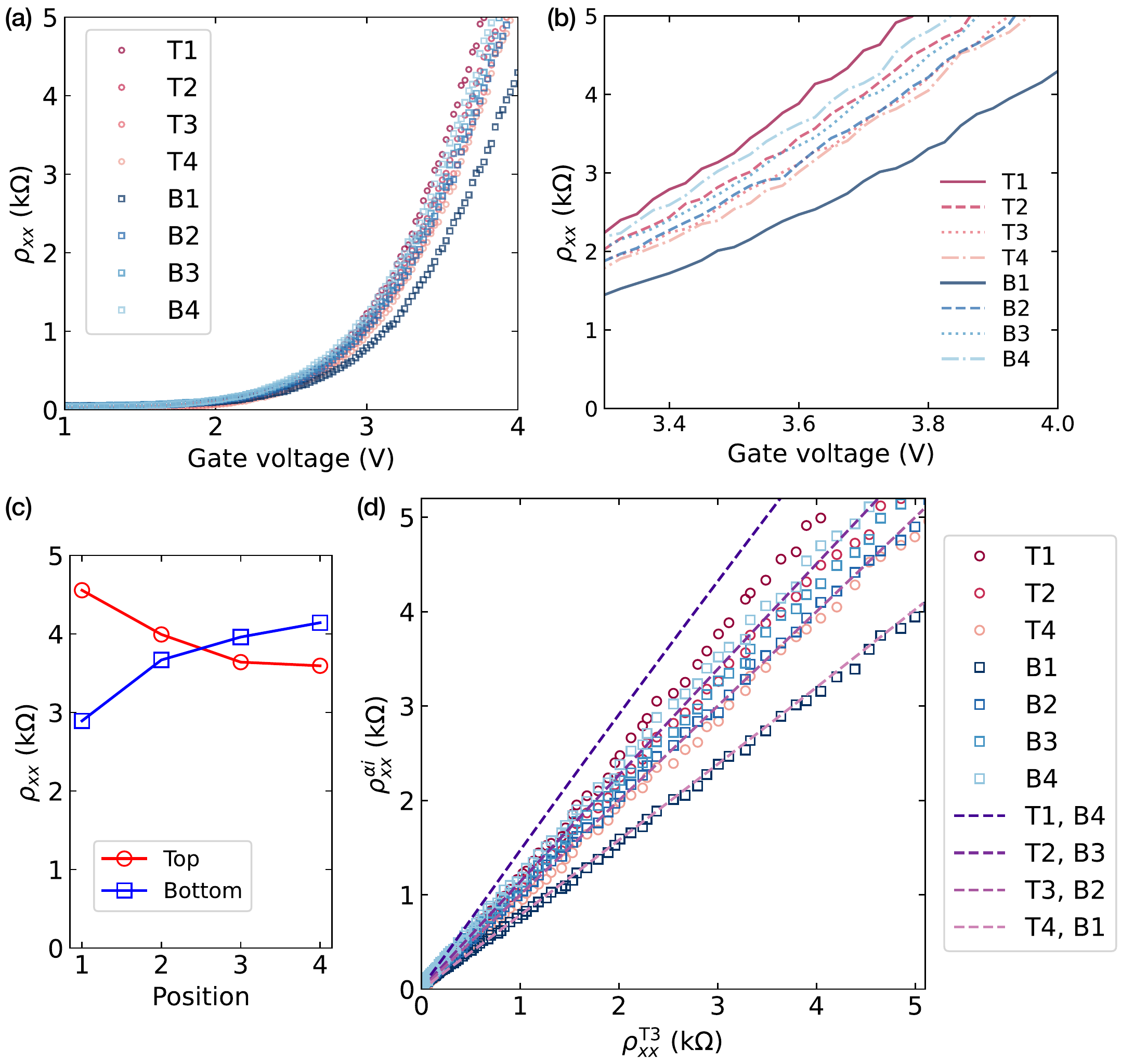}
	\caption{(a) Apparent resistivity as gate voltage is varied (same as Fig.~4 of the min text). (b) Detail of the same data for clarity. (c) Apparent resistivity versus position along the top an bottom edges at $V_g=3.7$~V. (d) Apparent resistivity as gate voltage is varied, plotted parametrically, along with (dashed lines) simulations of the Laplace equation.}
	\label{sfig_gate_base}
\end{figure}

While this approximate symmetry is preserved as the temperature is increased, as in Figs.~2 and 3 of the main text, it is explicitly broken by changing the gate voltage. In this section, we explore this effect. In Figure~4(b) of the main text, we presented the apparent resistivity as a function of gate voltage at base temperature. The same data is presented again in Fig.~\ref{sfig_gate_base} for more clarity, along with a parametric plot.

\begin{figure}[h!]
\centering
	\includegraphics[width=0.8\textwidth]{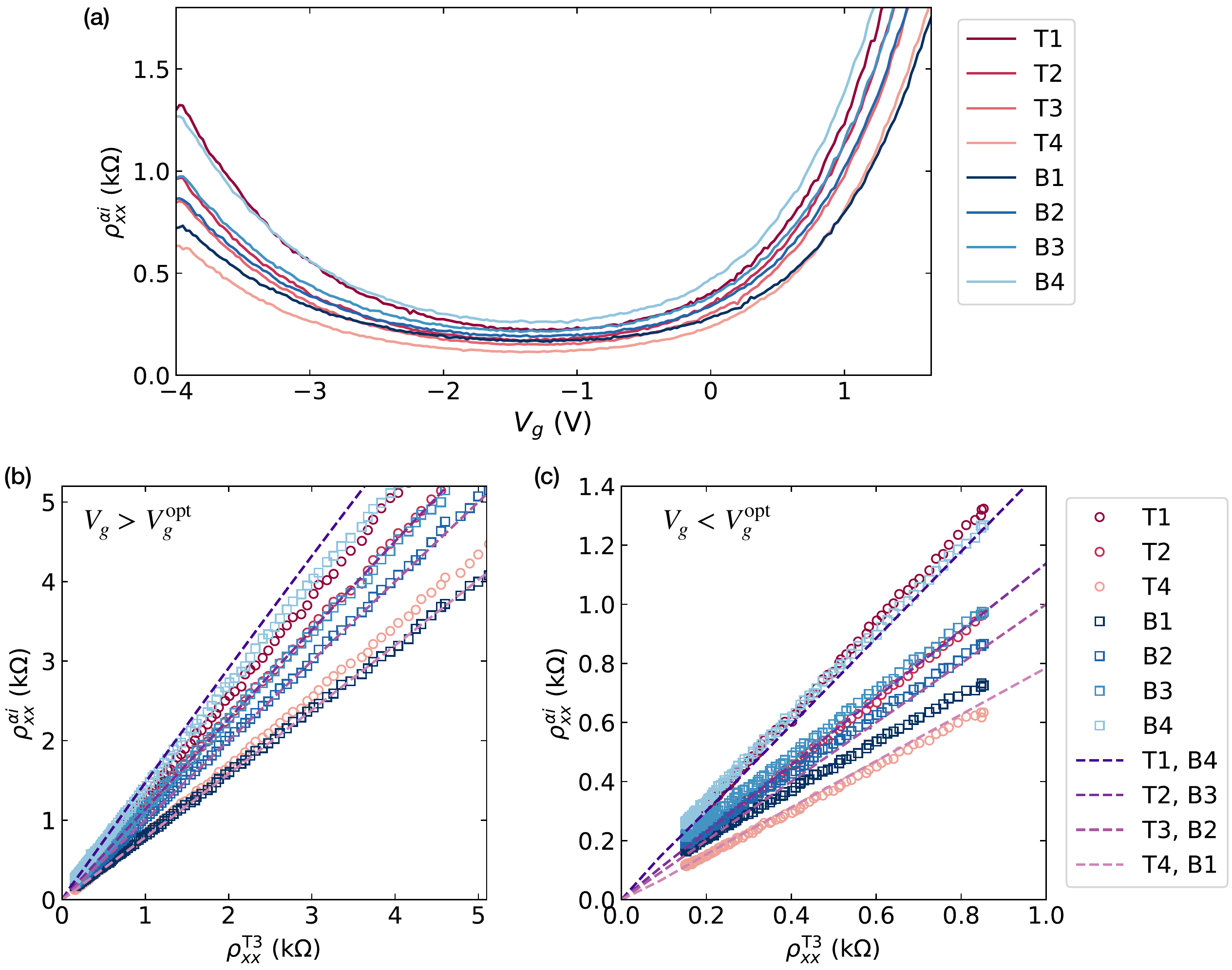}
	\caption{Apparent resistivity as gate voltage is varied, when the device is slightly heated. Data is plotted (a) against gate voltage and (b,~c) parametrically, for (b) $V_g>V_g^\mathrm{opt}$ and (c) $V_g<V_g^\mathrm{opt}$.}
	\label{sfig_above_below}
\end{figure}

We next present data taken with a small heater current through the on-chip (lattice) heater (see Section~\ref{sec:heaters}). The optimal gate voltage $V_g^\mathrm{opt}=-1.18$~V is offset from the operational range of the gate $-4~\mathrm{V}<V_g<4~\mathrm{V}$. At base temperature, there is a wide plateau of nearly-zero dissipation around $V_g^\mathrm{opt}$, and dissipation therefore is seen only at very high gate voltages. By measuring at a slightly elevated temperature, we may contrast dissipation in electron-doped ($V_g>V_g^\mathrm{opt}$) and hole-doped ($V_g<V_g^\mathrm{opt}$) regimes. The apparent resistivity with a small heater current is shown in Fig.~\ref{sfig_above_below}, along with parametric plots of the data restricted to the electron- and hole-doped regimes.

The broken electron-hole symmetry is borne out in the measurements. On the electron-doped side $V_g>V_g^\mathrm{opt}$, $\rho_{xx}^\mathrm{T1}$ is slightly less than $\rho_{xx}^\mathrm{B4}$, while $\rho_{xx}^\mathrm{T4}$ is slightly greater than $\rho_{xx}^\mathrm{B1}$. Yet on the hole-doped side $V_g<V_g^\mathrm{opt}$, this trend is reversed: $\rho_{xx}^\mathrm{T1}$ is slightly greater than $\rho_{xx}^\mathrm{B4}$, while $\rho_{xx}^\mathrm{T4}$ is slightly less than $\rho_{xx}^\mathrm{B1}$. 

More work is needed to understand the mechanism by which breaking electron-hole symmetry shifts the potential profile away from that predicted by the Laplace equation.

\clearpage
\section{Temperature dependence and different heaters}\label{sec:heaters}

In the main text, we presented resistance measurements of our device at various temperatures. The temperature of a device, however, is not always single valued: the lattice temperature, or temperature of the phonon bath, may depart from the electron temperature. The lattice is cooled through mechanical coupling to the source of the cooling power (in a dilution refrigerator, the mixing chamber), and the electronic system is cooled through the device's electrical leads, which are thermalized through electronic filtering coupled to the mixing chamber stage. Additionally, the lattice and electronic system exchange energy through electron-phonon scattering. The electron temperature itself may also be multipartite if, for example, the electron temperature of the CEM departs from the electron temperature in the bulk.

In the data presented in the main text, we assume equivalence of the electron and lattice temperatures. This is because heating was accomplished using a resistive heater on the mixing chamber plate, so that the lattice and electronic system are thermalized to the mixing chamber plate temperature regardless of electron-phonon coupling. It is intriguing to ask what happens when the assumption of equal electron and lattice temperature is broken, and in particular, if the chiral edge mode carries a heat current. To address this question, we designed heating elements intended to preferentially heat one system.

To primarily heat the lattice, we deposited a resistive gold meander surrounding the device, but not in electrical contact. The topological insulator was etched underneath the meander, so that the meander lay directly on the GaAs substrate. This ``lattice heater" has resistance of 400~$\Omega$ at base temperature, and sending a current through the meander produces Joule power, which is deposited into the substrate. Note that the lattice heater can heat the electronic system through electron-phonon scattering in the device (both in the gold contacts and in the topological insulator film).

To primarily heat the electronic system, we deposited two co-meandering resistive gold wires on a separate chip as an ``electron heater". One of the wires was connected in series with the source terminal of the device. A current was driven through the second wire to heat the first wire, thereby raising the electron temperature at the source terminal. The drain terminal was not connected through an electron heater, but was connected directly to the electrical filters at the mixing chamber stage. Because the electron heater was on a different chip than the device, we expect parasitic heating of the lattice from the electron heater to be small. Note that the electron heater can heat the electronic system through electron-phonon scattering in the device (both in the gold contacts and in the topological insulator film), and should do so asymmetrically, because only the source reservoir is heated whereas the drain reservoir should remain well-thermalized to the mixing chamber.

\begin{figure}[h!]
\centering
	\includegraphics[width=0.85\textwidth]{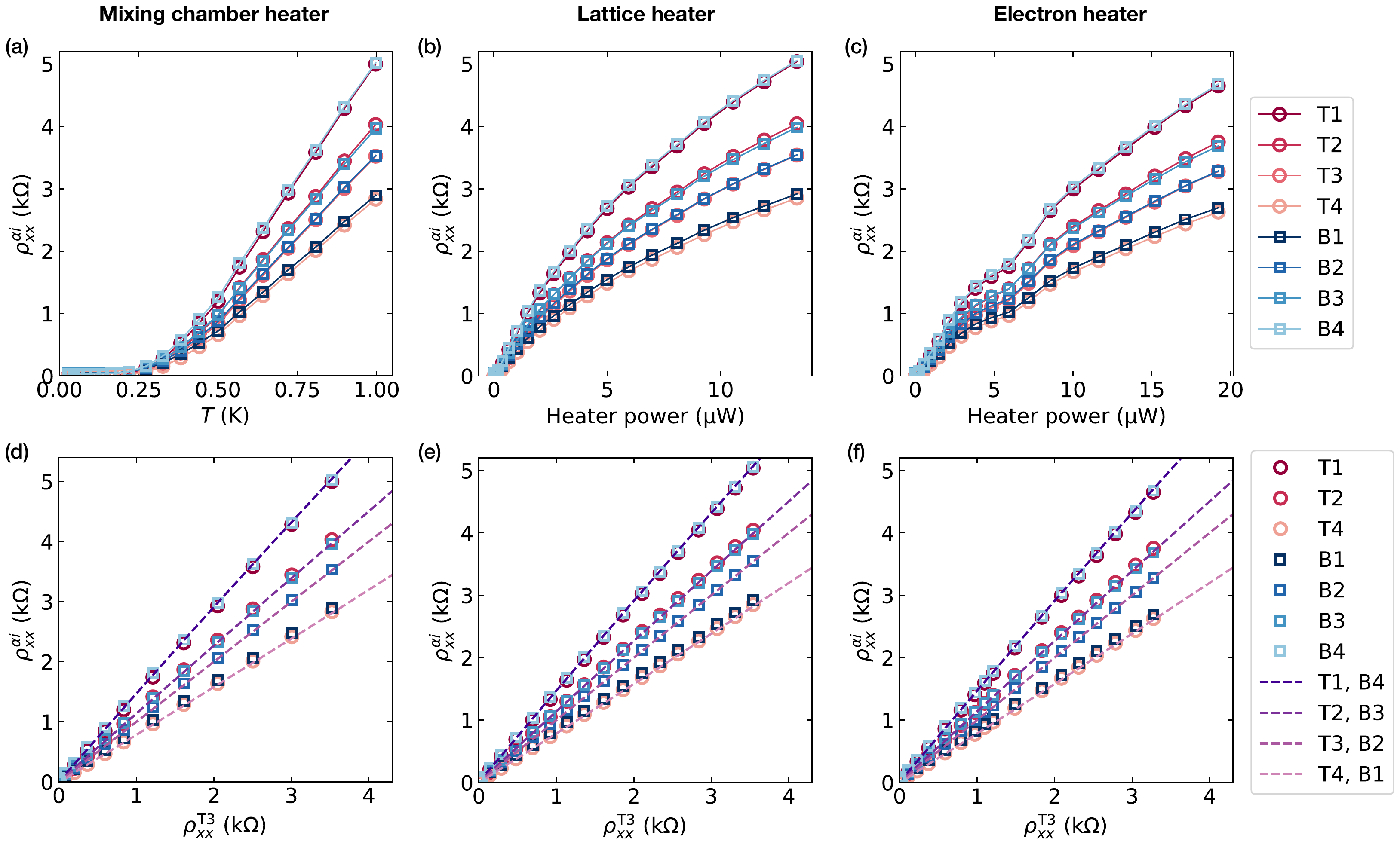}
	\caption{Apparent resistivity as temperature is varied with three different types of heaters. (a,~d) Apparent resistivity (a) as a function of temperature and (d) plotted parametrically, where the mixing chamber heater tunes the device temperature. Reproduced from Figs.~1(b) and~2(c). (b,~e) Apparent resistivity (b) as a function of temperature and (e) plotted parametrically, where the on-chip lattice heater heats the device. (c,~f) Apparent resistivity (c) as a function of temperature and (f) plotted parametrically, where the electron heater heats the device.}
	\label{sfig_heaters}
\end{figure}

The apparent resistivities are shown in Fig.~\ref{sfig_heaters} while the device is heated with the mixing chamber, lattice, and electron heaters, individually. Parametric plots of the apparent resistivities reveal that the spatial distribution of the electric potential is the same in all cases. Furthermore, centrosymmetry of the apparent resistivities ($\rho_{xx}^\mathrm{T1}\approx \rho_{xx}^\mathrm{B4}$, etc.) is preserved at high electron heater power, where there should be a different electron temperature at the source than at the drain. Thus our device failed to detect a chiral heat current.

This result does not rule out chiral heat transport in the chiral edge mode. A likely explanation for our observations is that the electron-phonon scattering length in the chiral edge mode is much smaller than the distance between voltage terminals (40~$\mu$m), so that the electronic system and lattice reach thermal equilibrium at length scales too small for our experiment to detect. In the quantum Hall effect, the thermal decay length for hot electrons in the chiral edge mode has been measured to be tens of microns~\cite{granger2009}. Experiments more refined than ours are needed to further understand the heat transport and electron-phonon coupling in QAH systems.

Note that sending current through the lattice heater and electron heater did heat the mixing chamber stage, but not enough to substantially affect the device. During operation of the heaters, the thermometer at the mixing chamber indicated that the mixing chamber stage rose from base temperature (30~mK) only to 100~mK at maximum heater currents. Here, the resistivity of the device (roughly 5~k$\Omega$) matched that when the device was heated to about 1~K with the mixing chamber heater (Fig.~\ref{sfig_heaters}(a)).

\begin{figure}[h!]
\centering
	\includegraphics[width=0.65\textwidth]{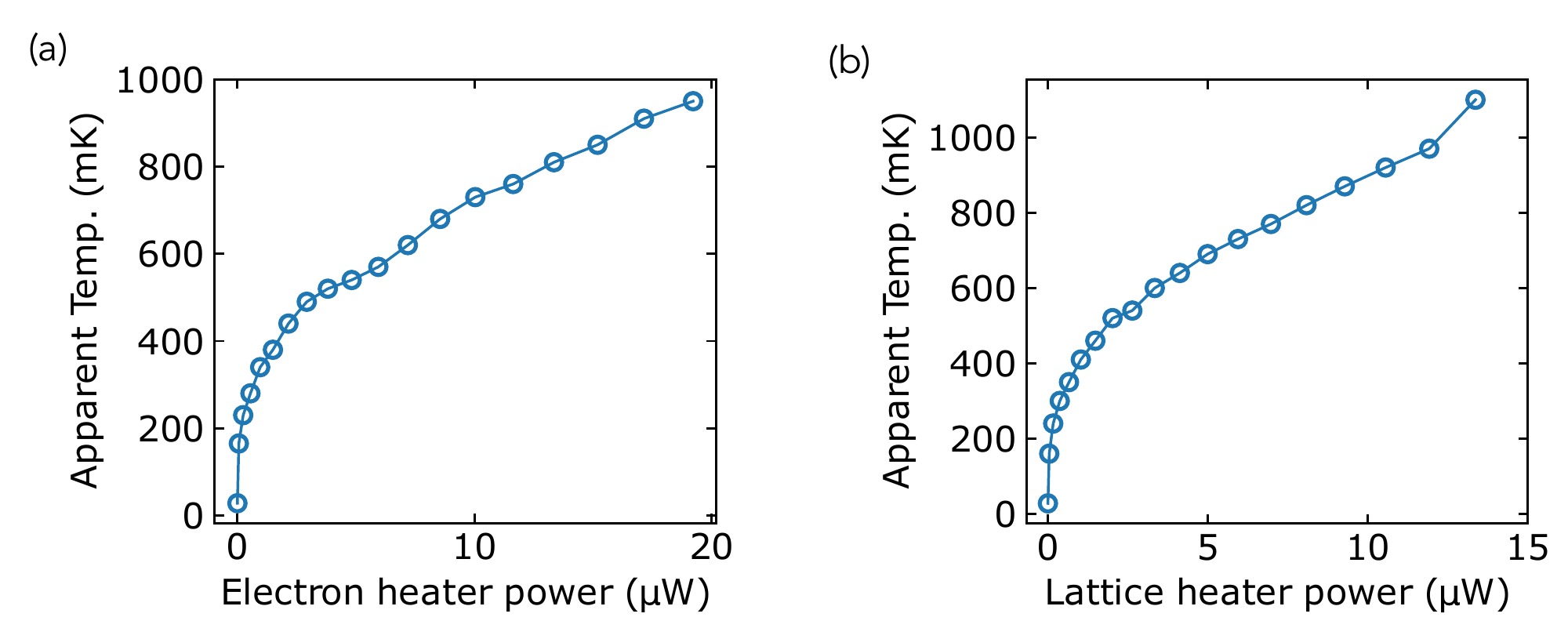}
	\caption{Apparent conversion between different heater powers and device temperature. (a) Apparent device temperature as a function of electron heater power. (b) Apparent device temperature as a function of lattice heater power. For each value of electron (lattice) heater power, device resistivities were compared to measurements of resistivity versus mixing chamber heater, for which fridge thermometry directly measures device temperature. The apparent temperature for each value of electron (lattice) heater power was taken as the measured temperature for which measured resistivity matches most closely between the electron (lattice) heater and mixing chamber heater data sets. Where the measured resistivities using the electron (lattice) heater fell between measured resistivities using the mixing chamber heater, a linear interpolation between mixing chamber heater data points was assumed.}
	\label{sfig_electron_heater_conversion}
\end{figure}

\clearpage
\section{Additional devices}

\begin{figure}[h!]
\centering
	\includegraphics[width=0.8\textwidth]{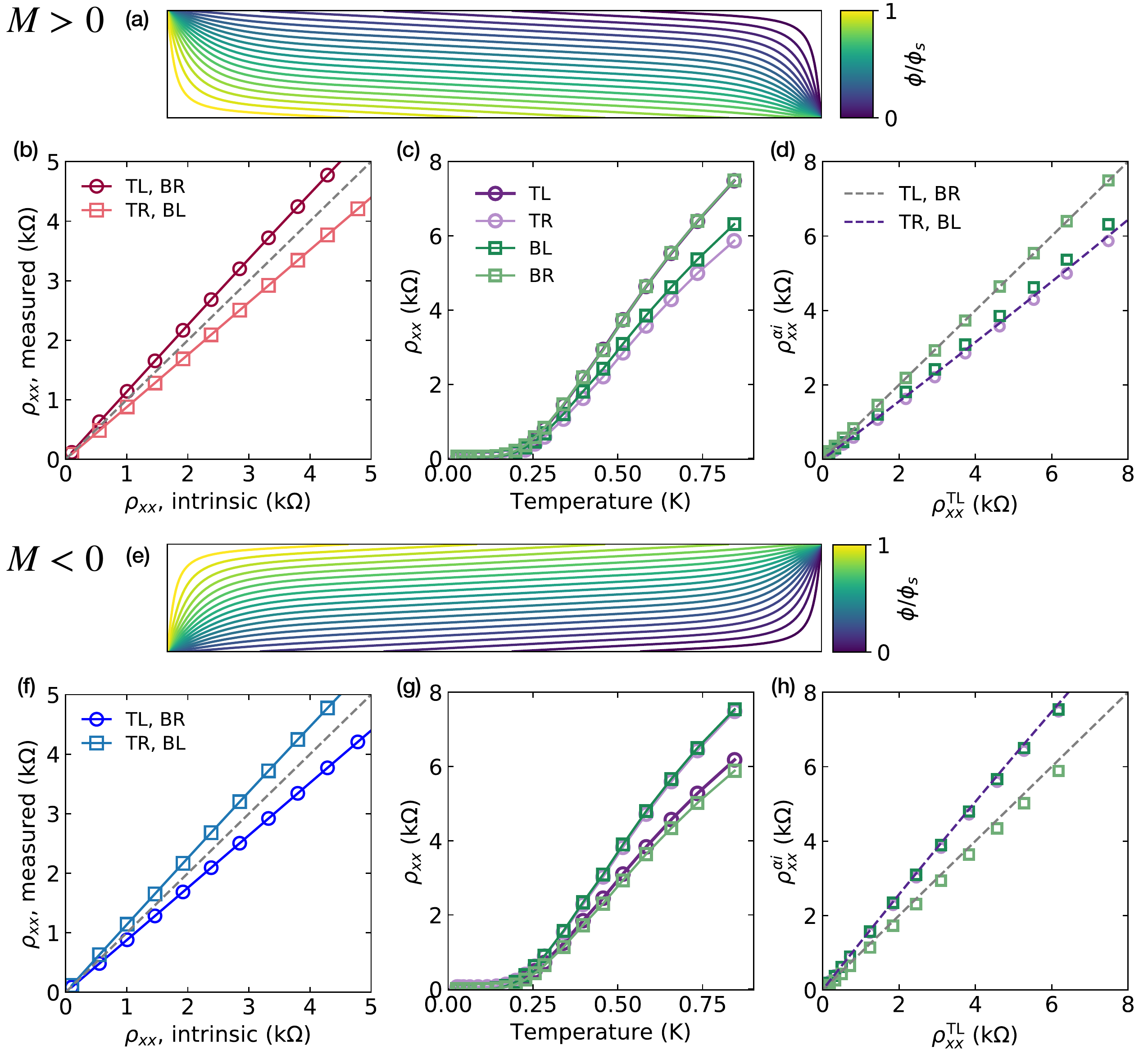}
	\caption{(a) Simulation of the electric potential in the geometry of Device~2 at $\sigma_{xx}/\sigma{xy}=0.05$. (b) Simulated apparent resistivities as a function of the intrinsic resistivity of the device. (c) Apparent resistivites as a function of temperature. (d) Apparent resistivites as a function of temperature, shown parametrically as a function of the apparent resistivity in the top left corner. (e-h) Same as (a-d), respectively, but for negative magnetization of the QAH film.}
	\label{sfig_HM_temp}
\end{figure}

\begin{figure}[h!]
\centering
	\includegraphics[width=0.6\textwidth]{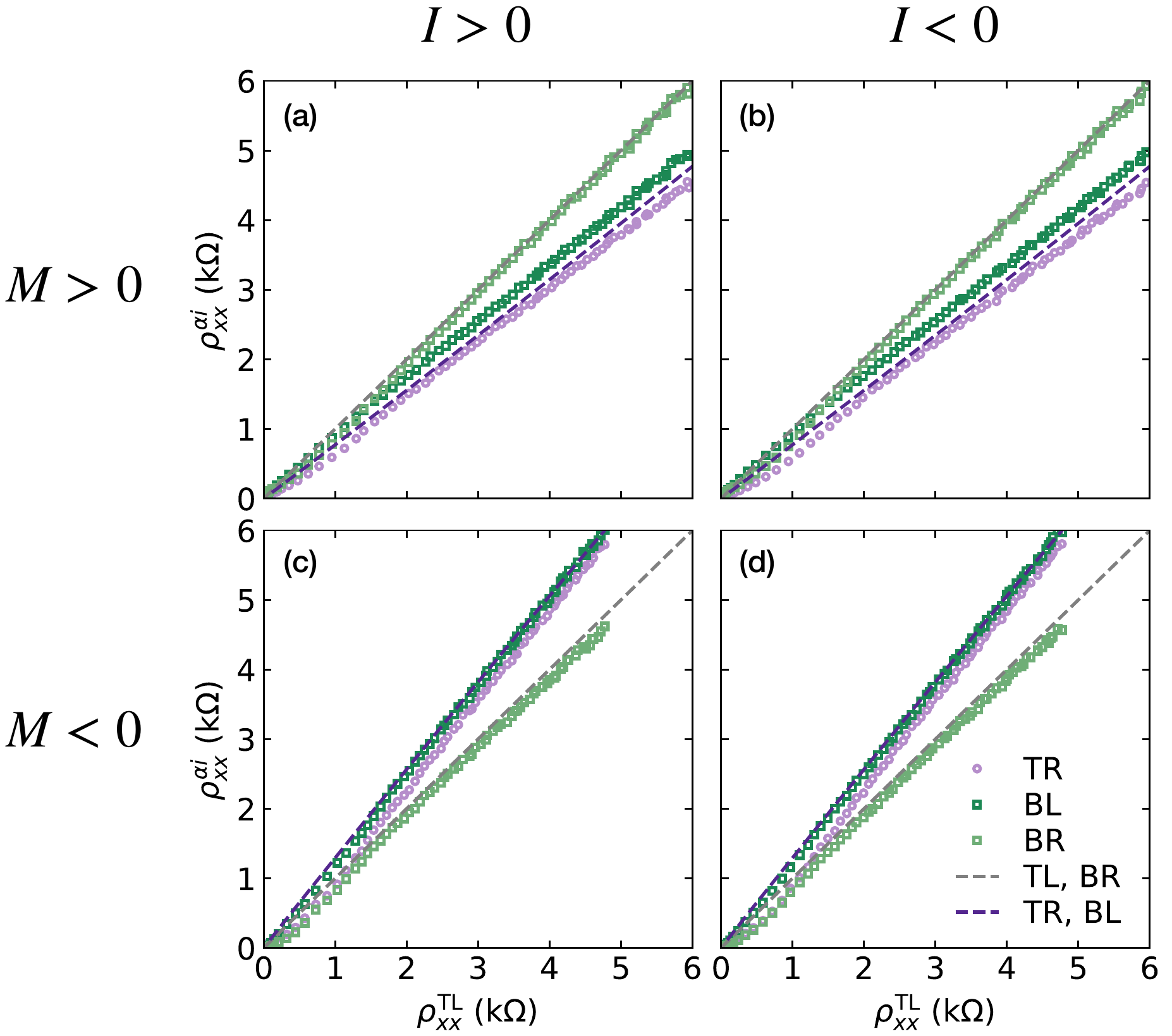}
	\caption{Apparent resistivities as a function of current bias, shown parametrically as a function of the apparent resistivity in the top left corner. (a, b) Positive magnetization of the QAH film. (c, d) Negative magnetization. (a, c) Positive-signed source-drain current bias. (b, d) Negative current bias.}
	\label{sfig_HM_bias}
\end{figure}

In this section, we present measurements similar to that shown in the main text taken in other Hall bar devices with slightly different geometry. Device~2
% HM
has 610~$\mu$m between source and drain contacts, is 100~$\mu$m in width, and has pairs of voltage contacts near each corner of the device. There is 50~$\mu$m (center-to-center) between the voltage contacts in each pair, and the contact in each pair closest to the corner is 65~$\mu$m (center-to-edge) from the source/drain contact. Like Device~1, voltage contacts have 6~$\mu$m width. Voltage measurements are indicated as T(B)R(L) for the measurement in the top (bottom right (left) corner. Data from Device~2 are shown in Figs.~\ref{sfig_HM_temp} and \ref{sfig_HM_bias}.

%%% HM2 %%%

\begin{figure}[h!]
\centering
	\includegraphics[width=0.75\textwidth]{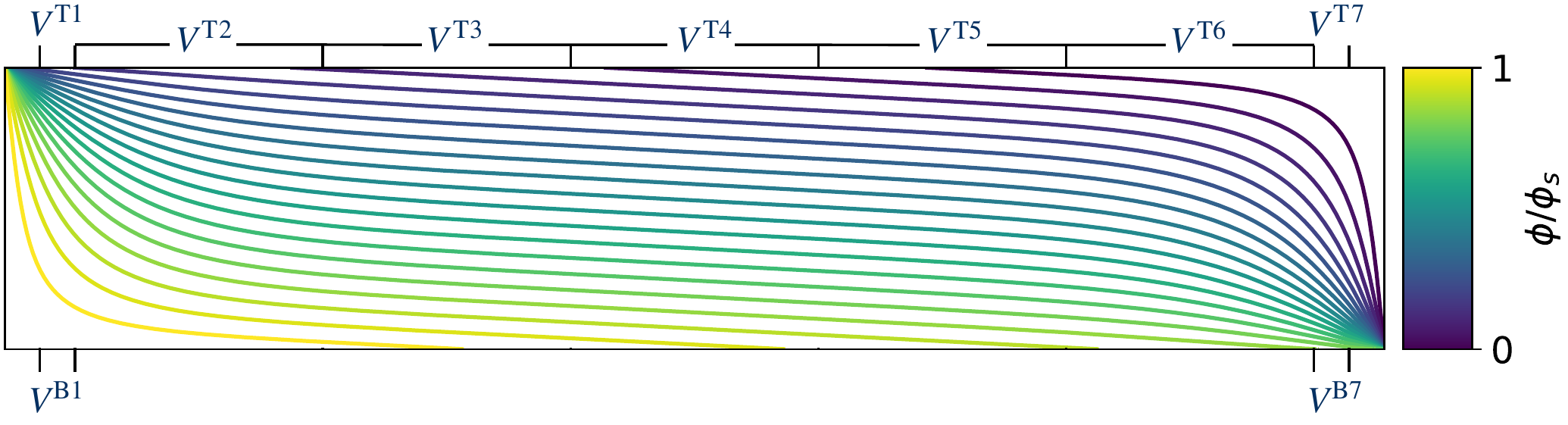}
	\caption{Simulation solution of the electric potential to the Laplace equation in Device~3 at $\sigma_{xx}/\sigma_{xy}=0.05$. The nine voltage measurements, made using the 12 voltage taps of the device, are indicated.}
	\label{sfig_HM2_contour}
\end{figure}

\begin{figure}[h!]
\centering
	\includegraphics[width=0.85\textwidth]{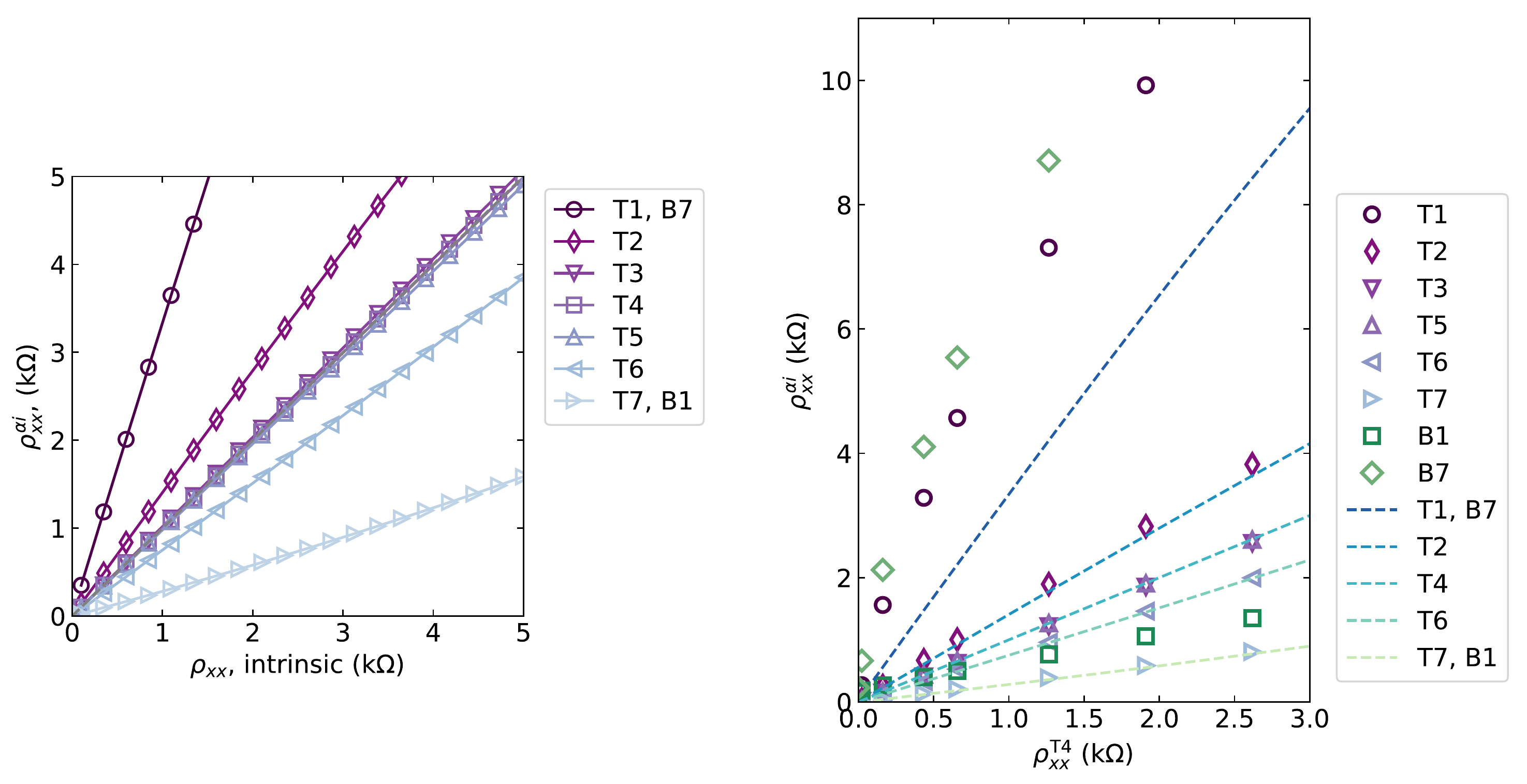}
	\caption{(a) Simulated apparent resistivities as a function of the intrinsic resistivity of the device. (d) Measured apparent resistivites as a function of temperature, shown parametrically as a function of the apparent resistivity in the center of the device (measurement T4). Note that measurements at T3 and T5 are overlapping. The simulated apparent resistivities are shown by dashed lines. Simulations for measurements T3 and T5 are omitted as they are virtually indistinguishable from measurement T4.}
	\label{sfig_HM2_data}
\end{figure}

Device~3 has 980~$\mu$m between source and drain terminals, and has pairs of voltage terminals close to each corner, plus four additional voltage terminals spaced across the device (see Fig.~\ref{sfig_HM2_contour}). This configuration provides seven measurements across the top edge (T1-7, where T1 is near the left corner and T7 is near the right corner) and two measurements on the bottom edge (B1 and B7, near the left and right corner, respectively). Measurements are compared to simulation in Fig.~\ref{sfig_HM2_data}. The data and simulations again match well, except very close to the hot spots (measurements T1 and B7). These contacts are much closer to the hot spots than the nearest contacts in Devices~1 and 2, and the electric field is highly concentrated around them. We speculate that the difference between measurement and simulation could indicate that the QAH effect is already in breakdown at these locations, even at the small ac current bias of 5~nA. Another possibility is that the simulations become inaccurate very close to the hot spots due to the finite-sized mesh.

\clearpage
\section{Simulation of alternative dissipation models}

In the main text, we argued that, in general, models of dissipative-regime non-equilibrium conduction driven by a chemical potential gradient along the edge mode do not give an edge mode chemical potential profile that satisfies solutions to the two-dimensional Laplace equation. To defend this statement, we present calculations of the chemical potential in edge-mode-based models of dissipation. To simulate transport driven by a chemical potential difference between the edge mode at the source and drain electrodes, we employ the Landauer-B\"utikker model
\begin{equation}
I_i = \frac{e^2}{h} \sum_j (T_{ji}V_i - T_{ij}V_j)
\end{equation}
where $T_{ji}$ is the transmission probability from contact $i$ to contact $j$, and $V_i$ and $I_i$ are the voltage and net current, respectively, in contact $i$. The dissipationless chiral state of the quantum anomalous Hall system is characterized by the transmission coefficients $T_{i+1,i}=1$ and all other coefficients zero. The apparent resistivities of the model are found through solutions for $\mathbf{V}$ when there is nonzero current through the source and drain contacts.

Here, we present three models of dissipation. The first model, proposed by Ref.~\cite{wang2013}, considers the addition of ``quasi-helical" edge modes--counter-propagating edge modes that are not immune to backscattering. In this model, the transmission coefficients are given by $T_{i+1,i}=1+p$ and $T_{i-1,i}=p$, where $p$ is the transmission probability of the quasi-helical mode between contacts. The second model considers the case where the chiral edge mode itself may, instead of propagating to the next chiral-ordered contact, backscatter with probability $p$ to the next anti-chiral-ordered contact, yielding the transmission coefficients $T_{i+1,i}=1-p$ and $T_{i-1,i}=p$. The third model considers the case where the chiral edge mode may scatter across the bulk of the device with probability $p$. In this model, the transmission coefficients are $T_{i+1, i} = 1-p$ and $T_{i,j}=p$ where $j$ is the contact directly across the Hall bar from contact $i$ (for the source and drain contacts of the Hall bar, $j=i$). A more intricate version of this model is presented by Ref.~\cite{fox2018}.

\begin{figure}[h!]
\centering
	\includegraphics[width=0.8\textwidth]{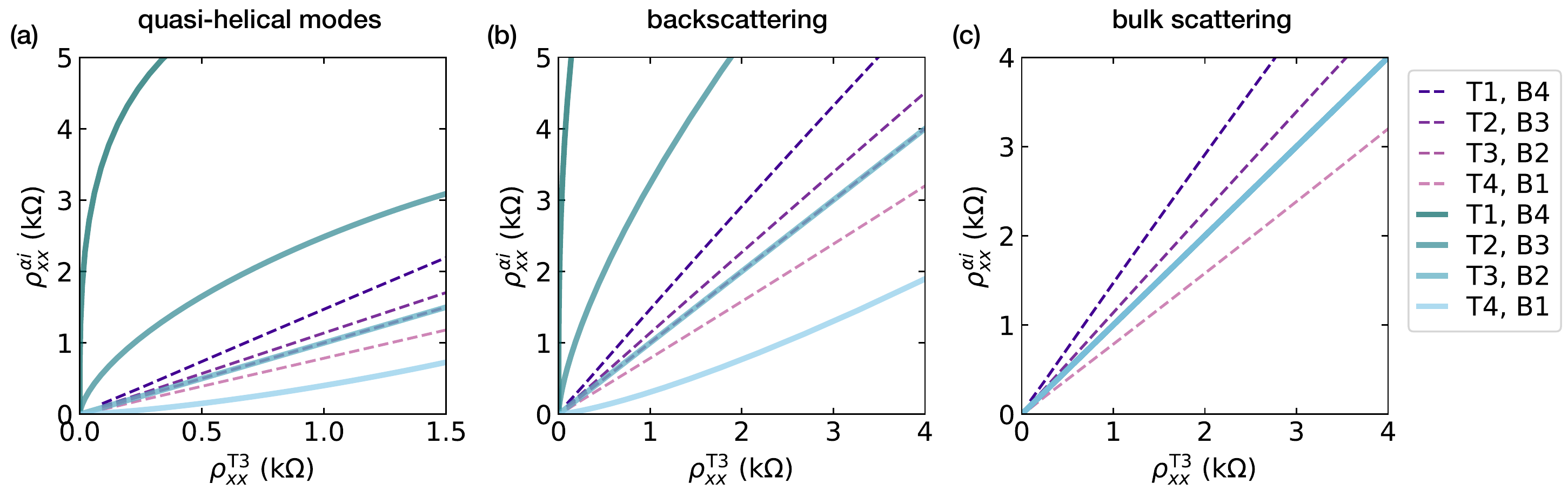}
	\caption{Simulation of three edge-mode models of dissipation in the quantum anomalous Hall system using the Landauer-B\"utikker formalism: (a) addition of quasi-helical edge modes, (b) backscattering of the chiral edge mode, and (c) scattering of the chiral edge mode across the bulk to the opposite edge of the Hall bar. Results from the edge-mode models (green solid lines) are plotted parametrically and compared to solutions to Laplace's equation (purple dashed lines), as in the main text. We note that in (c), the four apparent resistivities given by the bulk-scattering model are identical.}
	\label{sfig_LB_models}
\end{figure}

Results from these three models are presented in Fig.~\ref{sfig_LB_models}, where the parametric behavior of the model is compared with that of the Laplace simulation presented in the main text. These three models yield dramatically different results but, unlike our data, do not conform to solutions of Laplace's equation (solved throughout the two-dimensional bulk of the device). We do not claim that these unsophisticated models are well-justified by the particular microscopic behavior one might expect in a quantum anomalous Hall insulator; rather, we present these models to argue that edge-mode models of the dissipative-regime quantum anomalous Hall effect do not in general satisfy Laplace's equation. Finding an edge-mode model that produces a chemical potential that does would require careful fine-tuning of many phenomenological parameters. In contrast, the model we propose in the main text--that the source-drain electrochemical potential difference is mainly effectuated as an electrostatic potential gradient--explains our measurements with no free parameters.

\clearpage
\section{Simulation of nonlocal measurements}

Nonlocal measurement geometries are often used to study materials with edge modes. The idea of nonlocal measurements is to apply a bias between two contacts on one section of a device, and measuring the voltage at other points of the device.

\begin{figure}[h!]
\centering
	\includegraphics[width=0.8\textwidth]{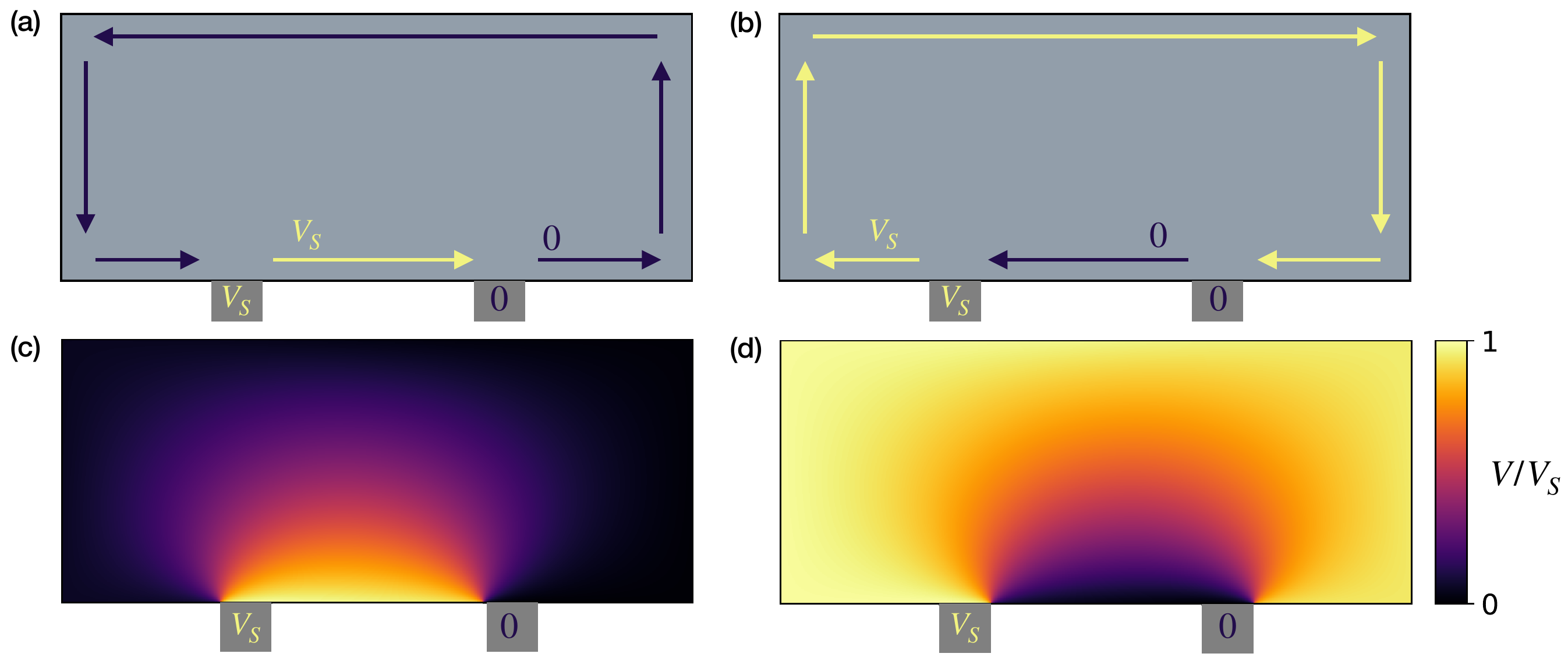}
	\caption{Simulation of nonlocal measurements. (a, b) Schematic of the QAH system in the edge-current picture, for (a) out-of-plane magnetization (counter-clockwise circulation of the chiral edge mode) and (b) into-plane magnetization (clockwise circulation). The QAH Hall bar is indicated by the light grey rectangle, and the source and drain contacts by dark grey squares. The potential of edge modes leaving the source terminal matches its potential, $V_S$ (yellow), while that of edge modes leaving the drain terminal have zero (ground) potential (dark purple). (c, d) The bulk current flow picture, for (c) out-of-plane magnetization ($\sigma_{xy}>0$) and (d) into-plane magnetization ($\sigma_{xy}<0$). The electric potential normalized to the potential of the source terminal, $V/V_S$, solving the Laplace equation at $\sigma_{xx}/\abs{\sigma_{xy}}=0.05$ is shown. The potential at the edge of the device matches that in the edge-current picture.}
	\label{sfig_nonlocal_sim}
\end{figure}

In the edge-current picture of the QAH system, a single CEM carries current. The electrochemical potential of the CEM at the measurement points, and therefore the measured voltage, depends on the chirality of the system (for QAH films, the sign of magnetization), as depicted in Fig.~\ref{sfig_nonlocal_sim}(a,~b). Without dissipation, this measurement is invariant to continuous deformation of the device geometry.

Yet these effects are not unique to systems with edge modes. In Fig.~\ref{sfig_nonlocal_sim}(c,~d) we simulate a nonlocal measurement geometry at a small (albeit nonzero) value of $\sigma_{xx}/\sigma_{xy}$. The potential at remote edges of the device tracks the source or drain potential, depending on the sign of the Hall conductivity. These simulations demonstrate that the nonlocal measurement patterns of the QAH effect do not imply that current flows through a CEM. Instead such patterns also naturally arise from the classical Hall effect when the $\sigma_{xx}$ becomes much smaller than $\abs{\sigma_{xy}}$.

\clearpage
\section{Simulation of devices with disjoint edges}

The Corbino geometry consists of an annulus of the material under study, with contacts along the inner and outer circumferences. As no edge of the material connects the two contacts, the Corbino geometry is often used to isolate edge conduction from bulk conduction.

\begin{figure}[h!]
\centering
	\includegraphics[width=0.4\textwidth]{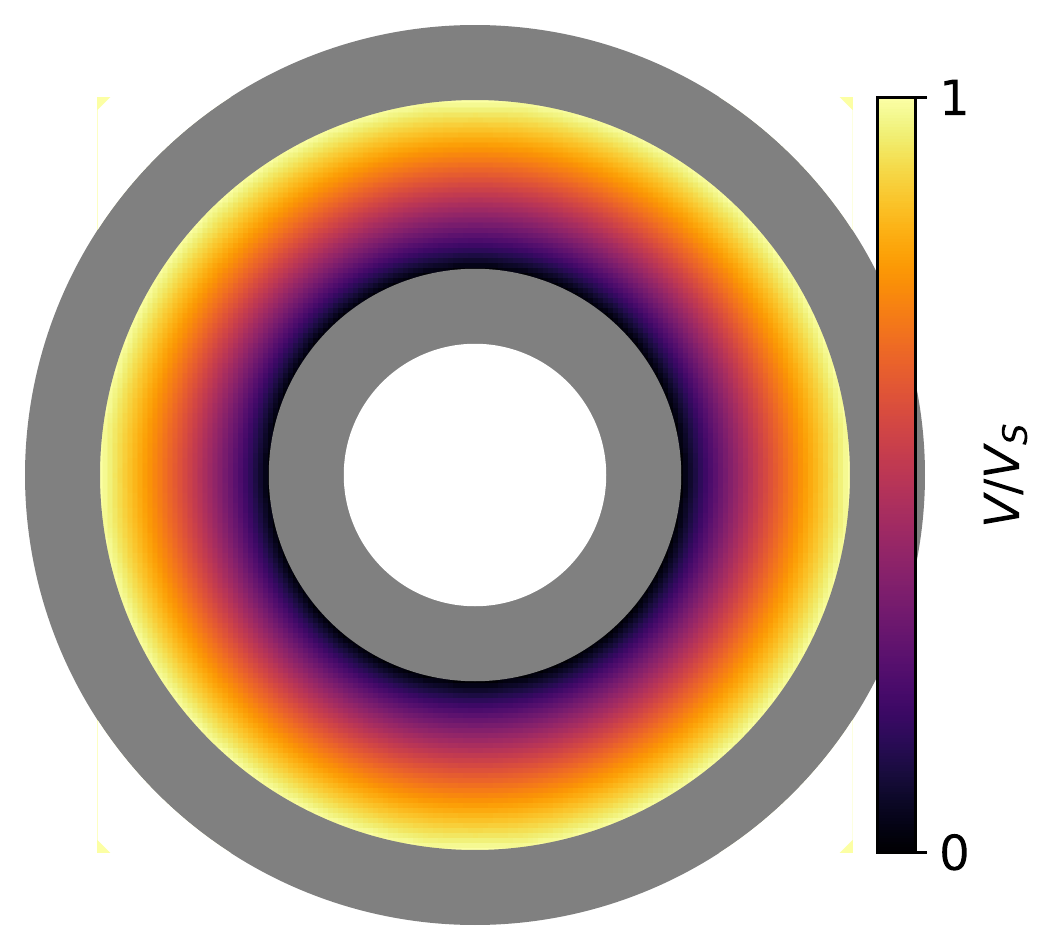}
	\caption{Simulation of the potential in a Corbino disk at $\sigma_{xx}/\sigma_{xy}=0.05$, with a bias sourced at the outer annular contact and drain at the inner annular contact. Contacts are indicated as grey annuli.}
	\label{sfig_corbino}
\end{figure}

We simulate a Corbino geometry using the numerical Laplace equation solver at high $\sigma_{xy}/\sigma_{xx}$ in Fig.~\ref{sfig_corbino}. As expected, the potential has rotational symmetry and drops monotonically as a function of distance from source to drain. This solution is invariant to the conductivity (so long as $\sigma_{xx}$ is finite), although the ratio of voltage to current is not.

A recent publication~\cite{fijalkowski2021} studied a Corbino-geometry device having four contacts spaced along the inner circumference of a Corbino disk, and four contacts along the outside circumference. One contact on the outside edge was used to source a bias $V_S$, and the nearest contact on the inside edge served as the drain; voltage was measured at the other six contacts. The study found that, in the near-dissipationless regime, contacts on the outside edge sat near the source potential $V=V_S$ and contacts on the inside edge sat at the drain potential $V=0$. When dissipation was induced, the potential at all voltage contacts moved towards $V_S/2$. For some but not all of the contacts, the transition from $V=V_S, 0$ to $V=V_S/2$ was nonmonotonic. These results were interpreted by the authors as evidence of spatial separation of edge and bulk current paths.

\begin{figure}[h!]
\centering
	\includegraphics[width=0.98\textwidth]{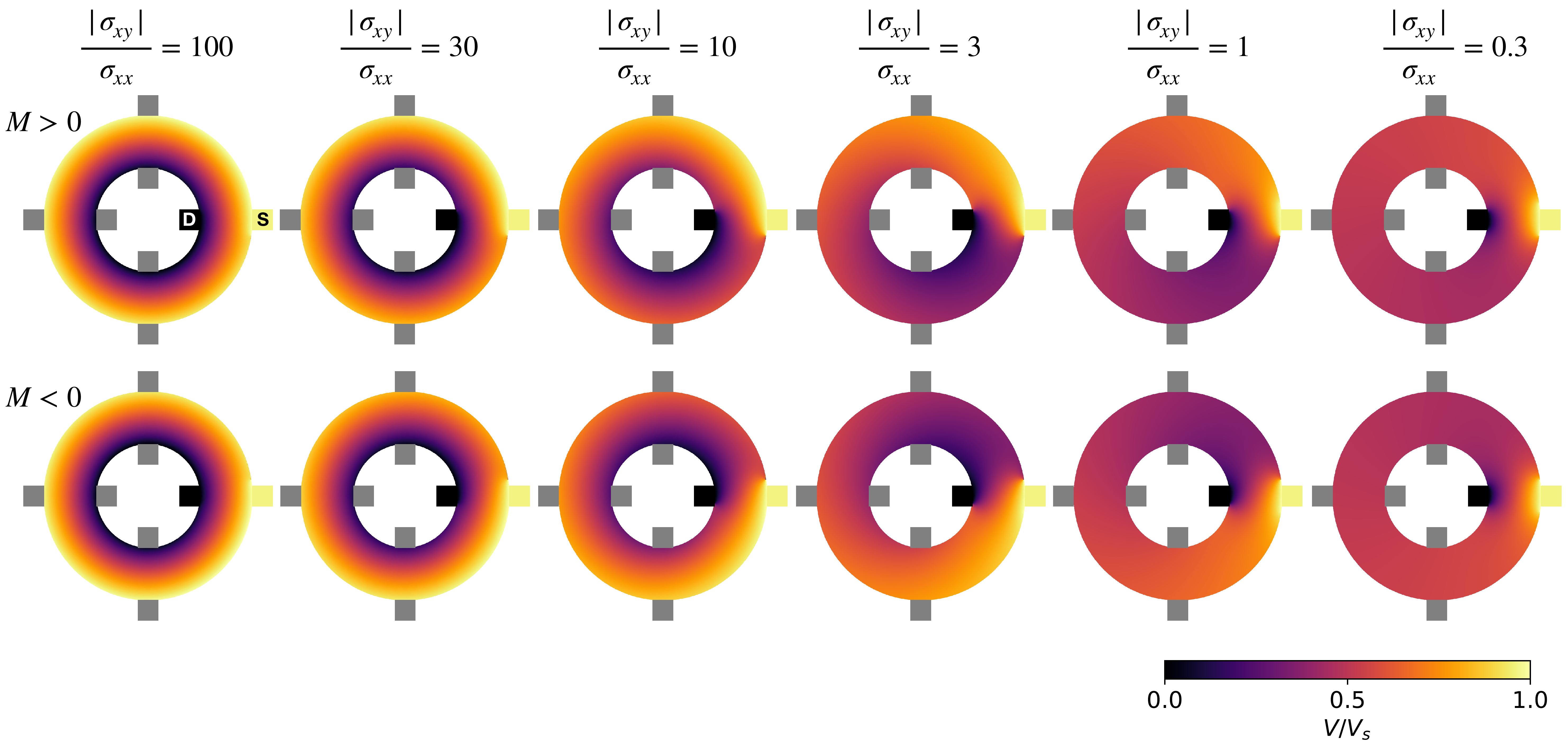}
	\caption{Simulations of multi-terminal Corbino devices. A bias is sourced at the outer right terminal (black, labeled ``S"), and drained at the inner right terminal (yellow, labeled ``D"). The six voltage contacts are shown (grey). The potential normalized to the source potential is plotted. The top (bottom) row of figures corresponds to positive (negative) magnetization. The columns correspond to different values of $\sigma_{xx}/\sigma_{xy}$, as indicated. The inner radius is half the outer radius.}
	\label{sfig_mcorb_potentials}
\end{figure}

Our simulations of the Laplace equation reproduce the salient features of this study, even though no edge modes are present in our simulations. The potential throughout a simulated device is shown in Fig.~\ref{sfig_mcorb_potentials} for various values of $\sigma_{xx}/\sigma_{xy}$. In the QAH regime $\sigma_{xx}/\sigma_{xy}\ll 1$, the outer (inner) edge tends approaches the potential of the source (drain) contact. The potential profile in this regime smoothly interpolates to that in the strongly dissipative regime $\sigma_{xx}/\sigma_{xy}\gtrsim 1$, where the potential throughout the device tends toward $V_S/2$ (excluding, of course, near the source and drain).

\begin{figure}[h!]
\centering
	\includegraphics[width=0.95\textwidth]{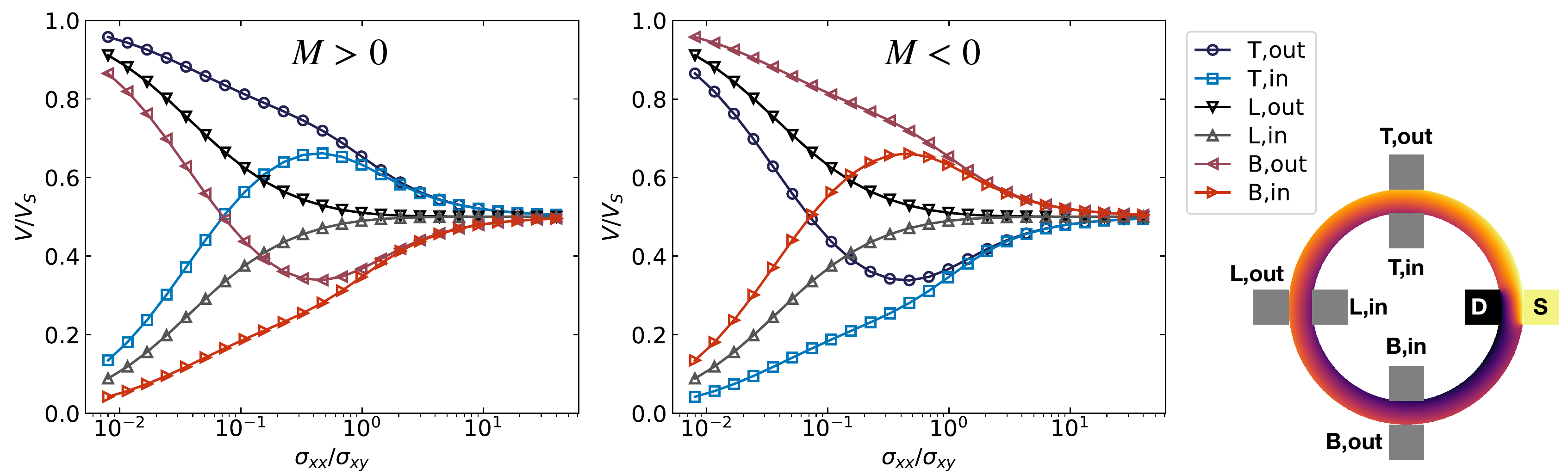}
	\caption{Simulated measured potential of multi-terminal Corbino devices, as a function of $\sigma_{xx}/\sigma_{xy}$ and normalized to the source potential. The ratio of inner radius to outer radius, as in Ref.~\cite{fijalkowski2021}, is 0.8. Data with positive (negative) magnetization are shown at left (right). The voltage terminals are labeled by combinations of T(op), L(eft), B(ottom), in(ner edge), and out(er edge), as shown by the device schematic at far right.}
	\label{sfig_mcorb_traces}
\end{figure}

Simulated measurements at the six voltage terminals are shown in Fig.~\ref{sfig_mcorb_traces}. The voltage terminals are labeled by their quadrant and by the edge of the device they contact. A symmetry is intact: $V_\mathrm{T,out} = V_S - V_\mathrm{B,in}$, $V_\mathrm{L,out} = V_S - V_\mathrm{L,in}$, and $V_\mathrm{B,out} = V_S - V_\mathrm{T,in}$. This symmetry is in essence equivalent to the centrosymmetry found in Hall geometry devices noted in the main text. Its consequence is that, in this multi-terminal Corbino geometry, there are only three independent values out of the six voltage measurements. Note that this symmetry is approximate due to the finite inner radius, and becomes exact in the limit of small $\sigma_{xx}/\sigma_{xy}$ as well as the limit of large inner radius compared to annulus width.

%remove from here if we're not using their data directly

\begin{figure}[h!]
\centering
	\includegraphics[width=0.95\textwidth]{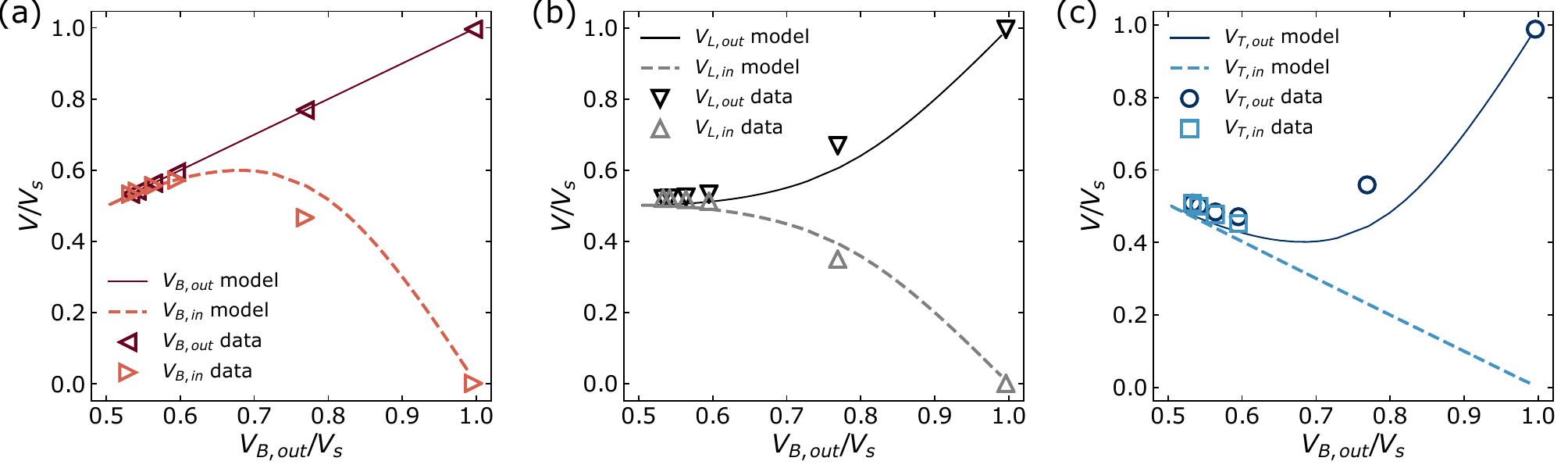}
	\caption{Comparisons between simulated (Fig.~\ref{sfig_mcorb_potentials}) and measured\protect\footnotemark[1] potentials in multi-terminal Corbino geometries for negative magnetization. Measured data were reproduced from Figs. 2(a)-(f) of Ref.~\cite{fijalkowski2021}; data points shown here were acquired as temperature is increased at the optimum applied gate voltage $V_g = 5.5~V$. Both simulations and data are shown parametrically as a function of the (measured or simulated) potential at the bottom outer contact. Data are shown with markers, and simulations are shown as solid or dashed lines. (a) Measured and simulated potentials at the bottom outer (dark red) and inner (light red) contacts.  (b) Measured and simulated potentials at the left outer (black) and inner (grey) contacts.  (c) Measured and simulated potentials at the bottom outer (dark blue) and inner (light blue) contacts. $V_{T,in}$ was only measured at four points, instead of six points as for each of the other contacts~\cite{fijalkowski2021}.}.
	\label{sfig_molenkamp_comparisons}
\end{figure}

In Fig.~\ref{sfig_molenkamp_comparisons} we directly compare the results of our simulations to the data of Ref.~\cite{fijalkowski2021} for each contact shown in Fig.~\ref{sfig_mcorb_potentials}. The experimental data were measured at $V_g = 5.5$ V at six different temperatures, ranging from 25~mK to 14~K. To directly compare simulations to measurements, the potential at each contact is plotted parametrically as a function of the potential on the bottom outer contact (relative to the source potential, i.e. normalized to 1). Our simulations quantitatively reproduce the measurements. This remarkable agreement suggests that their data are further evidence for bulk-dominated non-equilibrium current transport, rather than support for CEM-mediated current flow. In contrast with the Landauer-B\"uttiker-based model presented in Ref.~\cite{fijalkowski2021}, which utilizes three separate fit parameters for each temperature and gate voltage combination, our model has zero fit parameters (only the ratio $\sigma_{xx}/\sigma_{xy}$ is varied). This result highlights the predictive power of our simple model. 

\begin{figure}[h!]
\centering
	\includegraphics[width=0.95\textwidth]{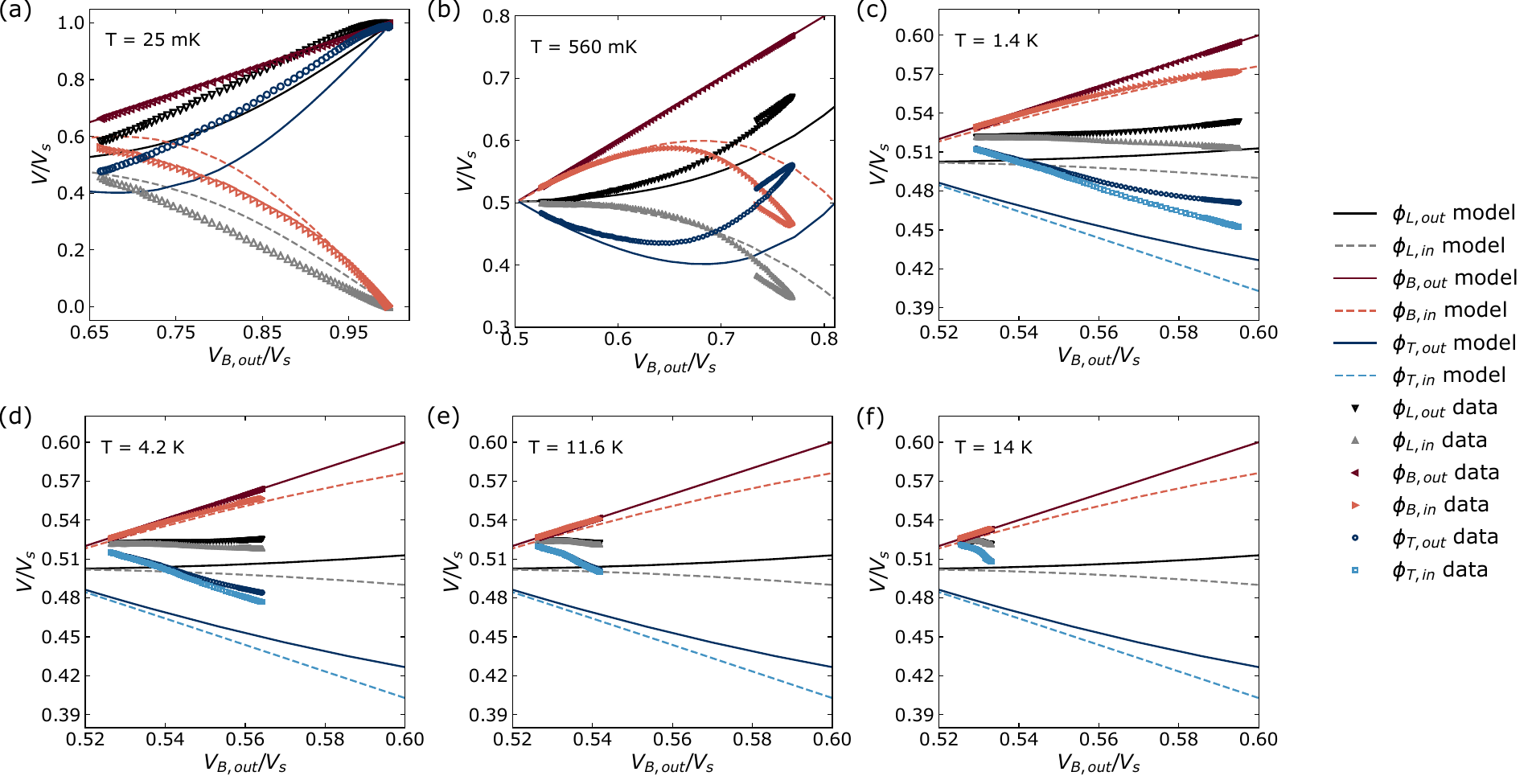}
	\caption{Comparisons between simulated (Fig.~\ref{sfig_mcorb_potentials}) and measured\protect \footnote[1]{The measured data points used in this figure are attributed to ``Quantum anomalous Hall edge channels survive up to the Curie temperature'' by K. M. Fijalkowski, {\em et al}., licensed under a  Creative Commons Attribution 4.0 International License. To view a copy of the original work, visit \url{https://doi.org/10.1038/s41467-021-25912-w}. To view a copy of the  license, visit \url{http://creativecommons.org/licenses/by/4.0/}} potentials in multi-terminal Corbino geometries for negative magnetization. Measured data were reproduced from Figs. 2(a)-(f) of Ref.~\cite{fijalkowski2021}. Each subplot shows the normalized potential at each contact for each gate voltage ($-9~V < V_g < 9~V$) at a specific temperature. Both simulations and data are shown parametrically as a function of the (measured or simulated) potential at the bottom outer contact. Data are shown with markers, and simulations are shown as solid or dashed lines. Left outer contacts are shown in black. Left inner contacts are shown in grey. Bottom inner (outer) contacts are shown in (dark) red. Top inner (outer) contacts are shown in (dark) blue.(a) T=25 mK. $V_{T,in}$ was not measured at this temperature, and the corresponding simulation is not plotted. (b) T=560 mK.  $V_{T,in}$ was not measured at this temperature, and the corresponding simulation is not plotted. (c) T=1.4 K. (d) T=4.2 K. (e) T=11.6 K. (f) T=14 K. }.
	\label{sfig_molenkamp_gatetemp}
\end{figure}

We additionally compare our simulations to the data of Ref.~\cite{fijalkowski2021} at fixed temperatures as the gate voltage is varied (Fig.~\ref{sfig_molenkamp_gatetemp}). As observed in our own measurements (Fig.~\ref{sfig_gate_base}), simulations agree qualitatively with measurements. Measurements are ordered from highest to lowest potential in the same order as the simulations, and both have the same functional form. Because Fig.~\ref{sfig_molenkamp_gatetemp} is plotted parametrically as a function of the potential on the bottom outer contact, an offset in the measured data is apparent; the measurements collapse to  $V/V_{s}>0.5$, while the simulations approach $V/V_{s}=0.5$. The authors of Ref.~\cite{fijalkowski2021} attribute this difference to varied contact resistances between the source and drain terminals. Sharp kinks in the measured data (most obvious in Fig.~\ref{sfig_molenkamp_gatetemp}(b)) occur where $V_g = V_{opt}$. Similar behavior was observed and discussed in our own devices (Fig.~\ref{sfig_above_below}). 

\clearpage
\section{Simulation details}

Here, we have simulated the Laplace equation $\nabla^2 V=0$ in two dimensions with boundary conditions discussed in the main text. The potential in the classical Hall effect is described by the Poisson equation $\nabla^2 V=\rho$ with the same boundary conditions, where $\rho$ is the accumulated charge density. Because the QAH system is a bulk insulator, we expect $\rho=0$.

We simulated only the primary rectangular area of the Hall bar devices and did not include the voltage taps in simulations. Simulations for Device~1 uses a square mesh with 481~points in the $x$ direction and 201~points in the $y$ direction. To accurately reflect device measurements, the voltages at the terminals are taken to be the average of the voltages at mesh points along the edge of the Hall bar that are within the 6~$\mu$m width of the device's voltage terminal channels~\cite{van1988,delahaye2003}.

Simulations use second-order finite difference operators to numerically approximate derivatives over a rectangular mesh. The current flowing through the Hall bar is determined by using finite difference operators to compute the electric field along a transverse line across the Hall bar at its midpoint, and then numerically integrating $j_x = \sigma_{xx}E_x + \sigma_{xy}E_y$ along this line to find the total longitudinal current.

Simulation of the multi-terminal Corbino device uses the polar coordinate system.

The QAH effect is known to have the same renormalization group properties as the QH effect and follows a semicircular trajectory in conductivity space~\cite{checkelsky2014,dolan1999}. Conversion between the parameter $\sigma_{xx}/\sigma_{xy}$ and the intrinsic conductivity of the device assumes this semicircular trajectory:
\begin{equation}
    \sigma_{xx}^2 + \left( \sigma_{xy}-\frac{e^2}{2h}\right)^2 = \left(\frac{e^2}{2h}\right)^2.
\end{equation}
Conversion between intrinsic conductivity and intrinsic resistivity uses the relations
\begin{equation}
    \sigma_{xx, xy} = \frac{\rho_{xx, xy}}{\rho_{xx}^2 + \rho_{xy}^2},
\end{equation}
\begin{equation}
    \rho_{xx, xy} = \frac{\sigma_{xx, xy}}{\sigma_{xx}^2 + \sigma_{xy}^2}.
\end{equation}
These conversions were used to convert the parameter $\sigma_{xx}/\sigma_{xy}$ to values of the microscopic resistivity $\rho_{xx}$ when plotting simulation results in Figs.~2(c) and 3(b) of the main text. Such conversion is not needed for the parametric visualization of Fig.~3(a).

\end{document}